\documentclass[
	reprint,
	superscriptaddress,
	showkeys,
	aps,
	pra,
	longbibliography,
	nofootinbib
]{revtex4-2}

\usepackage[utf8]{inputenc}

\usepackage[hyperref]{xcolor}
	\definecolor{goethe-blau}{cmyk}{1.0,0.2,0.0,0.4}
	\definecolor{hellgrau}{cmyk}{0.04,0.04,0.05,0.02}
	\definecolor{sandgrau}{cmyk}{0.12,0.09,0.13,0.0}
	\definecolor{dunkelgrau}{cmyk}{0.25,0.25,0.30,0.75}
	\definecolor{emo-rot}{cmyk}{0.04,1.0,0.8,0.07}
	\definecolor{purple}{cmyk}{0.08,1.0,0.3,0.36}
	\definecolor{senfgelb}{cmyk}{0.01,0.25,1.0,0.05}
	\definecolor{gruen}{cmyk}{0.62,0.4,0.87,0.09}
	\definecolor{magenta}{cmyk}{0.08,0.86,0.12,0.12}
	\definecolor{orange}{cmyk}{0.0,0.7,1.0,0.04}
	\definecolor{sonnengelb}{cmyk}{0.0,0.12,0.95,0.0}
	\definecolor{helles-gruen}{cmyk}{0.4,0.17,0.81,0.07}
	\definecolor{lichtblau}{cmyk}{0.8,0.0,0.06,0.04}

\usepackage[
	colorlinks,
	pdfpagelabels,
	breaklinks,
	pdfstartview=FitH,
	bookmarksopen=true,
	bookmarksnumbered=true,
	bookmarksopenlevel=2,
	plainpages=false,
	hypertexnames=false,
	citecolor=emo-rot,
	linkcolor=goethe-blau,
	urlcolor=purple,
	pdftitle={Bosonic fluctuations in the (1 + 1)-dimensional Gross-Neveu(-Yukawa) model at varying mu and T and finite N},
	pdfauthor={Jonas Stoll, Niklas Zorbach, Adrian Koenigstein, Martin J. Steil and Stefan Rechenberger},
	pdfkeywords={Gross-Neveu model, Gross-Neveu-Yukawa model, bosonic fluctuations, phase diagram, Z2 symmetry, chiral symmetry, symmetry breaking/restoration, finite N, FRG, numerical fluid dynamics}
]{hyperref}

\usepackage{amsmath} % AMS mathematical facilities for LaTeX
\usepackage{amssymb} % Additional math symbols
\usepackage{bm} % Access bold symbols in maths mode -- \mathbb{R}, \mathbb{Z} etc.
\usepackage{slashed} % Put a slash through characters -- Used for ‘Feynman slashed character’ notation
\usepackage{tensor} % Typeset tensors
\usepackage{bookmark} % A new bookmark (outline) organization for hyperref
\usepackage{graphicx} % Enhanced support for graphics -- include figure files
\usepackage{placeins} % Control float placement -- defines \FloatBarrier

\usepackage{booktabs} % Publication quality tables in LaTeX
\usepackage{dcolumn} % Align on the decimal point of numbers in tabular columns -- defines \extrarowheight (among other things)

\allowdisplaybreaks % Allow breaking of formulas across multiple pages

\begin{document}

%\preprint{}

\title{
	Bosonic fluctuations in the \texorpdfstring{$( 1 + 1 )$}{(1 + 1)}-dimensional Gross-Neveu(-Yukawa) model\texorpdfstring{\\}{ }at varying \texorpdfstring{$\mu$}{mu} and \texorpdfstring{$T$}{T} and finite \texorpdfstring{$N$}{N}
}

\thanks{
	J.~Stoll, N.~Zorbach, A.~Koenigstein, and M.~J.~Steil contributed in equal shares to this work. S.~Rechenberger was involved in early stages of this project before leaving academia and contributed to the final draft.
}

\author{Jonas Stoll}
	\email{stoll@itp.uni-frankfurt.de}		
	\affiliation{
		Institut für Theoretische Physik, Johann Wolfgang Goethe-Universität,\\
		Max-von-Laue-Straße 1, D-60438 Frankfurt am Main, Germany
	}

\author{Niklas Zorbach}
	\email{nzorbach@itp.uni-frankfurt.de}
	\affiliation{
		Institut für Theoretische Physik, Johann Wolfgang Goethe-Universität,\\
		Max-von-Laue-Straße 1, D-60438 Frankfurt am Main, Germany
	}

\author{Adrian Koenigstein}
	\email{koenigstein@itp.uni-frankfurt.de}
	\affiliation{
		Institut für Theoretische Physik, Johann Wolfgang Goethe-Universität,\\
		Max-von-Laue-Straße 1, D-60438 Frankfurt am Main, Germany
	}

\author{Martin J. Steil}
	\email{msteil@theorie.ikp.physik.tu-darmstadt.de}
	\affiliation{
		Technische Universität Darmstadt, Department of Physics, Institut für Kernphysik, Theoriezentrum,\\
		Schlossgartenstraße 2, D-64289 Darmstadt, Germany
	}
	
\author{Stefan Rechenberger}
	\email{stefanrechenberger@yahoo.de}
	\affiliation{
		Institut für Theoretische Physik, Johann Wolfgang Goethe-Universität,\\
		Max-von-Laue-Straße 1, D-60438 Frankfurt am Main, Germany
	}

\date{\today}

\begin{abstract}
	Using analogies between flow equations from the Functional Renormalization Group and flow equations from (numerical) fluid dynamics we investigate the effects of bosonic fluctuations in a bosonized Gross-Neveu model -- namely the Gross-Neveu-Yukawa model. We study this model for finite numbers of fermions at varying chemical potential and temperature in the local potential approximation. Thereby we numerically demonstrate that for any finite number of fermions and as long as the temperature is non-zero, there is no $\mathbb{Z}_2$ symmetry breaking for arbitrary chemical potentials.
\end{abstract}


\maketitle

\tableofcontents

\section{Introduction}
\label{sec:introduction}

	Since its original publication in 1974, the Gross-Neveu (GN) model \cite{Gross:1974jv} -- a relativistic QFT of $N$ massless Dirac fermions that are self-interacting via the scalar channel of the four-Fermi interaction -- was subject of intensive research \textit{w.r.t.}\ various aspects of strongly interacting systems. In $1 + 1$ space-time dimensions it was studied as a (perturbatively) renormalizable prototype model for asymptotic freedom \cite{Gross:1974jv,Wetzel:1984nw,Rosenstein:1990nm,Gracey:1990sx,Gracey:1990wi,Gracey:1991vy,Luperini:1991sv,ZinnJustin:1991yn,Peskin:1995ev,ZinnJustin:2002ru,Quinto:2021lqn}, while for $2 + 1$ space-time dimensions it served as a toy model for asymptotically safe QFTs \cite{Braun:2010tt} and is renormalizable with non-perurbative methods \cite{Rosenstein:1988pt,Rosenstein:1988dj,Rosa:2000ju,Hofling:2002hj}. In $3 + 1$ dimensions it mimics the dynamics of the important scalar channel of the Nambu-Jona-Lassinio model \cite{Nambu:1961fr,Nambu:1961tp} and is non-renormalizable, but can be seen as being part of a low-energy effective model of QCD, \textit{cf.}\ the reviews \cite{Klevansky:1992qe,Buballa:2003qv}. However, within this work and from now on, we exclusively focus on its original formulation in $1 + 1$ space-time dimensions, if not explicitly stated otherwise.\\
	
	 Furthermore, there are various connections to non-relativistic models in solid-state physics. Indeed, GN-type models can arise in the continuum limit (describing large distance physics) of one-dimensional solid-state systems. Examples of such one-dimensional systems are quantum antiferromagnets (described by spin-$\tfrac{1}{2}$ Heisenberg models \cite{Heisenberg:1928mqa}), interacting electrons in a one-dimensional conductor (described by a Tomonaga–Luttinger liquid \cite{Tomonaga:1950,Luttinger:1963zz}), and polyacetylene polymer chains \cite{Chodos:1993mf} -- ${(\mathrm{C}\mathrm{H})_x}$ --(described in the limit $x\rightarrow\infty$ with the Su-Schrieffer-Heeger model \cite{Su:1979ua} or the subsequent Takayama-Lin-Liu-Maki model \cite{Takayama:1980zz}). Recently, a one-dimensional probabilistic cellular automaton, where classical bits can be interpreted as Ising spins, was shown to be equivalent to a relativistic fermionic quantum field theory \cite{Wetterich:2021exk}. A central concept behind the emergence of fermionic QFTs in two dimensions from spin-systems in one dimension is the mapping of spin operators onto fermionic creation and annihilation operators by means of the Jordan–Wigner transformation \cite{Jordan:1928wi}, for a pedagogic discussion see Ref.~\cite{Fradkin:2013}. GN type models also arise naturally in the continuum limit of two-dimensional spin-systems like the Ashkin-Teller model \cite{Ashkin:1943}, which is related to the well-known Potts model \cite{Potts:1951rk} and as such used to study various phenomena of solid-state physics, see, \textit{e.g.}, Ref.~\cite{Wu:1982} for a general overview. In the following paragraphs we will list some explicit connections between the GN model and models used in solid-state physics.
	
	The lattice field theoretical formulation of the GN model in the limit $N\rightarrow\infty$ is equivalent to an Ising model \cite{Affleck:1981bn}.
	
	At finite $N$ the GN model can be considered as the continuum limit of the $N$-color ($N$ Ising spin) Ashkin-Teller model \cite{Ashkin:1943,Fradkin:1984,Shankar:1985zc}, which describes $N$ coupled Ising spins on a two-dimensional lattice \cite{Shankar:1985zc}.
	
	For $N=1$ the GN model is equivalent to the Thirring model \cite{Thirring:1958in,Witten:1978qu} due to Fierz identities, \textit{cf.}, Refs.~\cite{Peskin:1995ev,ZinnJustin:2002ru,Fradkin:2013}. In the continuum limit the one-dimensional spin-$\tfrac{1}{2}$ Heisenberg model is equivalent to a $N = 1$ GN (Thirring) model \cite{Fradkin:2013}. The Thirring model also arises in the infinite volume limit of the Luttinger model \cite{Luttinger:1963zz} with strictly local interactions \cite{Heidenreich:1980}, see Ref.~\cite{Fradkin:2013} for further details. The massive Thirring model is equivalent to the sine-Gordon model \cite{Coleman:1974bu,Delepine:1997bz}, which in turn is (among its other application in mathematical physics) the continuum-limit of the Frenkel–Kontorova model \cite{Frenkel:431595}. The latter is a simple model of a harmonic chain in a periodic potential known from solid-sate physics \cite{Kivshar:2000}. The equivalence of the Thirring model and sine-Gordon model is based on an Abelian bosonization transformation, see Refs.~\cite{Fradkin:2013,Delepine:1997bz} and references therein, which connects equivalent bosonic and fermionic two-dimensional quantum field theories.
	
	Apart form research efforts in high energy and solid-state physics the GN model is also of interest in the context of holographic methods \cite{Maldacena:1997re,Witten:1998qj} especially in the study of the AdS/CFT correspondence involving higher spin fields, see Ref.~\cite{Giombi:2016ejx} for a recent review.\\
	
	Despite this broad range of applications and interconnections as well as intensive research, there are still open questions. Some of these are addressed within this work.

\subsection{Phenomenology of the Gross-Neveu model}\label{subsec:GN_pheno}
	
	A peculiar feature of the massless GN model is, that at leading order of an $\tfrac{1}{N}$-expansion, thus in the infinite-$N$ limit (sometimes also 't~Hooft limit), the GN model dynamically develops a mass gap for the fermions, which is associated with an anti-fermion-fermion condensate $\langle \bar{\psi} \, \psi \rangle \neq 0$. In turn, this results in the breakdown of the discrete chiral symmetry of the initial microscopic ultraviolet (UV) theory. The formation of a mass gap is a purely non-perturbative effect, see, \textit{e.g.}, the Refs.~\cite{ZinnJustin:1991yn,ZinnJustin:2002ru} for details, and is a prime example for \textit{dimensional transmutation} -- the emergence of a dimensionful scale in a theory which has only dimensionless couplings in its UV classical action, see, \textit{e.g.}, Refs.~\cite{Kleinert:2016,Peskin:1995ev,ZinnJustin:2002ru}. Hence, by summing up loop-contributions of all orders in the four-Fermi coupling in a $\tfrac{1}{N}$-expansion the discrete chiral symmetry spontaneously breaks down and is absent in the macroscopic theory in the infrared (IR). In the partially bosonized version, this amounts to integrating out the fermion-loop contribution to the bosonic effective potential, which develops a non-trivial minimum in the IR -- the condensate \cite{ZinnJustin:1991yn,ZinnJustin:2002ru,Rosenstein:1990nm,Luperini:1991sv,Quinto:2021lqn}.\\

	Shortly after D.~J.~Gross and A.~Neveu had published their results \cite{Gross:1974jv}, the question came up, to what extent condensate formation is stable against thermal effects due to non-zero temperature $T$ or an increase in density, induced by a non-zero quark chemical potential $\mu$. Within the infinite-$N$ limit and allowing only for spatially homogeneous condensates, the answer to these questions was quickly settled and is remarkable \cite{Dolan:1973qd,Harrington:1974tf,Jacobs:1974ys,Dashen:1974xz,Dashen:1975xh,Wolff:1985av,Treml:1989,Pausch:1991ee,Karbstein:2006er}: The phase diagram of the GN model in the $\mu$-$T$-plane is actually unique. It consists of a region, where the discrete chiral symmetry is broken and a region of vanishing chiral condensate, \textit{cf.}\ Fig.~\ref{fig:GNlargeN_PD}. The phase-transition line between these regions splits up into a second-order phase transition (starting at $\mu = 0$ and some critical temperature $T_\mathrm{C} \neq 0$ and ending in a critical point $\mu_\mathrm{CP} \neq 0$ and $T_\mathrm{CP} \neq 0$) and a first order phase transition (starting at the critical point and ending on the $T = 0$ axis and some non-zero chemical potential $\mu_1$). Here, being ``unique'' means that the entire phase diagram solely depends on a single dimensionful parameter, which is related to a renormalization condition, and can be chosen freely. All other dimensionful quantities are fixed multiples of this parameter and choosing a different renormalization condition (fixing some other parameter) corresponds to simple rescalings of all dimensionful quantities, but does not change their ratios, the phenomenology and the phase diagram.\\
	
	Notwithstanding these successes, the discussion on the physics of the GN model did not stop. One of the assumptions, which has lead to the above results, is the assumption of a spatially homogeneous condensation of the fermions. Relaxing this assumption, but retaining the $N \rightarrow \infty$ limit, it was shown in Refs.~\cite{Thies:2003kk,Schnetz:2004vr,deForcrand:2006zz} that there are regions in the $\mu$-$T$-plane, where a spatially inhomogeneous but static condensate $\langle \bar{\psi} \, \psi \rangle ( x ) \neq \mathrm{const}.$\ is energetically favored over homogeneous condensation including a vanishing of the condensate. These so-called inhomogeneous phases attracted a lot of attention within the last decades in the high energy physics community and seem to be a rather robust feature of a lot of effective models of strongly correlated systems at least in the infinite-$N$ limit \cite{Buballa:2014tba,Braun:2014fga,Heinz:2015lua}. We are not going to focus and discuss the possibility for spatially inhomogeneous condensation in this work, but instead concentrate on the other main assumption, which underlies the above discussions -- namely the infinite-$N$ limit. This limit for the GN model leads to a complete suppression of bosonic quantum fluctuations.
	
\subsection{To break or not to break -- \texorpdfstring{$\mathbb{Z}_2$}{Z2} symmetry at finite \texorpdfstring{$N$}{N}}\label{subsec:z2breaking}
	
	The big question that immediately comes to mind if one relaxes the infinite-$N$ limit and studies the GN model for finite $N$ is, if spontaneous symmetry breaking (SSB) and condensation still takes place, especially when medium effects (non-zero $\mu$ and/or $T$) are included.
	
	At first sight, it seems as if already the Coleman-Mermin-Wagner-Hohenberg theorem \cite{Mermin:1966,Hohenberg:1967,Coleman:1973ci} forbids the formation of a condensate at non-zero temperatures. Though, the theorem strongly relies on the presence of massless Nambu-Goldstone bosons \cite{Nambu:1960tm,Goldstone:1961eq,Goldstone:1962es}, which are only included in extensions of the GN model, \textit{e.g.}, the chiral GN model \cite{Schon:2000qy,Basar:2009fg,Furuya:1982fh} with continuous chiral symmetry or other related models \cite{Witten:1978qu,Rosenstein:1990nm}. We therefore believe that one should exercise caution, when arguing directly with this theorem.
	
	Nevertheless, a qualitative argumentation was put forward already by L.~D.~Landau in 1950 \cite{Landau:1980mil}, which should in principle forbid also discrete chiral symmetry breaking in one-spatial dimension at non-zero temperature. L.~D.~Landau and E.~M.~Lifshitz argue in Chap.~163 of Ref.~\cite{Landau:1980mil} that for systems in one dimension of infinite extend only one phase can exist at $T>0$, since coexistence of more than one phase is energetically disfavored. To be concrete, they considered a bistable system in one dimension of infinite extend at non-zero temperature with $n$ interfaces between the two possible phases per length $L$ and showed that the thermodynamic potential of the system can be decreased by increasing the concentration of interfaces $n/L$, which is directly related to the entropy, assuming a finite interface energy. Thus the system breaks down into a macroscopic number of domains, which rendered macroscopic phase coexistence at non-zero temperature impossible. Landau's argument can be applied to a broad range of effectively one-dimensional systems including the Ising model in one spatial dimension \cite{Ising:1925em} at $T \neq 0$, which is always in its $\mathbb{Z}_2$ symmetric phase, \textit{cf.}\ Refs.~\cite{ZinnJustin:2002ru,Rosenstein:1990nm}. A dedicated discussion of Landau's argument in the context of one-dimensional systems can be found in Ref.~\cite{Theodorakopoulos:2006}.
	
	 In their study \cite{Dashen:1974xz} of the GN model based on a large-$N$ expansion R.~F.~Dashen, S.~Ma, and R.~Rajaraman were able to confirm Landau's argument, see also Ref.~\cite{Barducci:1994cb}. They found no SSB for any small but non-zero temperature and finite $N$. Using in parts heuristic arguments, they showed that the entropic gain of a field configuration of alternating kinks is large enough at finite $N$ and $1/T$ to be energetically preferred over a homogeneous configuration\footnote{Homogeneous and inhomogeneous field configurations in this context refer to the configurations used to evaluate the partition function in a saddle point approximation, see Ref.~\cite{Dashen:1974xz} for details on their computation, and they are not to be confused with homogeneous and inhomogeneous classical field configurations $\langle \bar\psi\psi\rangle(x)$ discussed in the previous Sub.Sec.~\ref{subsec:GN_pheno}.}. Those field configurations alternating in kinks have a vanishing chiral condensate $\langle \bar{\psi} \, \psi \rangle = 0$. In the infinite-$N$-limit (mean-field), the energy per kink becomes infinite and consequently the density of kinks approaches zero realizing a homogeneous field configuration compatible with $\langle \bar{\psi} \, \psi \rangle > 0$ allowing for SSB. This is no contradiction to Landau's argument since the latter only holds assuming finite interface energies \cite{Landau:1980mil,Theodorakopoulos:2006}. Although R.~F.~Dashen, S.~Ma, and R.~Rajaraman argue that $T_\mathrm{C} = 0$ at finite $N$, they do not discuss the situation for $T = 0$ and $\mu \geq 0$ for finite $N$ in Ref.~\cite{Dashen:1974xz}.
	
	Another discussion on the absence of symmetry breaking in one-spatial dimension at non-zero $T$ can be found in Ref.~\cite{Witten:1978qu} by E.~Witten, which discusses the absence of a phase with spontaneous symmetry breaking and long-range order in accordance with the Coleman-Mermin-Wagner-Hohenberg theorem \cite{Mermin:1966,Hohenberg:1967,Coleman:1973ci} and the related possibility for a phase of Berezinski-Kosterlitz-Thouless type \cite{Berezinsky:1970fr,Kosterlitz:1973xp} with quasi long-range order for the $\mathrm{SU}(N)$ Thirring model \cite{Thirring:1958in,Witten:1978qu}.
	
	Other authors, \textit{e.g.}, U.~Wolff \cite{Wolff:1985av}, argue based on the duality of the spatial and Euclidean time direction in $1+1$ dimensions for $T>0$ and $\mu=0$: The thermal GN model in $1+1$ dimensions is equivalent to a QFT with a finite spatial volume (but infinite Euclidean time direction) and hence no SSB takes place since it is canonically considered as an effect only present in systems with infinite volumes, see, \textit{e.g.}, Ref.~\cite{Weinberg:1996kr}. While this reasoning -- the general absence of SSB in finite systems -- might strictly speaking be sound, sufficiently large volumes, the inclusion of small (possibly infinitesimal) explicit symmetry breaking, or subtleties of the thermodynamic and/or infinite volume limit, \textit{cf.}\ Ref.~\cite{Landsman:2013aoa}, can lead to signatures reminiscent of SSB.
	
	Indeed, only recently some of our colleagues and collaborators found some indications via numerical lattice Monte-Carlo simulations \cite{Pannullo:2019bfn,Pannullo:2019prx,Lenz:2020bxk,Lenz:2020cuv}, that some (inhomogeneous) condensation phenomena in the massless bosonized GN model at finite $N$ and non-zero $\mu$ and $T$ still seem to be present. Similar results were already found in earlier lattice Monte-Carlo studies \cite{Cohen:1981qz,Cohen:1983nr,Karsch:1986hm} at finite $N$. However, in all of these works, the above arguments by Landau \textit{et al.}\ against condensation at non-zero $T$ could neither be completely ruled out nor be confirmed. In fact, most of the results suffer from the facts that proper continuum and infinite volume extrapolations were not performed. Consequently finite volume effects and discretization artifacts limit the predictive power of those results for the continuum theory in an infinite volume. The finite sized spatial domain (and the related boundary conditions) might have prevented a sufficient resolution of long-range fluctuations, which are however of uttermost importance for condensation and in this context especially vaporization phenomena in low-dimensional systems.
	
	Recent lattice results presented in Refs.~\cite{Pannullo:2019bfn,Pannullo:2019prx,Lenz:2020bxk,Lenz:2020cuv} have sparked further lattice studies of four-Fermi models in $1 + 1$ dimensions: We are aware of these parallel developments and computations using lattice Monte Carlo simulations in the GN model and related models (especially the chiral GN model) in Refs.~\cite{Mandl2021,Nonaka2021}, which are however not completed yet and therefore omitted in the following discussion. For the chiral GN model, we expect some interesting dynamics at finite $N$ due to competing effects from the Coleman-Mermin-Wagner-Hohenberg theorem and the $U ( 1 )_\mathrm{A}$ anomaly \cite{Furuya:1982fh}.\\
	
	All of this lead us to the idea to study the phenomenon of $\mathbb{Z}_2$ symmetry breaking and/or restoration in the GN model at finite \texorpdfstring{$N$}{N}, $T\geq0$ and also $\mu\geq0$ but in an infinite spatial volume within a different framework -- namely within the Functional Renormalization Group (FRG).\footnote{After completing this work, we became aware of Ref.~\cite{Blaizot:2002nh}, where next-to-leading order corrections of the $\tfrac{1}{N}$-expansion to the effective potential of the GN model were calculated. Finding that this expansion breaks down in the vicinity of mean-field critical temperature $T_\mathrm{C}$ the authors also suggest to analyze the GN model within the FRG framework, which is the main focus of this paper. We thank J.~Braun for drawing our attention to this interesting publication.} We wanted to find out, if it is possible to (numerically) confirm the arguments by Landau \textit{et al.}\ against symmetry breaking in the GN model or if there are some other competing effects, which are not captured in the aforementioned mostly qualitative/heuristic discussions, that allow for symmetry breaking or some long-range ordering.

\subsection{Main results}

	Using the FRG method in the local potential approximation (LPA) we find no spontaneous symmetry breaking in the Gross-Neveu-Yukawa (GNY) model, which is a bosonized version of the GN model that is used for practical computations, for any finite number of fermions, non-zero temperature and arbitrary chemical potential. Our results therefore support the qualitative arguments by Landau \textit{et al.}. While fermionic interactions dominate the dynamics during the RG-flow at high momenta and lead to the formation of a condensate -- similar to the situation at infinite-$N$ and finite $N$ lattice Monte-Carlo studies -- long-range bosonic quantum fluctuations restore the discrete chiral symmetry in the deep infrared regime -- at low momenta -- of the RG flow independent of the number of fermions and the value of the chemical potential for all non-zero temperatures.\footnote{This dynamics is analogous to the phenomenon of \textit{precondensation}, which was first discussed in Refs.~\cite{Boettcher:2012cm,Boettcher:2012dh,Boettcher:2013kia,Boettcher:2014tfa,Roscher:2015xha,Khan:2015puu} in the context of RG flows.} This is in accordance with the predictions of R.~F.~Dashen, S.~Ma, and R.~Rajaraman put forward in Ref.~\cite{Dashen:1974xz}. The result further supports the similarities between the one-dimensional Ising model \cite{Ising:1925em} and the GN model, because we find that the critical temperature for a phase transition to the symmetry broken phase is $T_\mathrm{C} = 0$ for all $\mu$ and finite $N$. The symmetry restoration deep in the IR might present a substantial challenge when trying to resolve infinite volume physics with simulations or computations in finite volumes.
	
	Computations in the vacuum ($T = \mu = 0$) limit indicate SSB even at finite $N$. Direct computations at $T=0$ and $\mu>0$ could not be performed at finite $N$ within this work but computations at small $T$ and $\mu>0$ indicate, after extrapolation to $T=0$ and under consideration of the vacuum results, a possible quantum phase transition at non-zero chemical potential between a phase of spontaneous broken symmetry at low chemical potentials and a restored phase at high chemical potentials at $T=0$ and finite $N$.\\

	Furthermore, as a consistency check, we show analytically that the mean-field approximation of the LPA flow equation in the FRG framework, \textit{i.e.} neglecting bosonic fluctuations in the limit $N \rightarrow \infty$, is equivalent to the canonical infinite-$N$ results in literature \textit{cf.} Refs.~\cite{Wolff:1985av,Thies:2006ti}. We use this fact to check our numerics by reproducing the infinite-$N$ phase diagram.\\

	To obtain numerical results at finite $N$, we use a novel reformulation of the LPA flow equation in terms of a conservation equation and solve it by using a finite-volume-method\footnote{We emphasize that finite-volume refers here to a discretization scheme for partial differential equations and not to a finite volume of the physical spatial direction of the GN model.}. In this setup fermions act like a sink/source term and introduce a moving discontinuity at zero (low) temperatures and non-zero chemical potentials, which makes it (practically) impossible to numerically calculate to arbitrary low RG-scales. This might indicate a fundamental conceptual problem and prevents practical numerical computations at $T = 0$ and $\mu > 0$ in this work.\\

\subsection{Structure}

	We tried to keep the discussion in this work as self-contained as possible. This should enable readers not familiar with all aspects of this work to follow the entire line of argumentation without resorting to the literature. This goal is the reason behind the substantial scope of this paper. In this subsection we provide an overview of the structure of this work.\\
	
	In Sec.~\ref{sec:GNYmodel} we introduce the GN model and its bosonized counterpart the GNY model in vacuum and medium as well as their differences. Readers familiar with the model may just skim over this section. In Sec.~\ref{sec:gny_frg} we give an introduction to and provide details on the FRG approach to the GN(Y) model. Readers familiar with the FRG can skip Sub.~Sec.~\ref{subsec:the_frg} and \ref{subsec:dSB} but might be interested in the remainder of the section. Readers who are not interested in the technical details of the FRG formalism in the context of this work may omit Sub.Secs.~\ref{subsec:comment_on_the_regulators} and \ref{subsec:comment_on_the_truncation} from their first reading. In Sec.~\ref{sec:frg_and_fluid_dynamics} we discuss the relation between RG flow equations and (numerical) fluid dynamics and our implementation. Readers who are not interested in technical details of the implementation or who are familiar with computational fluid dynamics can skip Sub.Sec.~\ref{subsec:boundary_conditions} and \ref{subsec:numerical_implementation}. Sub.Sec.~\ref{subsec:chemical_potential_shock_wave} mainly addresses readers who work with the FRG and can be bypassed by everyone else. The other parts of this section are central for this work. In Sec.~\ref{sec:mean-field} we recapitulate some results for the GN model in the infinite-$N$ limit. We further discuss initial conditions for the RG flows at finite $N$ and show consistency checks of our numeric implementation and choice of UV parameters. Readers familiar with the mean-field phenomenology and phase structure of the GN model may just skim over Sub.Sec.~\ref{subsec:mean-field_infinite_n_frg} - \ref{subsec:phase_diagram_mean_field}. In Sec.~\ref{sec:rg-flow-with-bosons} we discuss the explicit RG flows and the phase structure of the GNY model at finite $N$. This section comprises our main results. We conclude our work in Sec.~\ref{sec:conclusion_outlook} and provide an outlook to possible future research. Several sections are accompanied by appendices, which are supposed to facilitate a self-contained reading and give computational and conceptual details omitted in the main part of this paper.

\section{The Gross-Neveu(-Yukawa) model}
\label{sec:GNYmodel}
	
	In this section, we introduce the Gross-Neveu (GN) model, its bosonized counterpart and the Gross-Neveu-Yukawa (GNY) model in Euclidean space-time. We comment on its symmetries and its in-medium generalization for non-zero chemical potentials $\mu$ and temperatures $T$.

\subsection{In vacuum}

	The Gross-Neveu model (GN model) in one spatial and one temporal dimension in Euclidean space-time is defined via the classical action, \textit{cf.}\ Ref.~\cite{Gross:1974jv},
		\begin{align}
			\mathcal{S}_\mathrm{GN} [ \bar{\psi}, \psi ] = \int \mathrm{d}^2 x \, \big[ \bar{\psi} \, \slashed{\partial} \, \psi - \tfrac{g^2}{2 N} \, ( \bar{\psi} \, \psi )^2 \big] \, ,	\label{eq:gn-model}
		\end{align}
	where $\psi$ is an $N$-component object in flavor space ($f = 1, 2, \dots, N$ and $N > 1$)\footnote{We explicitly exclude $N = 1$, where the GN model is identical to the Thirring model \cite{Thirring:1958in,Witten:1978qu}, which has a vanishing perturbative one-loop $\beta$-function and different phenomenology than the GN model at $N>1$ \cite{Peskin:1995ev,ZinnJustin:2002ru}. Some details and References concerning the Thirring model can be found in Sec.~\ref{sec:introduction}.} and two-component spinor in Dirac space (see App.~\ref{app:conventions} for details on our conventions including conventions for Euclidean space-time and the connection to the original formulation \cite{Gross:1974jv} of the model in Minkowski space-time). The action involves a kinetic term and a four-fermion-interaction term with the dimensionless positive coupling constant $g^2$.
	
	Apart from Euclidean space-time symmetries (translations and a rotation), the action \eqref{eq:gn-model} is invariant under transformations of the symmetry group ${G = U(1) \times SU(N) \times \mathbb{Z}_2}$. The group acts on the fermion fields as follows 
		\begin{align}
			G \times \mathbb{C} ^{2 N} \to \, & \mathbb{C}^{2 N} \, ,	\vphantom{\bigg(\bigg)}	\label{eq:gn-group-action}
			\\
			( ( \alpha, \theta, n ), \psi ) \mapsto \, & \alpha \, ( \theta \, \gamma_\mathrm{R} + n \, \theta \, \gamma_\mathrm{L} ) \, \psi \, ,		\vphantom{\bigg(\bigg)}	\nonumber
		\end{align}
	where $n \in \mathbb{Z}_2 \equiv \{-1, +1\}$, $\gamma_{\mathrm{L}/\mathrm{R}} = \tfrac{1}{2} \, ( \openone \mp \gamma_{\mathrm{ch}} )$ are the left- and right-handed chiral projection operators, $\alpha \in U ( 1 )$ and $\theta \in SU ( N )$. The full group $G$ is defined as the direct product of the groups $U(1)$, $SU(N)$ and $\mathbb{Z}_2$. Hereby, the $SU(N)$ symmetry is usually called flavor or color symmetry and causes, according to Noether's theorem, the conservation of a vector current. The $\mathbb{Z}_2$ symmetry is called discrete chiral symmetry. The $U(1)$ symmetry is called phase symmetry and leads to a conserved Noether-charge density $\bar{\psi} \, \gamma^2 \, \psi$, which is usually called Baryon number density \cite{Fitzner:2010nv,Dunne:2011wu,Thies:2017fkr,Lenz:2020cuv}.\footnote{It is also worth mentioning, that the $SU(N)$ symmetry group of the GN model in $1 + 1$ space-time dimensions is an $O(N)$ symmetry, which is why it is sometimes also denoted as an $O(N)$ symmetric model \cite{Jacobs:1974ys,ZinnJustin:2002ru}. Furthermore the GN model in two dimensions has an additional hidden $O(2N)$ symmetry between Majorana components of the fermion fields \cite{Dashen:1975xh} which prevents the appearance of different four-fermion interaction channels during renormalization \cite{Luperini:1991sv}.} 
	
	It can be shown, see App.~\ref{app:hubbard-stratonovich}, that the partition function of the GN model is equivalent to the partition function of the bosonized GN (bGN) model,
		\begin{align}
			\mathcal{S}_\mathrm{bGN} [ \bar{\psi}, \psi, \phi ] = \int \mathrm{d}^2 x \, \big[ \bar{\psi} \, \big( \slashed{\partial} + \tfrac{h}{\sqrt{N}} \, \phi \big) \, \psi + \tfrac{h^2}{2 g^2} \, \phi^2 \big] \, ,	\label{eq:bgn-model}
		\end{align}
	where $\phi$ is a (real) scalar ``auxiliary'' or ``constraint'' field \cite{Harrington:1974tf,Jacobs:1974ys,Luperini:1991sv,Braun:2011pp}. Being equivalent means that one obtains the same correlation functions as for the purely fermionic action \eqref{eq:gn-model}, \textit{cf.}\ App.~\ref{app:hubbard-stratonovich}, the discussion at the beginning of Sub.Sec.~\ref{subsec:comment_on_the_truncation} or the textbooks \cite{ZinnJustin:2002ru,Peskin:1995ev}. 
	
	In the bosonized version \eqref{eq:bgn-model}, the four-Fermi interaction is replaced by a Yukawa interaction term with coupling constant $h$ as well as a quadratic (mass) term $h^2/g^2$ for the auxiliary field $\phi$. If we postulate
		\begin{align}
			G \times \mathbb{R} \to \, & \mathbb{R} \, ,	\vphantom{\bigg(\bigg)}
			\\
			( ( \alpha, \theta, n ), \phi ) \mapsto \, & n \phi \, ,	\vphantom{\bigg(\bigg)}	\nonumber
		\end{align}
	then the action \eqref{eq:bgn-model} is invariant under the same symmetry group $G$ as the original action \eqref{eq:gn-model}.

	Within this work, we are especially interested in the discrete chiral symmetry transformation, which we understand as the group element $(\alpha, \theta, n) = (1, \mathbb{I}, -1) \in G$, \textit{i.e.},
		\begin{align}
			&	\psi \mapsto \psi^\prime = \gamma_\mathrm{ch} \, \psi \, ,	&&	\phi \mapsto \phi^\prime = - \phi \, ,	\vphantom{\bigg(\bigg)}	\label{eq:discrete-chiral-transformation}
			\\
			&	\bar{\psi} \mapsto \bar{\psi}^\prime = - \bar{\psi} \, \gamma_\mathrm{ch} \, ,	&&	\vphantom{\bigg(\bigg)}	\nonumber
		\end{align}
	It is this symmetry which prevents the GN model from perturbatively generating a mass gap, see, \textit{e.g.}\ Ref.~\cite{Gross:1974jv,ZinnJustin:2002ru}.
	
	By including an additional kinetic term for the bosonic field we obtain the Gross-Neveu-Yukawa model \cite{ZinnJustin:1991yn,ZinnJustin:2002ru,Rosa:2000ju}
	\begin{align}
		& \mathcal{S} [ \bar{\psi}, \psi, \phi ] =	\vphantom{\bigg(\bigg)}	\label{eq:gny-model}
		\\
		= \, & \int \mathrm{d}^2 x \, \big[ \bar{\psi} \, \big( \slashed{\partial} + \tfrac{h}{\sqrt{N}} \, \phi \big) \, \psi - \phi \, ( \Box \phi ) + \tfrac{h^2}{2 g^2} \, \phi^2 \big] \, ,	\vphantom{\bigg(\bigg)}	\nonumber
	\end{align}
	which we will use for practical computations in the FRG framework. We elaborate on the specific model choice and differences between the GN (bGN) and GNY model in Sub.Sec.~\ref{subsec:comment_on_the_truncation}.

\subsection{In medium} 

	In this work, we are mainly interested in the in-medium properties of the GNY model. In order to work at non-zero Baryon density, we fix the net Baryon number density by introducing a quark chemical potential $\mu$. Analogously to standard statistical physics, see, \textit{e.g.},\ Refs.~\cite{Freedman:1976xs,Kapusta:1981aa,Actor:1986zf,Dashen:1974xz,Wolff:1985av} for details, we subtract
		\begin{align}
			\mu N \equiv \mu \int \mathrm{d}^2 x \, \bar{\psi} \, \gamma^2 \, \psi
		\end{align}
	from the classical (UV) action, respectively the probability distribution in the grand canonical partition function\footnote{Note that $\gamma^2$ is $\gamma$-matrix associated with the Euclidean temporal direction in our conventions.},
		\begin{align}
			\mathcal{Z} \propto \int [ \mathrm{d} \bar{\psi}, \mathrm{d} \psi, \mathrm{d} \phi ] \, \mathrm{e}^{- \mathcal{S} [\bar{\psi}, \psi, \phi ] + \mu N} \, .	\label{eq:partition_function}
		\end{align}
	Furthermore, we introduce non-zero temperature via a compactification of the Euclidean time direction\footnote{Geometrically and technically the introduction of non-zero temperature via compactification of the Euclidean temporal direction is identical to defining the entire GNY model on a cylinder of infinite length and finite circumference $\beta$ right from the beginning. For details, see App.~\ref{app:qft_on_a_cylinder}.}, \textit{i.e.}, we replace $\int \mathrm{d}^2 x \mapsto \int \mathrm{d} x \int_{0}^{\beta} \mathrm{d} \tau$, where $T = \tfrac{1}{\beta}$ is the temperature, and implement periodic and anti-periodic boundary conditions in the Euclidean time direction for bosons and fermions, respectively. In total, we obtain the in-medium or thermal GNY model,
		\begin{align}
			\mathcal{S} [ \bar{\psi}, \psi, \phi ] = \, & \int \mathrm{d} x \int_{0}^{\beta} \mathrm{d} \tau \, \big[ \bar{\psi} \, \big( \slashed{\partial} - \mu \, \gamma^2 + \tfrac{h}{\sqrt{N}} \, \phi \big) \, \psi	\vphantom{\Bigg(\Bigg)}	\label{eq:thermal_gny_action}
			\\
			& - \phi \, ( \Box \phi ) + \tfrac{h^2}{2 g^2} \, \phi^2 \big] \, .	\vphantom{\Bigg(\Bigg)}	\nonumber
		\end{align}
	Note that the heat bath, introduced by the non-zero temperature, as well as the chemical potential explicitly break (Euclidean) Poincar\'e invariance, \textit{i.e.}, invariance under rotations.

\section{The Gross-Neveu-Yukawa model via the Functional Renormalization Group}
\label{sec:gny_frg}

	In this section, we introduce the general concept of the FRG as well as the Exact Renormalization Group (ERG) equation as our setup of choice. The ERG equation is used to investigate the research question outlined in the introduction -- the influence of (bosonic) quantum fluctuations on discrete chiral $\mathbb{Z}_2$ symmetry breaking and restoration in the GNY model. Furthermore, we present and discuss our truncation scheme (ansatz) for the ERG equation and the resulting RG flow equation used for practical computations in this work. Finally, we discuss how (dynamic) symmetry breaking realizes itself during RG flows.
	
\subsection{The Functional Renormalization Group}
\label{subsec:the_frg}

	If summarized in a slightly oversimplified manner, the FRG maps the problem of solving high-dimensional functional integrals like Eq.~\eqref{eq:partition_function} to the problem of solving a coupled system of highly non-linear partial integro differential equations with the classical action  $\mathcal{S} ( \Phi )$ as initial condition. For general reviews on the FRG, we refer to Refs.~\cite{Berges:2000ew,Pawlowski:2005xe,Kopietz:2010zz,Rosten:2010vm,Gies:2006wv,Delamotte:2007pf,Dupuis:2020fhh,Koenigstein:2021syz,Gies:2006wv,PawlowskiScript}.
	
	At the same time, the FRG is much more than a practical tool for the calculation of macroscopic observables, like expectation values, from microscopic interactions. The FRG in fact mathematically formalizes the abstract concept of Kadanoff's block-spin transformations from small to large scales (in position space) and Wilson's corresponding idea of gradually integrating over momentum shells from high energy scales (UV) to small energy scales (IR) (in momentum space respectively) \cite{Kadanoff:1966wm,Wilson:1971bg,Wilson:1971dh,Wilson:1979qg}.
	
	In the modern approach the FRG is formulated in terms of the \textit{exact RG equation}\footnote{Being ``exact'' means that its solution for a specific model or theory is completely equivalent to ``solving the functional integral'' or to ``knowing all correlation functions of the model or theory''.}\footnote{Earlier attempts to formalize Wilson's RG approach, using similar (functional) differential equations, can be found in Refs.~\cite{Wegner:1972ih,Polchinski:1983gv,Hasenfratz:1985dm,Felder:1987,Zumbach:1994vg}.} (oftentimes also referred to as \textit{Wetterich equation}) \cite{Wetterich:1991be,Wetterich:1992yh,Reuter:1993kw,Morris:1993qb,Tetradis:1993ts,Ellwanger:1993mw},
		\begin{align}
			\partial_t \bar\Gamma_t [ \Phi ] = \, & \mathrm{STr} \Big[ \big( \tfrac{1}{2} \, \partial_t R_t \big) \, \big( \bar{\Gamma}^{(2)}_t [ \Phi ] + R_t \big)^{-1} \Big] =	\vphantom{\Bigg(\Bigg)}	\label{eq:wetterich}
			\\
			= \, &
			\begin{gathered}
				\includegraphics{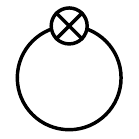}
			\end{gathered} \, .	\vphantom{\Bigg(\Bigg)}	\nonumber
		\end{align}
	Here, $t$ is the RG time (note our sign convention -- positive RG time),
		\begin{align}
			&	t = - \ln \big( \tfrac{k}{\Lambda} \big) \, ,	&&	t \in [ 0, \infty ) \, ,	\label{eq:def_rg_time}
		\end{align}
	which parameterizes the RG flow and $k ( t )$ is the corresponding RG scale with $\Lambda$ as a UV reference scale.

	The ERG equation \eqref{eq:wetterich} is formulated in terms of the effective average action $\bar{\Gamma}_t [ \Phi ]$, which is an RG-scale dependent version of the well-known effective action $\Gamma [ \Phi ]$ -- the generating functional of $1$PI-$n$-point correlation functions \cite{Goldstone:1962es,DeWitt:1965jb,Iliopoulos:1974ur,Greiner:1996zu,Weinberg:1996kr,Peskin:1995ev,ZinnJustin:2002ru,Wetterich:2001kra}. It can be shown \cite{Wetterich:1992yh,Wetterich:2001kra,Pawlowski:2005xe} that $\bar{\Gamma}_t [ \Phi ]$ approaches the classical microscopic action $\mathcal{S} [ \Phi ]$ in the UV at $t = 0$ ($k = \Lambda$). In the IR, for $t \rightarrow \infty$ ($k \rightarrow 0$), the effective average action $\bar{\Gamma}_t [ \Phi ]$ coincides with $\Gamma [ \Phi ]$. Thus, the ERG equation \eqref{eq:wetterich} describes the RG-time evolution of $\bar{\Gamma}_t [ \Phi ]$ from $\mathcal{S} [ \Phi ]$ to $\Gamma [ \Phi ]$ in terms of a (functional) differential equation, with $\mathcal{S} [ \Phi ]$ being the initial condition.
	
	The field content of the theory is summarized in the ``superfield'' variable $\Phi$, collecting all classical/mean fields under consideration. Thus the ``supertrace'' $\mathrm{STr}$ stands for traces over all internal indices (color, flavor \textit{etc.}) as well as a loop integration in momentum/position space \cite{Pawlowski:2005xe}. The ERG equation \eqref{eq:wetterich} can be applied to a broad range of QFTs ranging from gauge-theories, problems in solid-state theory to quantum gravity or effective QFTs of strong interactions \textit{etc.}. See, \textit{e.g.}, Refs.~\cite{Reuter:1993kw,Ellwanger:1995qf,Reuter:1996cp,Reuter:2001ag,Jungnickel:1995fp} for early works based on the ERG equation or Ref.~\cite{Dupuis:2020fhh} for a recent review. 
	
	The regulator $R_t$, that enters into the full scale dependent propagator (indicated by the solid black line in the Feynman graph) on the \textit{r.h.s.}\ of Eq.~\eqref{eq:wetterich} acts as a scale- and momentum-dependent IR-mode suppressor. It suppresses all quantum fluctuations at scales lower than the current RG scale $k ( t ) = \Lambda \, \mathrm{e}^{-t}$ and can be interpreted as a scale dependent mass. Hence, it ensures that only the quantum fluctuations from energy scales greater than $k(t)$ enter the effective average average action $\bar{\Gamma}_t [ \Phi ]$ at scale $k ( t )$. Lowering $k ( t )$, thus evolving in RG time $t$, integrates out and includes more and more quantum fluctuations in the effective average action $\bar{\Gamma}_t [ \Phi ]$, realizing an implementation of Wilson's concept of the RG. Considering the couplings (moments of $\bar{\Gamma}_t [ \Phi ]$) we note that in principle all couplings evolve as effective scale-dependent couplings and new couplings, which are in accordance with the symmetries of the path integral (partition function), are generated \cite{Polchinski:1983gv} in the RG flow.
	
	On the other hand, also the scale-derivative of the regulator $\partial_t R_t$ -- the so called \textit{regulator insertion} (indicated by the $\otimes$ symbol in the Feynman graph) plays a crucial role in the ERG equation \eqref{eq:wetterich}. For ``conventional'' regulators it ensures \textit{momentum locality} during RG-time evolution, meaning that predominantly fluctuating modes of similar energy to the energy of the RG scale $k ( t )$ enter during an integration step at the RG scale $k ( t )$. Formally $\partial_t R_t$ peaks the integrand of the momentum integrals in the supertrace $\mathrm{STr}$ around $p \approx k ( t )$, where $p$ is the loop momentum. (For qualitative sketches of the functional courses of these typical regulators, see, \textit{e.g.}\ Refs.~\cite{Rennecke:2015lur,Gies:2006wv,Dupuis:2020fhh}.)\\
	
	Although having a one-loop structure, the ERG equation \eqref{eq:wetterich} incorporates effects from all loop orders from a perturbative point of view, \textit{cf.}\ Refs.~\cite{Papenbrock:1994kf,Litim:2002xm,Baldazzi:2020vxk}. Consequently the FRG is a non-perturbative method, which allows for its application to systems including strong interactions -- even at non-vanishing chemical potentials\footnote{At least in certain truncations the FRG does not suffer from obvious complications at $\mu>0$ like the sign-problems encountered in lattice Monte-Carlo simulations. Computations with involved truncations for fermionic couplings are however not free of conceptional or at the least significant practical/numerical challenges at $\mu>0$, see, \textit{e.g.}, Refs.~\cite{Pawlowski:2014zaa,Braun:2020bhy}. Within this work we encountered practical complications at low temperatures (including $T=0$) and non-zero quark chemical potentials $\mu>0$ within the LPA truncation with one-dimensional, optimized Litim regulators see Sub.Sec.~\ref{subsec:chemical_potential_shock_wave} and Sec.~\ref{sec:rg-flow-with-bosons} for conceptual considerations and numerical results. We thank J.~Braun and N. Wink for various discussions on this issue.} and temperatures.\\
	
	However, the ERG equation \eqref{eq:wetterich} cannot be used directly for practical calculations -- and there is the rub. For any application so called truncations (approximations) are needed\footnote{There are some exceptional systems, where the ERG equation can be solved without any truncations, see \textit{e.g.}\ Refs.~\cite{DAttanasio:1997yph,Grossi:2019urj,Koenigstein:2021syz,Koenigstein:2021rxj,Steil:2021cbu,Steil:2021partIV}.}, which artificially restricts the entire theory space, in which $\bar{\Gamma}_t [ \Phi ]$ can evolve, to a limited subspace. This is necessary in order to convert the functional differential equation \eqref{eq:wetterich} to a finite set of ordinary and/or partial differential equations. Choosing appropriate and reasonable truncations for a specific problem under consideration, which do not alter the physics and do not ignore significant effects of the systems under consideration, is a key -- but also a very involved -- task, when performing practical computations with the FRG. Nevertheless, within the last three decades of usage of the ERG equation \eqref{eq:wetterich} specific truncation schemes were established for various applications, see Ref.~\cite{Dupuis:2020fhh} and references therein. While the effective (average) action in the IR ${\bar{\Gamma}_{t\rightarrow\infty} [ \Phi ]= \Gamma[ \Phi ]}$ is independent of the specific choice of regulator $R_t$ when considering the untruncated ERG equation \eqref{eq:wetterich}, IR results using truncations are no longer guaranteed to be independent of the regulator choice, see, \textit{e.g.}, Ref.~\cite{Pawlowski:2015mlf}. Checking for independence of physical results from truncation schemes and regulator choice is a tedious but in principle necessary task. We come back to this issue at several points during our work.\\ 

	We are aware that the FRG (at least in simple truncations like ours, see below) is usually not the method of choice for quantitative high precision predictions. But it has several advantages when it comes to the research questions addressed in this work. The FRG naturally resolves fluctuation effects -- including fermionic and bosonic quantum and thermal fluctuations -- at different energy scales and provides direct access to our observables of interest (the condensate, the curvature masses and the effective potential) at all scales $k ( t )$. The FRG as a continuum method can be used for direct computations in infinite volumes, incorporating quantum and thermal fluctuations non-perturbatively over a huge range of energies (wavelengths). It therefore complements 
	lattice Monte-Carlo simulations which are based on finite sized space-time boxes.
	
	Computations in finite volumes are also possible within the FRG framework. Actually, we plan to repeat the analysis of this work for the GNY in a finite spatial volume along the lines of Refs.~\cite{Braun:2004yk,Braun:2005gy,Braun:2005fj,Braun:2011iz,Braun:2011uq} elsewhere, in order to directly analyze the effects of a finite sized spatial volume and to compare our results to the ones obtained with lattice Monte-Carlo simulations \cite{Cohen:1981qz,Cohen:1983nr,Karsch:1986hm,Lenz:2020bxk,Lenz:2020cuv,Pannullo:2019bfn,Pannullo:2019prx}.
	
	The advantages and shortcomings of our FRG setup will become clear within the next (sub)sections, where we introduce our truncation scheme and the explicit RG flow equation, elaborate on the limitations of our approach, and discuss symmetry restoration/breaking within the FRG approach.

\subsection{Truncation and Renormalization Group flow equation(s)}
\label{subsec:truncation_and_rg_flow_equation}

	In the context of this work, we use the local potential approximation (LPA) as a truncation for Eq.~\eqref{eq:wetterich}. For the GNY model this means that the effective average action is approximated as
		\begin{align}
			\bar{\Gamma}_t [ \bar{\psi}, \psi, \varphi ] = \, & \int \mathrm{d} x\int_{0}^{\beta} \mathrm{d} \tau \, \big[ \bar{\psi} \, \big( \slashed{\partial} - \mu \gamma^2 + \tfrac{h}{\sqrt{N}} \, \varphi \big) \, \psi -	\vphantom{\Bigg(\Bigg)}	\nonumber
			\\
			& - \tfrac{1}{2} \, \varphi \, ( \Box \varphi ) + U ( t, \varphi ) \big] \ .	\vphantom{\Bigg(\Bigg)}	\label{eq:ansatz}
		\end{align}
	In this ansatz exclusively the so-called \textit{scale-dependent effective potential} $U ( t, \varphi )$ is evolving with RG time $t$ (along RG scale $k ( t )$), while all other couplings (\textit{e.g.}, the Yukawa coupling $h$) are kept constant.
	
	Although being a rather simplistic ansatz, the LPA has turned out as a common and powerful truncation scheme and is widely -- arguably sometimes wildly --	used in the FRG community, especially in the context of strongly interacting systems, \textit{e.g.}\ low-energy effective models, \textit{cf.}\ Ref.~\cite{Dupuis:2020fhh} and references therein. In the absence of fermions, the LPA can be viewed as the lowest order contribution of a derivative expansion \cite{Canet:2003qd,Canet:2002gs,Berges:2000ew,Dupuis:2020fhh}. The LPA is assumed to be (and is for certain setups proven to be \cite{DAttanasio:1997yph,Grossi:2019urj,Grossi:2021ksl,Koenigstein:2021syz,Koenigstein:2021rxj,Steil:2021cbu,Keitel:2011pn}) a good option for systems that are strongly coupled in field space and systems, where interactions with low-momentum transfer dominate the dynamics. For systems, where a high resolution in momentum space at all scales $k ( t )$ becomes relevant, a vertex expansion in powers of fields might be a better choice, \textit{cf.}\ Ref.~\cite{Dupuis:2020fhh} and references therein. For further discussions on the quality or the comparison of truncation schemes, see, \textit{e.g.}\ Refs.~\cite{Balog:2019rrg,Eser:2018jqo,Eser:2019pvd,Divotgey:2019xea,Cichutek:2020bli} and  references therein.
	
	When dealing with the effects of long-range interactions (low momenta) in a low-dimensional model at non-zero $\mu$ and $T$ (strong coupling and complicated dynamics in field space), that we are heading at, the LPA presents as a natural starting point for an analysis beyond the mean-field approximations.
	
	We are aware, that the inclusion of scale-dependent and potentially even field-dependent wave function renormalizations and couplings (especially a scale- and potentially field-dependent Yukawa coupling $h ( t, \varphi )$) would result in a significantly improved truncation. We will come back to this in Sub.Sec.~\ref{subsec:comment_on_the_truncation}. For the moment we will just start off with the LPA, which is already an improvement compared to the commonly used mean-field approximations for the GN(Y) model, where the effects of bosonic quantum fluctuations are usually completely ignored or compared to ``improved'' mean-field approximations, see \textit{e.g.}, Ref.~\cite{Dashen:1974xz}, which include only specific (effects of) bosonic modes.\\
	
	The RG flow equation for the effective potential in the LPA is obtained, by inserting the ansatz for $\bar{\Gamma}_t [ \bar{\psi}, \psi, \varphi ]$, Eq.~\eqref{eq:ansatz}, into the ERG equation \eqref{eq:wetterich} followed by a projection onto a suited (here constant) background field configuration ($\varphi ( \tau, x ) = \sigma$, $\bar{\psi} ( \tau, x ) = 0$, $\psi ( \tau, x ) = 0$), \textit{cf.} Chap. 23 of Ref.~\cite{Kleinert:2016}. Additionally one has to specify proper, explicit regulators. For a discussion on suitable choice and influences of different regulators on the RG flow and the IR results within a truncation, we refer to Refs.~\cite{Litim:2000ci,Litim:2001up,Pawlowski:2015mlf,Braun:2020bhy} and the discussion in the next Sub.Sec.~\ref{subsec:comment_on_the_regulators}. In the context of our work, we use so-called one-dimensional LPA-optimized momentum-space regulators for fermions and bosons \cite{Litim:2000ci,Litim:2001up}, see App.~\ref{app:frg}.
	
	Additionally and analogously to mean-field studies, we perform a rescaling of the bosonic (background) field and the scale-dependent effective potential,
		\begin{align}
			\varphi \mapsto \, & \tilde{\varphi} = \tfrac{1}{\sqrt{N}} \, \varphi \, ,	\label{eq:rescaling_with_n}	\vphantom{\bigg(\bigg)}
			\\
			U ( t, \varphi ) \mapsto \, & \tilde{U} ( t, \tilde{\varphi} ) = \tfrac{1}{N} \, U ( t, \varphi ) \, .	\vphantom{\bigg(\bigg)} \label{eq:rescaling_with_n_U}
		\end{align}
	This rescaling allows for a comparison of calculations at different finite values of $N$ and the infinite-$N$ limit. In the following we are exclusively working in rescaled quantities $\tilde{\varphi}$ and $\tilde{U} ( t, \tilde{\varphi} )$, such that we do not maintain the ``tilde'' in our notation.
	
	We obtain the RG flow equation for the scale dependent effective potential $U ( t, \sigma )$ in rescaled quantities,	
	\begin{widetext}
		\begin{align}
			\partial_t U ( t,  \sigma ) = \frac{1}{N} \, \Bigg(
			\begin{gathered}
				\includegraphics{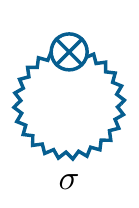}
			\end{gathered} +
			\begin{gathered}
				\includegraphics{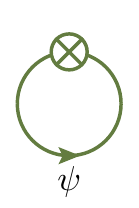}
			\end{gathered} \Bigg) = \, & - \frac{1}{\pi N} \, \frac{k^3 ( t )}{2 E_{\textrm{b}} ( t, \partial^2_\sigma U )} \, \big[ 1 + 2 \, n_\mathrm{b} ( \beta E_{\textrm{b}} ( t, \partial^2_\sigma U ) ) \big] +	\vphantom{\Bigg(\Bigg)}	\label{eq:pdeq-U}
			\\
			& + \frac{d_\gamma}{\pi} \, \frac{k^3 ( t )}{2 E_{\textrm{f}}  ( t, \sigma )} \, \big[ 1 - n_\mathrm{f} ( \beta [ E_{\textrm{f}}  ( t, \sigma ) + \mu ] ) - n_\mathrm{f} ( \beta [ E_{\textrm{f}}  ( t, \sigma ) - \mu ] ) \big] \, ,	\vphantom{\Bigg(\Bigg)}	\nonumber
		\end{align}
	\end{widetext}
	where we introduced the abbreviations
		\begin{align}
			E_\mathrm{b} ( t, \partial^2_\sigma U ) \equiv \, & \sqrt{ k^2 ( t ) + \partial_\sigma^2 U ( t, \sigma ) } \, ,	\vphantom{\bigg(\bigg)}	\label{eq:Eb}
			\\
			E_\mathrm{f} ( t, \sigma ) \equiv \, & \sqrt{ k^2 ( t ) + ( h \, \sigma )^2 } \, ,	\vphantom{\bigg(\bigg)}\label{eq:Ef}
		\end{align}
	for the Euclidean bosonic and fermionic ``energies'' (dispersion relations) and used
		\begin{align}
			&	n_\mathrm{b} ( x ) \equiv \frac{1}{\mathrm{e}^x - 1} \, ,	&&	n_\mathrm{f} ( x ) \equiv \frac{1}{\mathrm{e}^x + 1} \, .	\label{eq:distribution_function}
		\end{align}
	to denote the Bose-Einstein \cite{Bose:1924mk} and Fermi-Dirac \cite{Fermi:1926,Dirac:1926jz} distribution functions. Note the prefactor $\frac{1}{N}$ of the bosonic contributions (the first term on the \textit{r.h.s.}\ of Eq.~\eqref{eq:pdeq-U}) realizing the aforementioned suppression of bosonic fluctuations in the large-$N$ limit.
	
	For the sake of completeness, all further details on the derivation of the flow equation are provided in App.~\ref{app:frg}.\\

	Although looking complicated at first sight, the RG flow equation \eqref{eq:pdeq-U} for the effective potential $U ( t, \sigma )$ has a rather simple overall structure. It is a non-linear partial differential equation in only two variables $\sigma$ and $t$. Similar equations also appear in hydrodynamic problems. In the hydrodynamic language $\sigma$ plays the role of a spatial coordinate and $t$ is a temporal coordinate (details are provided in Sec.~\ref{sec:frg_and_fluid_dynamics}). In the following we adapt this language. The temporal and spatial domain of the PDE should not be confused with the Euclidean space-time coordinates. The spatial domain of the PDE is $\mathbb{R}$, hence $\sigma \in ( -\infty, + \infty)$, while the temporal domain is $\mathbb{R}^+$, hence $t \in [ 0, \infty )$. The PDE is first order in temporal derivatives and second order in spatial derivatives. Additionally, there are explicit time-($t$)-dependent factors and position-($\sigma$)-dependent contributions. We will continue the systematic analysis of this PDE in Sec.~\ref{sec:frg_and_fluid_dynamics}, when we draw some connections between RG flow equations and flow equations in their literal sense.\\
	
	Before we continue our main discussion, we present the zero-temperature limit of the RG flow equation \eqref{eq:pdeq-U},
		\begin{align}
			& \partial_t U ( t,  \sigma ) =	\vphantom{\bigg(\bigg)}	\label{eq:zero_t_flow_equation}
			\\
			= \, & - \frac{1}{\pi N} \, \frac{k^3 ( t )}{2 E_\mathrm{b} ( t, \sigma )}  + \frac{d_\gamma}{\pi} \, \frac{k^3 ( t )}{2 E_\mathrm{f} ( t, \sigma )} \, \Theta \Big( 1 - \tfrac{\mu^2}{E^2_\psi ( t, \sigma )} \Big)	\vphantom{\bigg(\bigg)}	\nonumber
		\end{align}
	as well as the vacuum limit for $T \rightarrow 0$ and $\mu \rightarrow 0$,
		\begin{align}
			\partial_t U ( t,  \sigma ) = \, & - \frac{1}{\pi N} \, \frac{k^3 ( t )}{2 E_\mathrm{b} ( t, \sigma )} + \frac{d_\gamma}{\pi} \, \frac{k^3 ( t )}{2 E_\mathrm{f} ( t, \sigma )} \, .	\label{eq:vacuum_limit_flow_equation}
		\end{align}
	The RG flow equation in vacuum is needed to fix the initial condition for all RG flows -- with and without the effects of a medium.
	
	Note that Eq.~\eqref{eq:vacuum_limit_flow_equation} differs from the popular LPA flow equation of the effective potential in vacuum, \textit{cf.}\ Ref.~\cite{Braun:2010tt,Rosa:2000ju},
		\begin{align}
			\partial_t U ( t,  \sigma ) = \, & - \frac{1}{4 \pi N} \, \frac{k^4 ( t )}{2 E_\mathrm{b}^2 ( t, \sigma )} + \frac{d_\gamma}{4 \pi} \, \frac{k^4 ( t )}{2 E_\mathrm{f}^2 ( t, \sigma )} \, ,	\label{eq:2dim_litim_flow_equation}
		\end{align}	
	due to the fact that we are using one-dimensional purely spatial LPA-optimized regulators in Eq.~\eqref{eq:vacuum_limit_flow_equation} that do not regulate the Matsubara summation (before taking the limit $T \rightarrow 0$), instead of using two-dimensional LPA-optimized regulators employed in the derivation of Eq.~\eqref{eq:2dim_litim_flow_equation}. We comment on this issue and its consequences for our work in the next Sub.Sec.~\ref{subsec:comment_on_the_regulators}, Sub.Sec.~\ref{subsec:vacFiniteN_LD1}, and App.~\ref{app:vac_flow_L2D}.
	
\subsection{Comment on regulators}
\label{subsec:comment_on_the_regulators}
	
	At this point we ought to comment on the choice of our regulators and the caveats that go hand in hand with this choice. The regulator shape functions \eqref{eq:regulator_shape_function_boson} and \eqref{eq:regulator_shape_function_fermion} that specify the regulators \eqref{eq:reg_boson} and \eqref{eq:reg_fermion} only regulate the spatial momentum (direction). Hence, if the spatial (loop) momenta are smaller than the RG scale $k$, they are suppressed. However, the Matsubara sums are not affected by the regulator at all and frequencies of all orders of magnitude enter the RG flow at all scales $k ( t )$. This has several direct consequences. Although regulators should in principle be in accordance with the symmetries of the theory, the regulators \eqref{eq:reg_boson} and \eqref{eq:reg_fermion} explicitly break (Euclidean) Poincar\'e invariance. Hence, we cannot expect to recover full (Euclidean) Poincar\'e symmetry in the limits $\mu \rightarrow 0$ and $T \rightarrow 0$, without the introduction of counter-terms (Ward identities) to account for this discrepancy, \textit{e.g.}, Refs.~\cite{Braun:2017srn,Steil:2021RGMF,Pawlowski:2017gxj}. This means that the IR results of the RG-flow equation \eqref{eq:vacuum_limit_flow_equation} do not necessarily coincide with IR results of the vacuum LPA flow equation \eqref{eq:2dim_litim_flow_equation} that can be derived with two-dimensional LPA optimized regulators, if one uses exactly the same UV initial condition. Whether or not these differences manifest in physical observables in vacuum or even medium, since computations at non-zero $T$ and $\mu$ usually use an UV initial condition fixated in vacuum, depends on the observable, model, and truncation under consideration, see, \textit{e.g.}, Refs.~\cite{Braun:2017srn,Steil:2021RGMF,Pawlowski:2017gxj}. Within the scope of this work we performed some test comparing results obtained with a two-dimensional LPA-optimized  regulator to results obtained using the one-dimensional LPA-optimized regulator in vacuum using identical initial conditions. For a brief discussion see App.~\ref{app:vac_flow_L2D}. The situation for the GN(Y) model might be discussed elsewhere in more detail especially regarding regulator dependencies at non-zero temperature \cite{Zorbach:2021thesis}.\\
	
	One might ask now, why we are -- regardless of these facts -- using one-dimensional LPA-optimized regulators? The answer to this question has several aspects:
	
	A first drawback of using two-dimensional regulators is that large classes of those regulators cause problems in the presence of chemical potentials and violate the so called \textit{Silver-Blaze} property \cite{Cohen:2003kd,Marko:2014hea,Khan:2015puu}. However, we are especially interested in calculations at non-zero $\mu$. Coping with this challenge is part of state of the art research, see, \textit{e.g.}, Refs.~\cite{Braun:2017srn,Braun:2018bik,Braun:2020bhy}, and we do not want to enter this discussion within this work.
	
	Secondly, the analytic evaluation of the Matsubara sums or the loop-momentum integrals in Eq.~\eqref{eq:flow_equation_after_traces} might become impossible or at least extremely challenging, which drastically complicates numerical computations. The presence of numerical sums and integrals in the flow equation would significantly increase computation time and hinder an in depth discussion at variable $T$, $\mu$ and $N$. 
	
	Note that for non-zero temperature and chemical potential (Euclidean) Poincar\'e invariance is broken anyhow, such that explicitly breaking this symmetry via the regulators might not spoil the results too drastically.

	Mainly to facilitate and speed up numerical computations and to avoid any conceptual issues at $\mu > 0$ we decided to use one-dimensional, spatial Litim regulators within this work as a first significant step beyond mean-field computations.\\
	
	Interestingly, the approach of exclusively regulating spatial momenta is similar to the common strategy employed in conventional mean-field studies at non-zero temperature including the mean-field computations for the GN(Y) model \cite{Dolan:1973qd,Harrington:1974tf,Jacobs:1974ys,Dashen:1974xz,Dashen:1975xh,Wolff:1985av}. There Matsubara summations are usually executed analytically before momentum integrals are regulated at all. Divergent contributions (usually associated with vacuum quantum fluctuations) to expressions are separated from convergent (usually thermal) contributions and only the divergent parts are regulated. Both approaches include all thermal/Matsubara modes independent of RG scale or in case of mean-field studies of the chosen regularization scheme.
	
\subsection{Truncations and initial conditions -- Differences and similarities of GN, bGN and GNY models}
\label{subsec:comment_on_the_truncation}

	In Sec.~\ref{sec:GNYmodel} we introduced three distinct models by specifying their actions: the GN model defined by $\mathcal{S}_\mathrm{GN}$ in Eq.~\eqref{eq:gn-model}, the bGN model defined by $\mathcal{S}_\mathrm{bGN}$ in Eq.~\eqref{eq:bgn-model} and ultimately the GNY model defined by $\mathcal{S}$ in Eq.~\eqref{eq:gny-model}. In this subsection we will elaborate on their differences and similarities in the FRG framework with special focus on the in this context central issues of truncations and initial conditions for their respective effective average actions $\bar{\Gamma}_t [ \bar{\psi}, \psi, \varphi ]$. As for all RG flows (and PDEs in general) their respective solutions depend on the corresponding initial and possible boundary condition(s), see, \textit{e.g.}, Refs.~\cite{LeVeque:1992,LeVeque:2002}. We discuss the (numerical) boundary conditions in Sub.Sec.~\ref{subsec:boundary_conditions} and solely focus on the initial condition for the moment.\\
	
	The GN model and its bosonized version -- the bGN model -- have an identical partition function and are thus physically equivalent. The bosonization procedure of App.~\ref{app:hubbard-stratonovich} is a merely technical reformulation with the goal to eliminate the four-Fermi coupling term $\propto ( \bar{\psi} \, \psi )^2$ of the GN model by introducing the auxiliary field $\phi$  with a mass term $\propto \phi^2$ and a Yukawa-Coupling term $\propto \phi \, \bar{\psi} \, \psi$. This reformulation facilitates computations especially for non-vanishing condensates $\langle \phi \rangle \propto \langle \bar{\psi} \, \psi \rangle$, \textit{cf.}~Ref.~\cite{Gross:1974jv,ZinnJustin:2002ru,Peskin:1995ev,Pannullo:2019}.\\
	
	As stated already in the introduction to the FRG method in Sub.Sec.~\ref{subsec:the_frg}, the initial condition for $\bar{\Gamma}_t [ \Phi ]$ for the ERG equation \eqref{eq:wetterich} is the classical action $\mathcal{S} [ \Phi ]$. Within the LPA truncation the only scale-dependent quantity is the potential $U ( t, \sigma )$ for which an appropriate initial condition at $k ( t = 0) = \Lambda$ can be read off directly form the classical action,
		\begin{align}
			U ( 0, \sigma ) = \, & \tfrac{h^2}{2 g^2} \, \sigma^2 \, ,	\label{eq:uv-initial}
		\end{align}
	while the Yukawa coupling $h$ keeps its initial value throughout the RG flow in the LPA. (Note the two rescalings in Eq.~\eqref{eq:rescaling_with_n}, which cancel for the rescaled initial potential \eqref{eq:uv-initial}). This initial condition $U ( 0, \sigma )$ is valid both for the bGN and GNY model and will be discussed further in Sub.Sec.~\ref{subsec:UUV}.
	
	The only difference between the bGN (and by proxy the GN) model and the GNY model is the kinetic term $-\phi \, ( \Box \phi )$ in the GNY model. A corresponding contribution $- \varphi \, ( \Box \varphi )$ (in terms of classical/mean fields $\varphi = \langle \phi \rangle$) is of course also present in the LPA ansatz \eqref{eq:ansatz} for $\bar{\Gamma}_t [ \bar{\psi}, \psi, \varphi ]$ of the GNY model to establish the proper initial condition $\bar{\Gamma}_{t = 0} = \mathcal{S}$. This kinetic term presents an obvious mismatch between the LPA formulation of the GNY model and a possible LPA formulation of the bGN model. Nevertheless, such a kinetic term is needed in the FRG framework to study the effect of bosonic quantum fluctuations of $\phi$ at finite $N$. In consequence, the LPA of the GNY model for $\bar{\Gamma}_t [ \bar{\psi}, \psi, \varphi ]$ seems not capable of resembling the bGN model (GN model by proxy) in the UV. One might argue that the results, which are obtained from the RG flow equation \eqref{eq:pdeq-U} with initial condition \eqref{eq:uv-initial}, are consequently not directly transferable to the GN model. It seems, as if bosonic quantum fluctuations, which are linked to the kinetic term, are artificially enhanced already at the beginning of the RG flow, while they should actually be strongly suppressed (strictly speaking vanishing directly in the UV when considering the limit $\Lambda\rightarrow\infty$) in the bGN model \eqref{eq:bgn-model}, which starts without the kinetic term in the UV and generates the term and bosonic fluctuations dynamically during the RG flow.
	
	From the FRG perspective, the intuitive way to cure this problem is the introduction of a bosonic wave function renormalization in the kinetic term,
		\begin{align}
			- \varphi \, ( \Box \varphi ) \rightarrow - Z_{\varphi} ( t ) \, \varphi \, ( \Box \varphi ) \, ,
		\end{align}
	and initializing ${1 \ggg Z_{\varphi} ( 0 )>0}$ in the UV\footnote{Initializing $Z_{\varphi} ( 0 )$ at exactly zero in the UV leads to complications in practical computation since, \textit{e.g.}, the renormalized mass term $\tfrac{1}{Z_\varphi ( t )} \, \partial_\sigma^2 U ( t, \sigma )$ in the dispersion relation \eqref{eq:Eb} would diverge. Besides technical problems $Z_{\varphi} ( 0 )=0$ at a finite UV initial scale $\Lambda$ would weakly violate RG-consistency \cite{Braun:2018svj} since fluctuations (especially fermionic ones) at scales $k>\Lambda$ would have already generated a small wave function renormalization.} to make direct contact with the action \eqref{eq:bgn-model} of the bGN model. Initializing the wave function renormalization with ${1 \ggg Z_{\varphi} ( 0 )>0}$ in the UV at a sufficiently large UV initial scale $\Lambda$ -- large enough  at the initial scale and in doing so realizing RG consistency \cite{Braun:2018svj} -- leads to a direct suppression of bosonic fluctuations at the beginning of the RG flow. This is directly seen on the level of $\bar{\Gamma}_t [ \bar{\psi}, \psi, \varphi ]$, where the bosonic kinetic term is suppressed by its small prefactor $Z_\varphi ( t = 0 )$ in the UV. Most likely, during the RG flow $Z_\varphi ( t )$ will turn non-zero already by its fermionic loop contribution, \textit{cf.}\ Ref.~\cite{Braun:2010tt,Braun:2011pp}, and we expect to recover the situation, where $Z_\varphi ( t )$ is non-zero and facilitates bosonic fluctuations in the IR.
	
	Hence, we already conclude at this point that a natural generalization of the LPA truncation in this publication is the inclusion of a bosonic wave function renormalization, such that the UV initial condition for $\bar{\Gamma}_t [ \bar{\psi}, \psi, \varphi ]$ in the GNY model really resembles the bGN and GN. This will be discussed elsewhere. (Of course, the additional inclusion of further scale- and potentially field-dependent couplings and wave function renormalizations, \textit{e.g.}, $h ( t, \varphi )$, $Z_\varphi ( t, \varphi )$, $Z_\psi ( t, \varphi )$, \textit{etc.}, is the natural step beyond the inclusion of just $Z_\varphi ( t)$.)
	
	However, for the sake of this work, we note that the effect of a strong (or even total) suppression of bosonic fluctuations in the UV (at fixed $N$) is present already in the LPA truncation anyhow due to peculiarities of the RG flow equation and the quadratic shape of the UV initial potential \eqref{eq:uv-initial}. The first effect relates to the choice of $h$ and $g^2$ at $k ( t = 0 ) = \Lambda$. We know that the  GN model is asymptotically free, meaning $\tfrac{1}{g^2} \sim \ln \big( \tfrac{\Lambda}{h} \big)$. Hence, for large $\Lambda$ and finite $h$ the bosonic mass stemming from Eq.~\eqref{eq:uv-initial} is large compared to the fermionic mass and bosonic fluctuations are expected to be of minor importance -- even if the bosonic wave function renormalization is fixed $Z_\mathrm{\varphi} ( t ) = 1$. The second effect relates to the quadratic shape of the initial potential \eqref{eq:uv-initial} and the (diffusive) character of the bosonic contribution to the RG flow equation. A purely quadratic self-interaction potential is the manifestation of a free, non-interacting bosonic sector, which decouples completely from the fermionic one and has no dynamics during RG flow. This holds true for the bosonic sector as long as the fermionic contributions do not significantly alter the self-interaction potential away from its initial quadratic shape. Both effects -- the large mass and the retention of a quadratic self-interaction potential in the UV -- are discussed in detail in Sub.Sec.~\ref{subsec:UUV}, after we firstly introduced the fluid-dynamic reformulation of the flow equation \eqref{eq:pdeq-U} in Sec.~\ref{sec:frg_and_fluid_dynamics} and secondly discussed asymptotic freedom and RG consistency in the context of the infinite-$N$ limit (mean-field) in Sec.~\eqref{sec:mean-field}.
	
	For the moment, we note that the LPA seems to turn out as a decent approximation to capture the qualitative dynamics, without spoiling the actual RG flow too much. This statement is refined in Sub.Sec.~\ref{subsec:UUV} and supported by our explicit numerical results in Sec.~\ref{sec:rg-flow-with-bosons}, where it is visible that the bosonic contributions to the RG flow are relevant at much later RG times (lower scales) than those of the fermions. Nevertheless, we plan to study higher order truncation effects elsewhere.\\
	
	Finally, we have to briefly comment on another issue, which also stems from the truncation scheme and the mismatch of the initial conditions for the bGN and GNY models. The original GN action \eqref{eq:gn-model} as well as its bosonized counterpart in Eq.~\eqref{eq:bgn-model} without the kinetic term $-\phi \, ( \Box \phi )$ have only a single coupling constant $g^2$. The artificial Yukawa coupling $h$ introduced during the bosonization procedure can actually be absorbed in the bosonic field. In mean-field calculations, see below, this is evident, because all observables can be fixed via a single dimensionful parameter (\textit{e.g.}, the IR bosonic curvature mass, the IR fermion mass, the critical temperature \textit{etc.}) for $\Lambda \rightarrow \infty$ and can be mapped into each other, if different renormalization schemes were chosen. Including bosonic fluctuations, \textit{i.e.}\ the kinetic term for the bosons, the Yukawa coupling $h$ can no longer be completely absorbed via appropriate substitutions. Hence, the model involves an additional parameter that needs to be fixed via some renormalization condition \cite{ZinnJustin:2002ru} -- even if we stick to quadratic initial potentials like \eqref{eq:uv-initial}.
	
	We also come back to this aspect in Sec.~\ref{sec:mean-field}, where we motivate our explicit initial conditions for the RG flows. Though, we have to recapitulate some mean-field results in advance.	

\subsection{Dynamical symmetry breaking and RG flows}
\label{subsec:dSB}

	Since the FRG is formulated in terms of the effective (average) action, we set the following criteria for (spontaneous) symmetry breaking.
	
	The vacuum of a QFT is defined as the field configuration with least energy, which has to be a field configuration that minimizes the IR effective action $\Gamma [ \Phi ]$, \textit{cf.}\ Refs.~\cite{Wetterich:2001kra,Weinberg:1996kr}. Hence it has to be a solution to the quantum equations of motion for the mean fields $\bar{\psi}$, $\psi$, and $\varphi$,
		\begin{align}
			&	0 \overset{!}{=} \frac{\delta \Gamma [ \bar{\psi}, \psi, \varphi ]}{\delta \bar{\psi}} \, ,	&&	0 \overset{!}{=} \frac{\delta \Gamma [ \bar{\psi}, \psi, \varphi ]}{\delta \varphi} \, .
		\end{align}
	Within our truncation \eqref{eq:ansatz}, these equations reduce to the following two classical PDEs
		\begin{align}
			0 = \, & ( \slashed{\partial} - \mu \gamma^2 + h \, \varphi ) \, \psi \, ,	\vphantom{\Bigg(\Bigg)}	\label{eq:qeqom1}
			\\
			0 = \, & \Box \, \varphi  - \partial_\varphi U ( t_\mathrm{IR}, \varphi ) - \tfrac{h}{N} \, \bar{\psi} \, \psi \, ,	\vphantom{\Bigg(\Bigg)}	\label{eq:qeqom2}
		\end{align}
	where we did not separately list the corresponding equations for the associated spinor $\bar{\psi}$.
	
	At this point, it is natural to consider vanishing fermion mean fields $\bar{\psi} ( \tau, x ) = 0$, $\psi ( \tau, x ) = 0$. Furthermore, if we assume the field configuration to be invariant under any kind of space-time translations, thus $\varphi ( \tau, x ) = \mathrm{const}.\equiv\sigma$, we arrive at
		\begin{align}
			0 \overset{!}{=} \partial_\sigma U ( t_\mathrm{IR}, \sigma ) \, .
		\end{align}
	In order to ensure that this extremum condition defines a minimum and not a maximum, we need the additional sufficient condition $\partial_\sigma^2 U ( t_\mathrm{IR}, \sigma ) > 0$ at the extremum, \textit{viz.} a positive IR curvature mass for $\sigma$.

	Consequently, in the FRG framework a minimum $\sigma_\mathrm{min}$ of the scale dependent effective potential $U ( t, \sigma )$ in the IR (for $t \rightarrow \infty$) is considered to be a (translational invariant) ground state of our system. The system is said to be in the symmetry broken phase, if $\sigma_\mathrm{min} \neq 0$, and in the symmetric phase if $\sigma_\mathrm{min} = 0$, because non-trivial $\sigma_\mathrm{min}$ break the $\mathbb{Z}_2$ symmetry of the vacuum.
	
	The breaking or restoration of the $\mathbb{Z}_2$ symmetry takes place \textit{dynamically}. By integrating out quantum fluctuations from short wavelength (UV) to long wavelength (IR), the fermion interactions may \textit{dynamically} form a condensate, which manifests as a non-trivial minimum $\sigma_\mathrm{min} ( t )$ of $U ( t, \sigma )$ during the RG-flow at finite $t$. On the other hand -- as discussed in the introduction -- there are arguments \cite{Landau:1980mil,Ising:1925em,Dashen:1974xz,Rosenstein:1988dj,ZinnJustin:2002ru} for the dynamical restoration of $\mathbb{Z}_2$ symmetry in the IR, which can be interpreted as a realization of the precondensation phenomenon found by Refs.~\cite{Boettcher:2012cm,Boettcher:2012dh,Boettcher:2013kia,Boettcher:2014tfa,Roscher:2015xha,Khan:2015puu} in the context of RG flows. The underlying dynamics can be attributed to long-range bosonic fluctuations in the FRG framework. Within this work, we can explicitly resolve this complicated dynamics during the RG flows, compare Sec.~\ref{sec:rg-flow-with-bosons}.\\
	
	We are aware, that there might be field configurations, which are not invariant under translations, solve the quantum equations of motion \eqref{eq:qeqom1} \& \eqref{eq:qeqom2}, and which might have an even smaller energy than the homogeneous solution $\sigma_\mathrm{min}$. However, these so called ``(spatially) inhomogeneous (chiral) condensates''\footnote{For further literature on ``(spatially) inhomogeneous (chiral) condensation'' we refer to the Review~\cite{Buballa:2014tba} and references therein and to the comprehensive work of M.~Thies and collaborators in the context of $(1 + 1)$-dimensional QFTs, see Tab.~1 of Ref.~\cite{Thies:2020ofv} and references therein.} are not part of this work and might be considered elsewhere.

\section{The FRG and (numerical) fluid dynamics}
\label{sec:frg_and_fluid_dynamics}

	This section is dedicated to the formalities of the (numerical) solution of the RG flow equation \eqref{eq:pdeq-U}. Our discussion is twofold. Firstly, we discuss analogies of the RG flow equation of the effective potential \eqref{eq:pdeq-U} and common PDEs from the field of (numerical) fluid dynamics. This leads to an intuitive understanding of the dynamics of the scale dependent effective potential $U ( t, \sigma )$ (or rather its derivative $u ( t, \sigma ) = \partial_\sigma U ( t, \sigma )$) during the RG flow in terms of fluid dynamic notions like diffusion or sink/source terms. Secondly, we introduce the explicit implementation of our numerical scheme, which originally stems from the research field of computational fluid dynamics.

\subsection{The LPA flow equation as a non-linear heat equation with sink/source term}

	Within the last decades a huge variety of different numerical schemes was used to solve field-dependent RG flow equations of similar type than Eq.~\eqref{eq:pdeq-U}. However, only recently it was found by some of the authors and their collaborators \cite{Grossi:2019urj,Grossi:2021ksl,Koenigstein:2021syz,Koenigstein:2021rxj,Steil:2021cbu,Steil:2021partIV} that the RG flow equation for the scale dependent effective potential for a large class of models from QFT can be recast as a conservation law in terms of a (advection-)diffusion-source/sink equation and be entirely understood in terms of fluid dynamic notions\footnote{Earlier works, which already revealed some analogies between RG flows and fluid dynamic PDEs are Refs.~\cite{Zamolodchikov:1986gt,Rosten:2010vm,Zumbach:1993zz,Zumbach:1994kc,Zumbach:1994vg}.}. A fluid dynamic reinterpretation is as advantageous for several reasons. On the one hand, the different contributions to the RG flow can easily be understood via their fluid dynamic analogues. On the other hand, this formulation provides access to the highly developed toolbox of computational fluid dynamics \cite{LeVeque:1992,LeVeque:2002,RezzollaZanotti:2013,KTO2-0}.
	
	Hence, the starting point of our discussion is the conversion of the PDE \eqref{eq:pdeq-U} into a non-linear diffusion equation. To this end, we act with $\frac{\mathrm{d}}{\mathrm{d}\sigma}$ on both sides of Eq.~\eqref{eq:pdeq-U} and obtain a flow equation in conservative form\footnote{Some authors, see, \textit{e.g.} Refs.~\cite{Tetradis:1995br,Litim:1995ex,Aoki:2014,Aoki:2017rjl}, already studied flow equations of this form in a weak formulation without the explicit identification and interpretation as a conservation law.}
		\begin{align}
			\partial_t u = \tfrac{\mathrm{d}}{\mathrm{d}\sigma} \, Q ( t, \partial_\sigma u ) + S ( t, \sigma ) \, ,	\label{eq:pdeq-u}
		\end{align}
	where $u = u ( t, \sigma ) \equiv \partial_\sigma U ( t, \sigma )$ acts as the conserved quantity. Henceforth, we refer to
		\begin{align}
			& Q ( t, \partial_\sigma u ) \equiv	\vphantom{\bigg(\bigg)}	\label{eq:diffusion_flux}
			\\
			\equiv \, & - \frac{1}{\pi N} \, \frac{k^3 ( t )}{2 E_{\textrm{b}} ( t, \partial_\sigma u )} \, \big[ 1 + 2 \, n_\mathrm{b} ( \beta E_{\textrm{b}} ( t, \partial_\sigma u ) ) \big] \, ,	\vphantom{\bigg(\bigg)}	\nonumber
		\end{align}
	as a non-linear diffusion flux and
		\begin{align}
			& S ( t, \sigma ) \equiv	\vphantom{\bigg(\bigg)}	\label{eq:source_sink}
			\\
			\equiv \, & \frac{\mathrm{d}}{\mathrm{d}\sigma} \bigg( \frac{d_\gamma}{\pi} \, \frac{k^3 ( t )}{2 E_{\textrm{f}} ( t, \sigma )} \, \big[ 1 - n_\mathrm{f} ( \beta [ E_{\textrm{f}} ( t, \sigma ) + \mu ] ) -	\vphantom{\bigg(\bigg)}	\nonumber
			\\
			& - n_\mathrm{f} ( \beta [ E_{\textrm{f}} ( t, \sigma ) - \mu ] ) \big] \bigg) \, ,	\vphantom{\bigg(\bigg)}	\nonumber
		\end{align}
	as a local, non-linear, internal sink/source term, \textit{cf.} Ref.~\cite{KTO2-0}. On the level of the PDE, especially from a fluid dynamic point of view, the RG time $t \in [ 0, \infty )$ naturally corresponds to an effective temporal variable (parameter), while the field space, spanned by $\sigma \in ( - \infty, \infty )$, presents as an effective (infinite) spatial domain. The derivative of the effective potential $u ( t, \sigma )$ plays the role of a fluid. The interpretation of $Q ( t, \partial_\sigma u )$ as a non-linear diffusion flux becomes apparent when the derivative in Eq.~\eqref{eq:pdeq-u} is performed,
		\begin{align}
			\tfrac{\mathrm{d}}{\mathrm{d} \sigma} \, Q ( t, \partial_\sigma u ) = D ( t, \partial_\sigma u ) \, \partial_\sigma^2 u ( t, \sigma ) \, .	\label{eq:heat_equation_analogy}
		\end{align}
	Without the source term, we find that the problem of solving the RG flow equation of the effective potential \eqref{eq:pdeq-U} has reduced to solving a heat equation, \textit{cf.} Ref.~\cite{Cannon:1984}, with a time- and gradient-dependent highly non-linear diffusion coefficient $D ( t, \partial_\sigma u )$.
	
	The interpretation of the purely fermionic contribution $S ( t, \sigma )$ in terms of a time dependent source/sink term is also rather natural. From Eq.~\eqref{eq:source_sink} we find that the fluid $u$ does not explicitly enter $S ( t, \sigma )$ and the contribution to the flow of the fluid $u$ solely depends on the time $t$ and position in the spatial direction $\sigma$. Actually it presents on a formal level in our flow equation as a classical potential (similar to a Newtonian gravitational potential) of a conservative (external) force field acting on the fluid, \textit{cf.}\ Ref.~\cite{KTO2-0}. Executing the $\sigma$-derivative in Eq.~\eqref{eq:source_sink} one finds that the term enters the flow equation \eqref{eq:pdeq-u} for $u$ with negative sign for $\mu=0$, which explains the denotation as a ``sink'' for $\sigma > 0$ and ``source'' for $\sigma < 0$ and is responsible for symmetry breaking during the flow at $\mu = 0$\footnote{Note that the dynamics of $u ( t, \sigma )$ is anti-symmetric in $\sigma$. Due to this anti-symmetry and \textit{w.l.o.g.}\, see Sub.Sec.~\ref{subsec:boundary_conditions}, we focus on $\sigma \in [ 0, \infty )$, where the ``sink'' interpretation is more natural. For negative $\sigma$ the dynamics has of course opposite sign.}. The bosonic contributions in terms of non-linear diffusion tend to fill the ``sink'' or distribute the ``source'' that is caused by the fermions and work against symmetry breaking. At non-zero $\mu$ the sink/source term \eqref{eq:source_sink} is not always manifest negative/positive: it has a rather intricate dynamic (especially at low and zero temperature) and manifests as either source or sink depending on the value for $\sigma$ at a given $\mu$ and $t$ for both -- negative and positive $\sigma$. This will be discussed at length in Sub.Sec.~\ref{subsec:chemical_potential_shock_wave}.

\subsection{Irreversibility and entropy production}

	Another direct consequence of the reinterpretation of the RG flow equation \eqref{eq:pdeq-U} in terms of a diffusion-type equation \eqref{eq:pdeq-u} is that the irreversible character of the RG as a semi group is directly hard-coded and understood on the level of the PDE: Diffusion is a dissipative and irreversible process, that introduces a ``thermodynamic arrow of time'' \cite{Lebowitz:2008} into the problem. This arrow of time clearly singles out the RG time (scale) as a temporal coordinate. Additionally, the dissipative character in field space (space of couplings) can be seen as a direct realization of Kadanoff's irreversible block spin-transformations (in position space) \cite{Zumbach:1994vg,Zamolodchikov:1986gt,Wilson:1979qg}. It follows that the reverse process of integrating from the IR towards the UV is generically impossible -- also directly on the level of the PDE.
	
	Though, the mean-field truncation (infinite-$N$ limit) seems to violate this fact, especially concerning the discussion on asymptotic freedom, see Sec.~\ref{sec:mean-field}. However, this has to be understood as an artifact of the oversimplified restrictions on the evolution of $\bar{\Gamma}_t [ \bar{\psi}, \psi, \varphi ]$ in theory space. For mean-field approximations the theory-space is entirely described via a single field-independent coupling constant $g^2$, which is evolving with $t$. Beyond the infinite-$N$ limit, this is an oversimplification of the real dynamics -- especially in the IR. However, via the inclusion of bosonic fluctuations, the RG flow can retain its correct irreversible diffusive character. Nevertheless, this is only possible if we do not restrict the bosonic effective potential to a finite set of couplings. Instead we have to resolve the dissipative dynamics in field space correctly -- also by using adequate numerical methods. For further discussions on these and related issues, we refer to Refs.~\cite{Zumbach:1994vg,Zamolodchikov:1986gt,Koenigstein:2021syz,Koenigstein:2021rxj,Steil:2021cbu,Steil:2021partIV}.

\subsection{Field space boundary conditions}
\label{subsec:boundary_conditions}

	In order to make the solution of a PDE a well-posed problem, one needs to specify initial and boundary conditions. Formally, our PDE \eqref{eq:pdeq-u} presents as a pure initial value problem on $\mathbb{R}$, because the initial condition for the PDE (the UV potential) $u ( 0, \sigma )$ is specified for $\sigma \in ( - \infty, \infty )$, \textit{cf.} Eq.~\eqref{eq:uv-initial}. Thus, on a formal mathematical level, boundary conditions are not needed, because the problem is completely defined via the asymptotics of the flow equation \eqref{eq:pdeq-u} and the initial condition, the $\sigma$-derivative of Eq.~\eqref{eq:uv-initial}, for $| \sigma | \rightarrow \infty$. Asymptotically, the $\sigma$-derivative of the diffusion flux \eqref{eq:diffusion_flux} vanishes. The same holds true for the sink term, which is suppressed for large $| \sigma |$. In total $u ( t, \sigma )$ does not change for $| \sigma | \rightarrow \infty$ and keeps its initial shape. But still, there is a lot of dynamics, which however takes place only within a small subdomain of the field space. Additionally, direct practical calculations are impossible on an infinite interval. To facilitate practical (numerical) computations, we artificially introduce boundary conditions to the problem. These boundary conditions have to be constructed in a way, that does not spoil the original dynamics of the system.
	
	In a first step, we take advantage of the $\mathbb{Z}_2$ symmetry $ U ( t, \sigma ) = U ( t, -\sigma )$, which implies a $\mathbb{Z}_2$ anti-symmetry of $u ( t, \sigma ) = \partial_\sigma U ( t, \sigma )$, \textit{i.e.}, $u ( t, \sigma ) = - u ( t, - \sigma )$ and therefore restrict our analysis to $\sigma \in [ 0, \infty )$ \textit{w.l.o.g.}. This amounts to the introduction of anti-symmetric reflective boundary conditions at $\sigma = 0$ on the level of the PDE \eqref{eq:pdeq-u}. On the other hand, we argued that the diffusion flux and sink-contribution fall off drastically for large $| \sigma |$. Therefore, it is feasible to truncate the spatial domain at some finite but large $\sigma_\mathrm{max}$, beyond which the $t$-evolution $u ( t, \sigma )$ is negligible. At this second artificial spatial boundary, we can use linear extrapolating boundary conditions for $u ( t, \sigma )$ due to the linear asymptotics of the initial condition for $u ( t, \sigma )$, the derivative of the UV potential \eqref{eq:uv-initial}, which does not change during the flow. Obviously, we have to test, which numerical values for $\sigma_\mathrm{max}$ are sufficiently large. This is done in App.~\ref{subsec:test_computational_domain}.\\
	
	For a detailed discussion on field space boundary conditions for RG flow equations (and their explicit numeric implementation), we refer to Refs.~\cite{Koenigstein:2021syz,Steil:2021cbu} and Sub.Sec.~\ref{subsec:numerical_implementation}.

\subsection{The chemical potential as a ``shock wave'' in field space}
\label{subsec:chemical_potential_shock_wave}
	
	Formulating the RG flow equation on the level of the derivative of the scale-dependent effective potential $u ( t,  \sigma ) = \partial_\sigma U ( t, \sigma )$ also allows for a better understanding of the dynamics induced by a non-vanishing chemical potential $\mu$. It turns out that the chemical potential induces a kind of shock wave in $u ( t, \sigma )$ in field space. This is understood as follows.
		\begin{figure}
		\centering
		\includegraphics{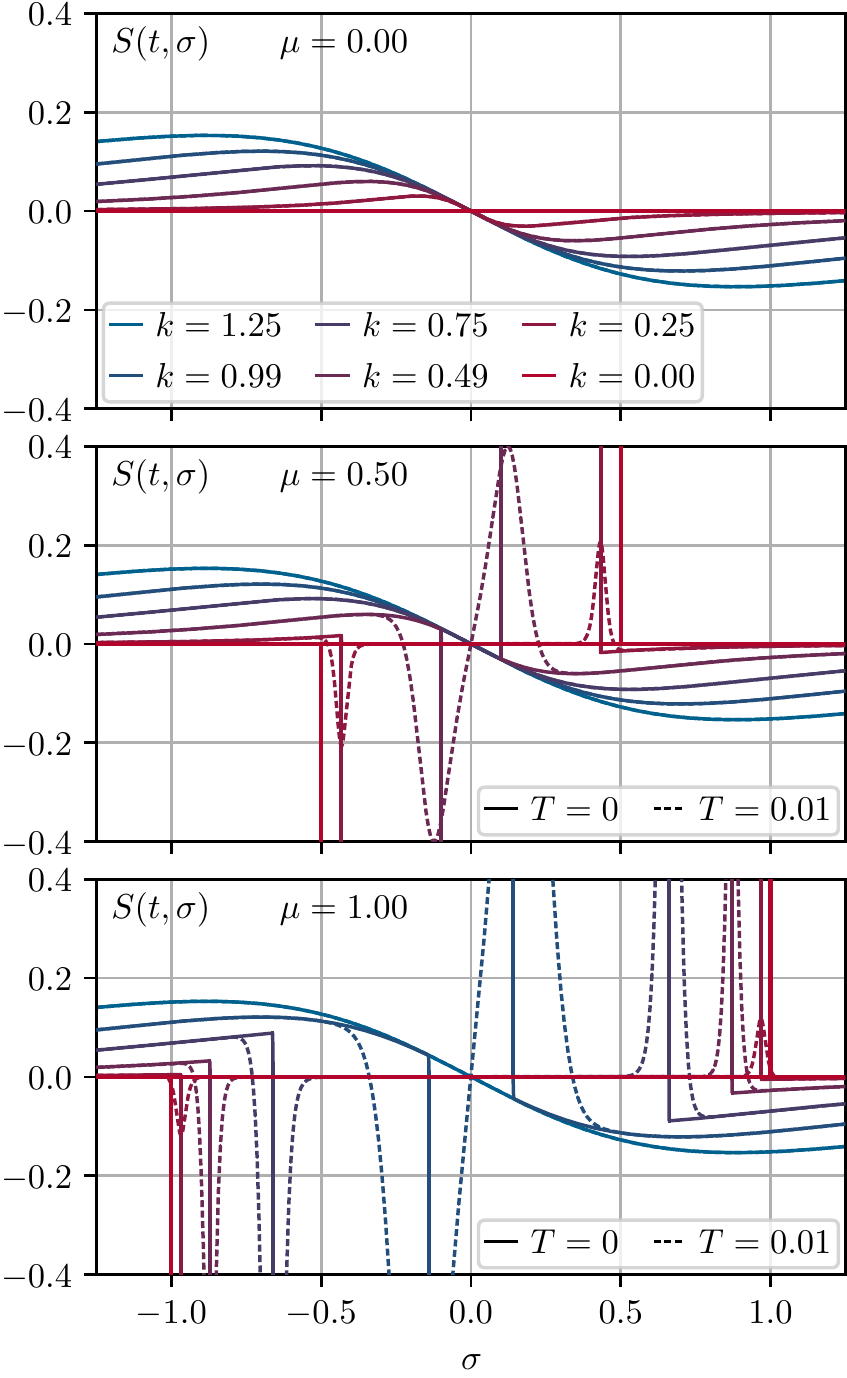}
		\caption{\label{fig:GN_sinkTerm}
			Source/sink term of Eq.~\eqref{eq:source_sink} with Yukawa coupling $h=1$ at zero (solid lines) and small temperature $T=0.01$ (dashed lines) at various RG scales $k$ (close to and below the model scales) between $k = 1.25$ in {blue} and $k = 0$ in {red} at $\mu = 0$ (upper panel), $\mu = 0.5$ (middle panel) and $\mu = 1.0$ (lower panel).
		}
	\end{figure}
	For the sake of simplicity, we analyze the RG flow equation \eqref{eq:pdeq-U} at vanishing (and small) temperature, hence Eqs.~\eqref{eq:zero_t_flow_equation} and \eqref{eq:vacuum_limit_flow_equation}, but on the level of $u ( t, \sigma )$. We find that the difference between the $T = 0$ flow equations for $u ( t, \sigma )$ at vanishing and non-vanishing chemical potential is the contributions stemming from the derivative of the Heaviside function $\Theta ( 1 - \mu^2/E_\mathrm{f}^2 ( t, \sigma ) )$ in the source/sink term. The sink/source term at different RG scales and chemical potentials is plotted in Fig.~\ref{fig:GN_sinkTerm} for $T=0$ and also small $T=0.01$ to support and illustrate the following discussion. For small $T$ the Heaviside function gets smeared out. We will discuss the dynamics induced by the sink/source term in the following paragraphs. The approximate signs in the rest of this subsection hold at small temperatures and become exact for $T=0$.
	
	As long as $E^2_\psi \gtrsim \mu^2$, effects due the chemical potential are not visible in the RG flow, because $\mu$ does not show up in the source/sink term, see Fig.~\ref{fig:GN_sinkTerm}. However, when $E_\mathrm{f}^2 \approx \mu^2$, the chemical potential becomes relevant and distinguishes RG flows with $\mu = 0$ and $\mu \neq 0$. Therefore, it is important to understand when (in terms of RG time $t$), where (in field space $\sigma$), and how this is going to happen. This can be done by analyzing the estimate
		\begin{align}
			E_\mathrm{f}^2 = k^2 + ( h \sigma )^2 \approx \mu^2	\label{eq:influence_of_mu}
		\end{align}
	for different positions in field space.
	
	The first time during the RG flow, when $\mu$ is going to influence the dynamics is, when $k^2 ( t ) \approx \mu^2$, \textit{cf.} Fig.~\ref{fig:GN_sinkTerm} at $k=0.99$ and $k=0.49$ in the middle and lower panel. At this RG time (scale), we find that the chemical potential will influence the dynamics at positions in field space close to $\sigma = 0$. At slightly later RG times, when $t$ is larger and the RG scale $k ( t )$ is lowered a little, also field space positions at slightly larger $| \sigma |$ will be influenced by $\mu$. The largest $| \sigma |$ that are directly affected by $\mu$ in the PDE are of the order $h^2 \sigma^2 \approx \mu^2$. This is the case when $t \rightarrow \infty$ and $k ( t ) \rightarrow 0$ in Eq.~\eqref{eq:influence_of_mu}, \textit{cf.} Fig.~\ref{fig:GN_sinkTerm} at $k=0.0$ in the middle and lower panel. The peaks induced by the chemical potential are located at ${\sigma\approx\pm \tfrac{1}{h}\sqrt{\mu^2-k(t)^2}}$ for $\mu \gtrsim k(t)$ and change the character of the sink/source term around their respective position. For $\sigma>0$ ($\sigma<0$) the peak presents as a source (sink).
	 
	Of course, the diffusive character of the $\sigma$-loop contribution will transport the effects of and the information about the chemical potential via diffusion also to larger $| \sigma |$. The same holds true for the non-zero temperature version of the RG flow equation, where the Heaviside function is smeared out in terms of a Fermi-Dirac distribution function \eqref{eq:distribution_function}. However, overall the RG time $t$ and scale $k ( t )$ as well as the relevant positions in field space $\sigma$ for the dynamics induced by the chemical potential are very similar at low temperatures where the dynamics is dominated by the chemical potential, \textit{cf.} Fig.~\ref{fig:GN_sinkTerm} for $T=0.01$.
	
	The remaining question, that needs to be answered is, how the chemical potential $\mu$ influences the dynamics of the RG flow.
	
	We have argued before, that the fermionic contributions to the RG flow of $u ( t, \sigma )$ enter as local time-dependent sink terms. Now we found that the sinking stops at certain positions in field space and certain RG times suddenly during the RG flow due to the chemical potential. Hence, we expect that the chemical potential introduces a sharp edge in $u ( t, \sigma )$ at $\sigma \approx 0$, when $k^2 ( t ) \approx \mu^2$. This edge will ``move'' in field space like a shock wave towards larger $| \sigma |$ until it reaches $| \sigma | \approx | \mu/h |$, where it comes to a halt. This dynamics is imprinted by the underlying dynamics of the sink/source term \eqref{eq:source_sink} visualized and discussed earlier.
	
	As already said before, this discontinuity is smeared by the diffusion an washed out right from the beginning at non-zero temperatures. However, its dynamics is clearly underlying all diffusive processes. In order to visualize the drastic effects of this discontinuity, we plot and contrast mean-field RG flows with RG flows involving bosonic fluctuations in Sub.Sec.~\ref{subsec:chemical_potential_mf_vs_with_bosons}. In mean-field, only the fermions are active (the sink term), such that the induced shock wave is not subject to diffusion effects and thus clearly visible.\\
	
	In fact, it is the chemical potential that introduces a jump (mean-field) or cusp (finite $N$) discontinuity in $u ( t, \sigma )$, which seems to render practical calculations at zero temperature involving bosonic fluctuations of the $\sigma$-mode impossible: The discontinuity in $u ( t, \sigma )$ corresponds to a large jump in the derivative $\partial_\sigma u ( t, \sigma )$ with negative sign. However, this happens at rather small $k^2 ( t ) \approx \mu^2$, such that the bosonic energy function,
		\begin{align}
			E_\mathrm{b}^2 ( t, \partial_\sigma u ( t, \sigma ) ) = k^2 ( t) + \partial_\sigma u ( t, \sigma ) \, ,
		\end{align}
	turns negative and one overshoots the pole of the bosonic propagator $\tfrac{1}{E_\mathrm{b}}$ in the RG flow equations. Preventing further numerical evolution towards the IR.
	
	When considering a model involving also Goldstone modes, which enter the RG flow as advection, \textit{cf.}\ Refs.~\cite{Grossi:2019urj,Grossi:2021ksl,Koenigstein:2021syz,Steil:2021cbu,Steil:2021partIV}, the huge gradients introduced via the chemical potential induce a shock wave that propagates towards smaller $|\sigma|$.\\
	
	A possible reason, why this subtle dynamics in field space was not discussed in detail in the past within FRG studies might be that it is hardly visible and understandable on the level of the scale dependent effective potential $U ( t, \sigma )$ itself, because it is the integral of $u ( t, \sigma )$, where jumps might only show up as tiny cusps, compare Sub.Sec.~\ref{subsec:chemical_potential_mf_vs_with_bosons}. Additionally, a lot of previous studies did not perform calculations at sufficiently low temperatures, where this effect is not completely washed out by the thermal distribution functions \eqref{eq:distribution_function}. At this point the underlying reason for and resolution of this practical problem are not clear to the authors. The disconnection between the fermionic sector and the dynamics of $u(t,\sigma)$ in LPA, the regulator choice or the formulation\footnote{A formal classification and proper weak formulation of the PDEs \eqref{eq:pdeq-U} and \eqref{eq:pdeq-u} might be paramount to understand the nature of the arising discontinuities and weak/physical solutions in their presence.} of the flow equation could be possible reasons for this intricate problem. Solving this problem is beyond the scope of the current work. Consequently, we claim that understanding and/or capturing this effect correctly will be one of the central challenges in FRG at non-zero $\mu$ within the next years -- independent of the specific models.
	
	For related discussions on these novel findings, we also refer to Refs.~\cite{Grossi:2021ksl,Steil:2021partIV,Koenigstein:2021}.

\subsection{Numerical implementation}
\label{subsec:numerical_implementation}

	In this subsection we present the explicit numeric implementation of the RG flow equation \eqref{eq:pdeq-u}. Thereby we make use of our findings of the previous sections and related publications \cite{Koenigstein:2021syz,Koenigstein:2021rxj,Steil:2021cbu,Steil:2021partIV} on the relation between RG flow equations and flow equations in the true sense of the word.
	
	To adequately capture highly non-linear diffusive effects as well as the position-dependent source/sink terms, we use the so called Kurganov-Tadmor (KT) scheme \cite{KTO2-0}. The KT scheme is an established numerical scheme, that can cope with the aforementioned challenges. Furthermore, the KT scheme will allow for a straightforward generalization of our numerics, if one extends the GNY model to a model with continuous chiral symmetry group and related Goldstone modes, which enter the LPA RG flow equation as non-linear advective contributions \cite{Grossi:2019urj,Grossi:2021ksl,Koenigstein:2021syz,Steil:2021cbu,Steil:2021partIV}. The original KT scheme, which was presented by A.~Kurganov and E.~Tadmor in Ref.~\cite{KTO2-0}, was in fact developed primarily for non-linear hyperbolic PDEs. However, Ref.~\cite{KTO2-0} also presented generalizations of their scheme for PDEs that involve source/sink and diffusion terms. For the scope of this work, we only use these extensions.\\
	
	The scheme is formulated in a semi-discrete manner. Hence, position (field) space is discretized via a finite volume discretization, while the temporal coordinate is formally continuous. This enables the use of up-to-date generic ODE time-steppers to solve the discretized equations.\footnote{Within this work, we use the time-stepper \texttt{solve\_ivp} with the \texttt{LSODA} option using an Adams/BDF method with automatic stiffness detection and switching from the \textit{SciPy~1.0} library \cite{2020SciPy-NMeth}. For further details, see App.~\ref{app:numerical_tests}. We also crosschecked some of our results with the code-basis of Refs.~\cite{Koenigstein:2021syz,Koenigstein:2021rxj,Steil:2021cbu,Steil:2021partIV}, which is implemented with \texttt{Mathematica} \cite{Mathematica:12.1} using \textit{NDSolve}.}
	
	The explicit discretization of Eqs.~\eqref{eq:pdeq-u}, \textit{i.e.} the diffusion flux \eqref{eq:diffusion_flux} and sink term \eqref{eq:source_sink} is as follows. The compact computational domain $[ 0, \sigma_\mathrm{max} ]$, which was introduced in Sub.Sec.~\ref{subsec:boundary_conditions}, is subdivided into $n \in \mathbb{N}$ equally sized (finite) volume cells of size $\Delta \sigma$ centered at positions $\sigma_i$, $i = 0, 1, \ldots, n - 1$. The zeroth volume cell is centered at ${\sigma = \sigma_0 = 0}$, while the last cell is centered at $\sigma_{n - 1} = \sigma_\mathrm{max}$. Within a single volume cell $\sigma_i$, the cell-average of the ``fluid'' $u ( t, \sigma)$ is denoted as $\bar{u}_i ( t )$. The actual computation and scheme is entirely formulated via these cell-averages that evolve with time.
	
	For the implementation of boundary conditions two (one) so-called \textit{ghost cells} are required at each interval boundary when considering a problem involving advective (solely diffusive) contributions in the semi-discrete KT-scheme. For the GNY model which has no advective contribution in its LPA flow equation one ghost cell at each interval boundary is sufficient. The ghost cells are of size $\Delta \sigma$ and centered at $\sigma_{-1} = - \Delta \sigma$ and $\sigma_{n} = \sigma_\mathrm{max} + \Delta \sigma$. As described in Sub.Sec.~\ref{subsec:boundary_conditions} we use the anti-symmetry of $u ( t, \sigma )$ as boundary condition to fix the cell-averages of the zeroth cell and first ghost cell to $\bar{u}_0 ( t ) = 0$ and $\bar{u}_{-1} ( t ) = - \bar{u}_{1} ( t )$ for all times $t$ respectively. For the ghost cell at $\sigma_n$ we use linear extrapolation, thus for the cell average $\bar{u}_{n} ( t ) = 2 \, \bar{u}_{n - 1} ( t ) - \bar{u}_{n - 2} ( t )$ for all $t$.
	
	Additionally, we introduce the grid of cell interfaces, which are positioned at $\sigma_{i + \frac{1}{2}} \equiv \sigma_i + \tfrac{\Delta \sigma}{2}$.
	
	Within this setup, the semi-discrete scheme for the non-linear diffusion-sink equation \eqref{eq:pdeq-u} and the evolution of the cell-averages $\bar{u}_i ( t )$ reads \cite{KTO2-0}
		\begin{align}
			\partial_t \bar{u}_i = \tfrac{1}{\Delta \sigma} \, \big( P_{i + \frac{1}{2}} - P_{i - \frac{1}{2}} \big) + S_i \, .
		\end{align}
	Here, $S_i = S ( t, \sigma_i )$, where $S ( t, \sigma )$ is the sink term \eqref{eq:source_sink} (after analytic evaluation of the $\sigma$-derivative). For a detailed comment on the implementation of the sink term, we refer to App.~\ref{app:source_sink_implementation}.
	
	Furthermore, according to Ref.~\cite{KTO2-0} we specify
		\begin{align}
			P_{i + \frac{1}{2}} = \, & \frac{Q \big( t, \bar{u}_i, \frac{\bar{u}_{i + 1} - \bar{u}_i}{\Delta \sigma} \big) + Q \big( t, \bar{u}_{i + 1}, \frac{\bar{u}_{i + 1} - \bar{u}_i}{\Delta \sigma} \big)}{2} \, ,
		\end{align}
	where for our purpose $Q ( t, u, \partial_\sigma u ) = Q ( t, \partial_\sigma u )$ is the diffusion flux \eqref{eq:diffusion_flux}, In fact, the independence of $u$ of our special diffusion flux leads to further simplification of the original $P_{i + \frac{1}{2}}$ from Ref.~\cite{KTO2-0}. For an alternative implementation of the diffusion flux in the KT scheme, see, \textit{e.g.},\ Ref.~\cite{Chertock2005}.\\
	
	For more details on the KT-scheme and its original design for advection driven problems, we refer to the original publication Ref.~\cite{KTO2-0}. For a comprehensive discussion of the KT-scheme in the context of RG flow equations, we refer to Refs.~\cite{Koenigstein:2021syz,Koenigstein:2021rxj,Steil:2021cbu,Steil:2021partIV}, where also references for further reading and extensions of the scheme are listed.

\section{Infinite-\texorpdfstring{$N$}{N} analysis within the FRG --\texorpdfstring{\\}{ }Consistency checks and the UV initial condition}
\label{sec:mean-field}

In this section, we recap some of the well-known (analytical) results for the infinite-$N$ limit (mean-field) of the GNY model. Furthermore, we demonstrate that the FRG is capable of reproducing the latter results analytically and numerically. Additionally, these mean-field calculations are used to motivate a proper UV initial condition for the flow equation \eqref{eq:pdeq-U}, but also serve as a consistency check of our numerical implementation in the limit $N \rightarrow \infty$, thus in the limit of vanishing bosonic fluctuations.

\subsection{Mean-field, infinite-\texorpdfstring{$N$}{N}, and FRG}
\label{subsec:mean-field_infinite_n_frg}

	Within the FRG framework (arguably even in general), the term ``mean-field approximation'' has no universal, agreed upon formal definition. Usually performing calculations on ``mean-field level'' in the context of fermionic models refers to computations disregarding bosonic quantum and thermal fluctuations and including only fermionic ones. Whether this includes fermionic vacuum fluctuations and/or fermionic contributions beyond the effective potential usually depends on the work under consideration. In the FRG one way to define a mean-field approximation is the usage of a LPA truncation disregarding the bosonic fluctuations. On a practical level, this results in simply ignoring the bosonic loop contribution(s). On LPA level this can be formally achieved by taking the infinite-$N$ limit for the LPA flow equation under consideration, in this work Eq.~\eqref{eq:pdeq-U}, after appropriate rescalings, like the ones introduced in Eqs.~\eqref{eq:rescaling_with_n} and \eqref{eq:rescaling_with_n_U}.
	
	However, it can be shown that in general the infinite-$N$ limit and ``ignoring the bosonic loop in a LPA truncation'' is not the same procedure. In fact, if starting with a more advanced FRG truncation scheme, like LPA${}^\prime$, which also includes wave function renormalizations, one finds, that even in the infinite-$N$ limit, there are fermionic loop contributions to the bosonic wave-function renormalization, \textit{e.g.}\ Ref.~\cite{Braun:2010tt,Braun:2011pp}. Hence, in general the order of ``limits'' plays a crucial role and ``choosing a truncation scheme in FRG'' and ``taking the infinite-$N$ limit'' do not commute. Furthermore in models including Goldstone bosons a large-$N$ limit for fermions, \textit{e.g.}, in the large-$N_f$ limit for chiral fermion flavors, does not lead to the desired suppression of bosonic modes, because the Goldstone modes do not form as flavor singlets. Another degree of freedom, \textit{e.g.}, the number of colors $N_c$, has to be used to facilitate the large-$N$ limit and the desired suppression of bosonic fluctuations.
	
	For our purpose and the sake of simplicity, we simply start by definition on the level of the LPA and take all limits like the infinite-$N$ limit or the zero-$T$ and zero-$\mu$ limit afterward. Hence, within our truncation, the mean-field limit and the infinite-$N$ limit are considered to be identical, which simplifies the discussion and allows to make direct contact with established conventional mean-field computations for the GN model, \textit{cf.} Refs.~\cite{Wolff:1985av,Thies:2006ti}, which consider only fermionic fluctuations on the level of the effective potential.\\
	
	Performing the large-$N$ limit For the flow equation \eqref{eq:pdeq-U}, yields
		\begin{align}
			& \lim\limits_{N \rightarrow \infty} \partial_t U ( t,  \sigma ) =	\vphantom{\Bigg(\Bigg)}	\label{eq:pdeq-U-MF}
			\\
			= \, & \frac{d_\gamma}{\pi} \, \frac{k^3 ( t )}{2 E_{\textrm{f}} ( t, \sigma )} \, \big( 1 - n_\mathrm{f} [ \beta ( E_{\textrm{f}} ( t, \sigma ) + \mu ) ] -	\vphantom{\Bigg(\Bigg)}	\nonumber
			\\
			& - n_\mathrm{f} [ \beta ( E_{\textrm{f}} ( t, \sigma ) - \mu ) ] \big) \, .	\vphantom{\Bigg(\Bigg)}	\nonumber
		\end{align}
	We find that the former PDE decouples in field space and reduces to a first order ordinary differential equation in $t$ for each $\sigma$ separately. In the fluid dynamic picture, on the level of $u ( t, \sigma ) = \partial_\sigma U ( t,  \sigma )$, this is rather intuitive, since the fermionic contribution to the flow equation \eqref{eq:pdeq-u} presents as a local time-dependent source/sink term \eqref{eq:source_sink} and the spatial movement of the fluid (via diffusion in field space) is totally suppressed.\\
	
	When disregarding bosonic fluctuations completely in mean-field approximation all three model variants -- GN, bGN and GNY -- discussed Sec.~\ref{sec:GNYmodel} and Sub.Sec.~\ref{subsec:comment_on_the_truncation} are equivalent. The bGN and GNY models are identical in mean-field approximation and the bGN model as the bosonized version of the GN Model is in general physically equivalent to the GN model as discussed in Sec.~\ref{sec:GNYmodel}.

\subsection{The mean-field potential and asymptotic freedom}
	
	Due to the decoupling in field space, the differential equation \eqref{eq:pdeq-U-MF} can be integrated analytically in $k ( t )$. Using the definition of the RG time \eqref{eq:def_rg_time}, which implies $\partial_t = - k \, \partial_k$, we find
	\begin{widetext}
		\begin{align}
			U_{k=0}(\sigma) = \, & U_{k=\Lambda} (\sigma ) + \frac{d_\gamma}{\pi} \int_{0}^{\Lambda} \mathrm{d} k \, \frac{k^2}{2 E_\mathrm{f}} \, \big( 1 - n_\mathrm{f} [ \beta ( E_\mathrm{f} + \mu ) ] - n_\mathrm{f} [ \beta ( E_\mathrm{f} - \mu ) ] \big) =	\vphantom{\Bigg(\Bigg)}	\label{eq:rg-U-MF}
			\\
			U_{0}(\sigma) = \, & U_{\Lambda} (\sigma ) + \frac{d_\gamma}{\pi} \, \bigg[ \frac{k}{2} \, \big( E_\mathrm{f} + \tfrac{1}{\beta} \ln \big[ 1 + \mathrm{e}^{- \beta ( E_\mathrm{f} + \mu )} \big] + \tfrac{1}{\beta} \ln \big[ 1 + \mathrm{e}^{- \beta ( E_\mathrm{f} - \mu )} \big] \big) \bigg]_{0}^{\Lambda}	\vphantom{\Bigg(\Bigg)}	\nonumber
			\\
			& - \frac{d_\gamma}{2 \pi} \int_{0}^{\Lambda} \mathrm{d} k \, \big( E_\mathrm{f} + \tfrac{1}{\beta} \, \ln \big[ 1 + \mathrm{e}^{- \beta ( E_\mathrm{f} + \mu )} \big] + \tfrac{1}{\beta} \, \ln \big[ 1 + \mathrm{e}^{- \beta ( E_\mathrm{f} - \mu )} \big] \big) \, ,	\nonumber
		\end{align}
	\end{widetext}
	where the UV-initial condition for this trivial integrable ``RG-flow'' is given by the classical UV-potential, see Sub.Sec.~\ref{subsec:comment_on_the_truncation},
		\begin{align}
			U_{\Lambda}\equiv U ( t=0, \sigma ) = \tfrac{1}{2 g^2} \, ( h \sigma )^2	\label{eq:initial_potential}
		\end{align}
	and where we introduced the notation 
		\begin{align}
			U_k(\sigma)\equiv U_{k(t)}(\sigma)\equiv U(t,\sigma) 
		\end{align}
	for the potential at a given RG scale $k$.\\
	
	In the second line of Eq.~\eqref{eq:rg-U-MF} we integrated by parts in order to recover the usual expression for the grand canonical potential density for $\Lambda \rightarrow \infty$ (up to an infinite constant $\sim k \, E_\mathrm{f} \big|_{k = \Lambda}$), \textit{cf.}\ Refs.~\cite{Harrington:1974tf,Jacobs:1974ys,Wolff:1985av,Thies:2006ti}. The first terms of the integrands in the first and third line of Eq.~\eqref{eq:rg-U-MF} lead to divergences. These divergences have to be canceled by ``renormalizing'' the coupling $g^2$ such that the IR observables are finite and could in principle be matched with experimental observations. The renormalized version of this so called \textit{vacuum contribution} (the only contribution, that does not depend on $\mu$ and $T$) is directly linked to the initial condition of our RG-flows, when solving the PDE \eqref{eq:pdeq-U} and ODE \eqref{eq:pdeq-U-MF} (numerically), see below.
	
	To this end, we turn to the $\mu = 0$ and $\beta \rightarrow \infty\Leftrightarrow T\rightarrow 0$ limit of Eq.~\eqref{eq:rg-U-MF},
		\begin{align}
			U_{0;\mathrm{vac}} (\sigma ) \equiv \, & \lim\limits_{\mu,T \rightarrow 0} U_0 ( \sigma ) =	\vphantom{\Bigg(\Bigg)}	\label{eq:ir_potential_mean-field_vac}
			\\
			= \, & \frac{h^2 \sigma^2}{2 g^2} + \frac{d_\gamma}{2 \pi} \int_{0}^{\Lambda} \mathrm{d} k \, \frac{k^2}{\sqrt{ k^2 + h^2 \sigma^2 }} \, .	\vphantom{\Bigg(\Bigg)}	\nonumber
		\end{align}
	and study the corresponding gap equation \eqref{eq:gapeq} at possible non-trivial minima $\sigma_0 \neq 0$, \textit{cf.}\ Refs.~\cite{Rosenstein:1990nm,Wolff:1985av,Thies:2006ti},
		\begin{align}
			0 \overset{!}{=} \, & \frac{1}{h^2\sigma_0} \, \partial_{ \sigma} U_{0;\mathrm{vac}} (\sigma ) \big|_{ \sigma = \sigma_0} =	\vphantom{\Bigg(\Bigg)} \label{eq:gapeq}
			\\
			= \, & \frac{1}{g^2} - \frac{d_\gamma}{2\pi} \int_{0}^{\Lambda} \mathrm{d} k \, \frac{k^2}{( k^2 + h^2 \sigma_0^2 )^\frac{3}{2}} =	\vphantom{\Bigg(\Bigg)}	\nonumber
			\\
			= \, & \frac{1}{g^2} + \frac{d_\gamma}{2 \pi} \, \bigg[ \frac{k}{\sqrt{ k^2 + h^2 \sigma_0^2 }} \bigg|_0^\Lambda - \int_{0}^{\Lambda} \mathrm{d} k \, \frac{1}{\sqrt{ k^2 + h^2 \sigma_0^2 }}\bigg]	= \vphantom{\Bigg(\Bigg)}	\nonumber
			\\
			= \, & \frac{1}{g^2} + \frac{d_\gamma}{2 \pi} \, \bigg[ \Big[ 1 + \big( \tfrac{h \sigma_0}{\Lambda} \big)^2 \Big]^{- \frac{1}{2}} - \vphantom{\Bigg(\Bigg)}	\nonumber
			\\
			& - \mathrm{artanh} \bigg( \Big[ 1 + \big( \tfrac{h \sigma_0}{\Lambda} \big)^2 \Big]^{- \frac{1}{2}} \bigg) \bigg] \, . \vphantom{\Bigg(\Bigg)}	\nonumber
		\end{align}
	Hence, as a first result, assuming that $\sigma_0$ is non-zero and finite, we can study the asymptotic behavior of the equation for $\Lambda \ggg h$,
		\begin{align}
			\tfrac{1}{g^2} = \, & \tfrac{d_\gamma}{2 \pi} \, \big[ - 1 + \tfrac{1}{2} \ln \big( \big( \tfrac{2 \Lambda}{h \sigma_0} \big)^2 \big) \big] + \mathcal{O} \big( \big( \tfrac{h \sigma_0}{\Lambda} \big)^2 \big) \, .	\label{eq:asymptotics_of_g}
		\end{align}
	This reflects the asymptotically free behavior of the four-Fermi coupling \cite{Gross:1974jv,ZinnJustin:2002ru,Peskin:1995ev} of the original Gross-Neveu model \eqref{eq:gn-model}, since
		\begin{align}
			\lim\limits_{\frac{\Lambda}{h} \rightarrow \infty} g^2 = 0 \, .
		\end{align}
	Furthermore, we can use \eqref{eq:asymptotics_of_g} and solve for $\sigma_0$,
		\begin{align}
			\sigma_0 = \pm \tfrac{2 \Lambda}{h} \, \mathrm{e}^{- \big( \frac{4 \pi}{d_\gamma g^2} + \frac{1}{2} \big)} \, .
		\end{align}
	Hence, due to the asymptotic free behavior of $g^2$, there is $\mathbb{Z}_2$ (discrete chiral) symmetry breaking via two non-trivial minima for all non-zero $g^2$ in vacuum and for infinite-$N$ -- a central feature of the GN model \cite{Gross:1974jv}.

	Furthermore, we can insert the results from the gap equation \eqref{eq:gapeq} in Eq.~\eqref{eq:initial_potential} and solve for the UV potential
		\begin{align}
			U_\Lambda( \sigma ) = \, & \tfrac{d_\gamma}{4 \pi} \, ( h \sigma )^2 \, \bigg[ \mathrm{artanh} \bigg( \Big[ 1 + \big( \tfrac{h \sigma_0}{\Lambda} \big)^2 \Big]^{- \frac{1}{2}} \bigg) -	\vphantom{\Bigg(\Bigg)}	\label{eq:initial-condition}
			\\
			& - \Big[ 1 + \big( \tfrac{h \sigma_0}{\Lambda} \big)^2 \Big]^{- \frac{1}{2}} \bigg] \, .	\vphantom{\Bigg(\Bigg)}	\nonumber
		\end{align}
	In the limit $\tfrac{\Lambda}{h} \to \infty$ we approach the Gaussian fixed-point (UV fixed-point) for the bosonized GN model \eqref{eq:bgn-model}, which becomes clear when considering dimensionless quantities
		\begin{align}
			&	\tilde h = \tfrac{1}{\Lambda} \, h \, ,	&&	\tilde{U}_\Lambda ( \sigma ) = \tfrac{1}{\Lambda^2} \, U_\Lambda (\sigma ) \, .	\label{eq:gaussian_fixed_point}
		\end{align}
	Both, $\tilde{h}$ and $\tilde{U}_\Lambda (\sigma )$, vanish in the limit $\tfrac{\Lambda}{h} \to \infty$. This implies -- as expected -- that the bGN model \eqref{eq:bgn-model} in vacuum in the infinite-$N$ limit, is also asymptotically free.\\
	
	Turning to the IR mean-field potential \eqref{eq:ir_potential_mean-field_vac} and using the previous results, we find
		\begin{align}
			& U_{0;\mathrm{vac}} ( \sigma ) =	\vphantom{\Bigg(\Bigg)}\label{eq:U0vac}
			\\
			= \, & \tfrac{d_\gamma}{4 \pi} \, ( h \sigma )^2 \, \bigg[ \mathrm{artanh} \bigg( \Big[ 1 + \big( \tfrac{h \sigma_0}{\Lambda} \big)^2 \Big]^{- \frac{1}{2}} \bigg) -	\vphantom{\Bigg(\Bigg)}	\nonumber
			\\
			& - \Big[ 1 + \big( \tfrac{h \sigma_0}{\Lambda} \big)^2 \Big]^{- \frac{1}{2}} - \mathrm{artanh} \bigg( \Big[ 1 + \big( \tfrac{h \sigma}{\Lambda} \big)^2 \Big]^{- \frac{1}{2}} \bigg) \bigg] +	\vphantom{\Bigg(\Bigg)}	\nonumber
			\\
			& + \tfrac{d_\gamma}{4 \pi} \, \Lambda^2 \, \sqrt{1 + \big( \tfrac{h \sigma}{\Lambda} \big)^2 } \, .	\vphantom{\Bigg(\Bigg)}	\nonumber
		\end{align}
	Considering the first derivative of Eq.~\eqref{eq:U0vac} one can verify ${0=\partial_\sigma U_{0;\mathrm{vac}}( \sigma )|_{\sigma_0}}$ which has to hold by construction and we note that
		\begin{align}
			\partial_\sigma^2 U_{0; \mathrm{vac}} ( \sigma ) |_{\sigma_0} = \frac{d_\gamma}{2 \pi} \, \frac{\Lambda^3}{[ \Lambda^2 + ( h \sigma_0 )^2 ]^{3/2}} \, h^2, \label{eq:msigma_MF_Lambda}
		\end{align}
	which is manifest positive -- again consistent with the notion of $\sigma_0$ as a non-trivial minimum by construction -- and corresponds to the squared curvature mass $m_\sigma^2$ of the sigma mode in vacuum.
	
	Considering the limit $\tfrac{\Lambda}{h} \rightarrow \infty$, the divergent contributions of the two $\mathrm{artanh}$ cancel exactly. For the last term, we use
		\begin{align}
			\Lambda^2 \, \sqrt{1 + \big( \tfrac{h \sigma}{\Lambda} \big)^2} = \Lambda^2 + \tfrac{1}{2} \, ( h \sigma )^2 + \mathcal{O} \big( \big( \tfrac{h \sigma}{\Lambda} \big)^2 \big) \, ,
		\end{align}
	which results in an unobservable infinite constant and a finite contribution. In total we find the well-known renormalized vacuum IR effective potential, \textit{cf.} Refs.~\cite{Wolff:1985av,Thies:2006ti},
		\begin{align}
			& U_{0;\mathrm{vac}} ( \sigma ) =	\vphantom{\bigg(\bigg)}	\label{eq:vacuum_potential}
			\\
			= \, & \tfrac{d_\gamma}{8 \pi} \, ( h \sigma )^2 \, \big( \big[ \ln \big( \tfrac{h \sigma}{h \sigma_0} \big)^2 \big] - 1 \big) - \tfrac{d_\gamma}{4 \pi} \, \Lambda^2 + \mathcal{O} \big( \big( \tfrac{h \sigma}{\Lambda} \big)^2 \big) \, ,	\vphantom{\bigg(\bigg)}	\nonumber
		\end{align}
	with its global minimum at $\pm\sigma_0$ and with a corresponding squared curvature mass of 
		\begin{align}
			m_\sigma^2 = \tfrac{d_\gamma}{2\pi} \, h^2 \, ,	\label{eq:msigma_MF}
		\end{align}
	\textit{cf.}\ Ref.~\cite{Gross:1974jv}. Finally, the result \eqref{eq:vacuum_potential} can be used to also simplify the full IR potential in medium \eqref{eq:rg-U-MF} by replacing the vacuum contributions,
	\begin{widetext}
		\begin{align}
			& U_0( \sigma ) =	\vphantom{\Bigg(\Bigg)}	\label{eq:medium_potential}
			\\
			= \, & \tfrac{d_\gamma}{8 \pi} \, ( h \sigma )^2 \, \big( \big[ \ln \big( \tfrac{h \sigma}{h \sigma_0} \big)^2 \big] - 1 \big) - \tfrac{d_\gamma}{4 \pi} \, \Lambda^2 + \mathcal{O} \big( \big( \tfrac{h \sigma}{\Lambda} \big)^2 \big) - \tfrac{d_\gamma}{2 \pi} \int_{0}^{\Lambda} \mathrm{d} p \, \big(\tfrac{1}{\beta} \, \ln \big[ 1 + \mathrm{e}^{- \beta ( E_\mathrm{f} + \mu )} \big] + \tfrac{1}{\beta} \, \ln \big[ 1 + \mathrm{e}^{- \beta ( E_\mathrm{f} - \mu )} \big] \big) \, .	\vphantom{\Bigg(\Bigg)}	\nonumber
		\end{align}
	\end{widetext}
	Let us again point out that mean-field IR potentials \eqref{eq:vacuum_potential} and \eqref{eq:medium_potential} of the bGN/GNY model \eqref{eq:bgn-model}/\eqref{eq:gny-model} are unique. For $\Lambda \rightarrow \infty$ we can read off from Eq.~\eqref{eq:vacuum_potential} that the model contains only a single internal dimensionful parameter, for instance $h$, because the bosonic field $\sigma$ is dimensionless and $h$ and $\sigma$ only appear in combination. All other dimensionful quantities ($\mu$, $T$, $\Lambda$, $U$) can be expressed via this single parameter and results for different specifications of the reference parameter can be mapped into each other via simple rescaling. A direct consequence is that also the mean-field phase diagram is unique, as is discussed in Sub.Sec.~\ref{subsec:phase_diagram_mean_field}.

\subsection{Analytical results for the mean-field phase diagram in the renormalized limit \texorpdfstring{$\Lambda\rightarrow\infty$}{}}
\label{subsec:phase_diagram_mean_field}
	
	In this subsection we will discuss the mean-field potential $U_0 ( \sigma )$ of Eq.~\eqref{eq:medium_potential} in the limit $\Lambda\rightarrow\infty$. We will compare our analytical results of this subsection with the existing renormalized mean field results of specifically Refs.~\cite{Wolff:1985av,Thies:2006ti} for the Gross-Neveu model at vanishing bare fermion mass. Of special interest is the phase-diagram (see Figure~1 of Ref.~\cite{Thies:2006ti} or Fig.~3 of the preceding publication~\cite{Wolff:1985av}) and the underlying renormalized grand canonical potential density, see, \textit{e.g.} Eq.~(4) of Ref.~\cite{Thies:2006ti}. The results of this section serve as a reference in the consistence check of our numerical implementation in Sub.Sec.~\ref{subsec:numeric_consistency_check_mean-field}.\\

	Consider the mean-field potential $U_0(\sigma)$ of Eq.~\eqref{eq:medium_potential} in the renormalized limit $\Lambda\rightarrow\infty$,
	\begin{widetext}
		\begin{align}
			& \lim_{\Lambda \rightarrow \infty} U_0(\sigma) - \tfrac{d_\gamma }{4\pi} \, \Lambda^2 =	\vphantom{\Bigg(\Bigg)}	\label{eq:rg-U-MF-largeLambda}
			\\
			= \, & \tfrac{d_\gamma}{8 \pi} \, ( h \sigma )^2 \, \big( \ln \big[ \big( \tfrac{h \sigma}{h \sigma_0} \big)^2 \big] - 1 \big) - \tfrac{d_\gamma}{4 \pi} \int_{-\infty}^{\infty} \mathrm{d} k \, \big( \tfrac{1}{\beta} \, \ln \big[ 1 + \mathrm{e}^{- \beta ( E_\mathrm{f} + \mu )} \big] + \tfrac{1}{\beta} \, \ln \big[ 1 + \mathrm{e}^{- \beta ( E_\mathrm{f} - \mu )} \big] \big) \, ,	\vphantom{\Bigg(\Bigg)}	\nonumber
		\end{align}
	\end{widetext}
	where we used the symmetry ($k \rightarrow - k$) of the remaining integral. We identify several noteworthy terms in Eq.~\eqref{eq:rg-U-MF-largeLambda}: the remaining integral is the canonical one-loop, convergent medium contribution of fermions in $1+1$ dimensions, the vacuum contribution carries a, for $1 + 1$ dimensions typical \cite{Actor:1985vh,Actor:1985xp,Actor:1986zf}, term proportional to $( h \sigma )^2 \, \log ( h \sigma )$ and we subtracted a (in $\sigma$, $\mu$ and $\beta$) constant but otherwise divergent contribution proportional $\Lambda^2$ to perform the limit $\Lambda \rightarrow \infty$. In observables this divergent but constant contribution cancels (since $U_0(\sigma)$ is only defined up to a constant) and therefore it can be subtracted from $U_0(\sigma)$ without further consequences. For the following we therefore consider
		\begin{align}
			V ( \sigma ) \equiv \, & \lim_{\Lambda \rightarrow \infty} U_0(\sigma) - \tfrac{d_\gamma }{4\pi} \, \Lambda^2  =	\vphantom{\bigg(\bigg)}	\label{eq:rgMFV}
			\\
			= \, & \tfrac{1}{4 \pi} \, \sigma^2 \, [ \log( \sigma^2 ) - 1 ] + I_V ( T, \mu ) \, ,	\vphantom{\bigg(\bigg)}	\nonumber
		\end{align}
	where we used $d_\gamma=2$ as well as \textit{w.l.o.g.} $h = 1$ and $\sigma_0 = 1$ and abbreviate the medium contribution with 
		\begin{align}
			I_V ( T, \mu ) \equiv \, & - \tfrac{T}{2 \pi} \int_{-\infty}^{\infty} \mathrm{d} k \, \ln \big[ 1 + \mathrm{e}^{- \tfrac{1}{T} ( E_\mathrm{f} + \mu )} \big] -	\vphantom{\Bigg(\Bigg)}
			\\
			& - \tfrac{T}{2 \pi} \int_{-\infty}^{\infty} \mathrm{d} k \, \ln \big[ 1 + \mathrm{e}^{- \tfrac{1}{T} ( E_\mathrm{f} - \mu )} \big] \, .	\vphantom{\Bigg(\Bigg)}	\nonumber
		\end{align}
	For a given $\mu$ and $T$ the physical value of the condensate $\sigma$ is found by minimization of $V ( \sigma )$, \textit{cf.} Sub.Sec.~\ref{subsec:dSB}. When evaluated at its global minimum $V ( \sigma )$ is the renormalized grand canonical potential per spatial volume. The renormalized (FRG) mean-field result of Eq.~\eqref{eq:rgMFV} for $V ( \sigma )$ obtained in the infinite-$N$ limit of the LPA flow equation with a one-dimensional LPA-optimized regulator \eqref{eq:pdeq-U-MF} coincides with the renormalized mean-field results in literature, see, \textit{e.g.}, Refs.~\cite{Wolff:1985av,Thies:2006ti}, and explicitly Eq.~(4) of Ref.~\cite{Thies:2006ti}. In the rest of this subsection we derive and recapitulate known MF results for the homogeneous phase diagram of the GN model, see, \textit{e.g.}, \cite{Wolff:1985av,Thies:2006ti}. This discussion may be skipped by readers familiar with these results.\\
	
	\begin{figure}
		\centering
		\includegraphics{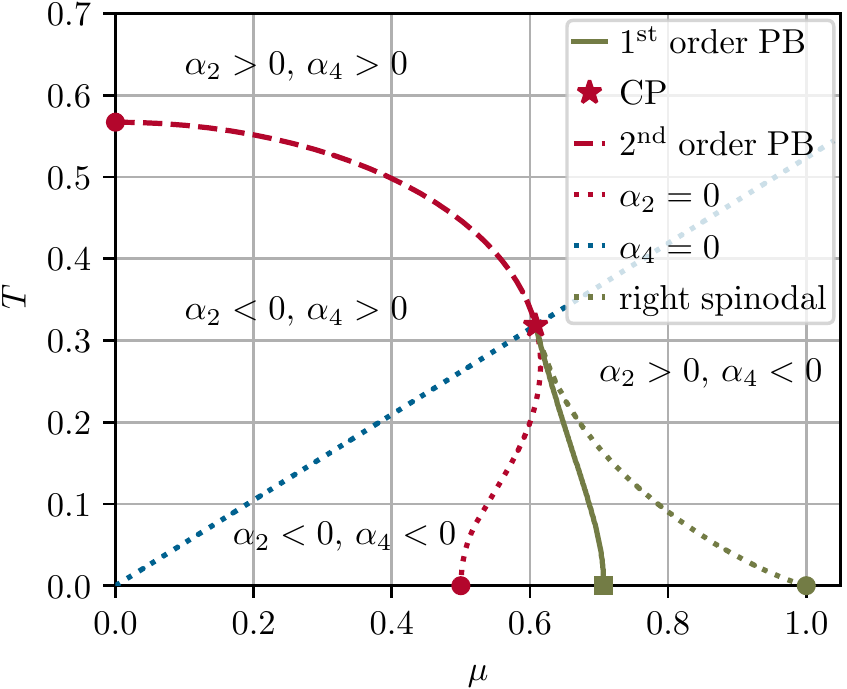}
		\caption{\label{fig:GNlargeN_PD}
			Phase diagram of the renormalized GN model in the infinite-$N$ limit. The {blue}\ ($\alpha_4=0$) and  {red}\ ($\alpha_2=0$) lines are obtained from the Ginzburg-Landau type expansion and Eq. \eqref{eq:alpha2} and Eq.~\eqref{eq:alpha2n} for $n=2$ respectively. The critical point denoted by the {red} star separates the second order phase boundary (red-dashed line) from the first order phase boundary (solid green line). The red-dotted line and the green-dotted line are the left and right spinodal lines. The first order phase boundary and the right spinodal have been obtained by explicit numerical integration \cite{cubature:2020} and subsequent repeated, local, numerical minimization \cite{Nelder:1965} of the renormalized grand canonical potential per spatial volume \eqref{eq:rgMFV}. The critical temperature $T_\mathrm{C}$ is marked on the $T$-axis with a red disk, while the end points at $\mu_\mathrm{L}$ and $\mu_\mathrm{R}$ of the spinodal lines are marked on the $\mu$-axis with red and green disks respectively. The first-order phase transition at zero temperature is marked with a green square at $\mu_1$.
		}
	\end{figure}
	
	The phase diagram of Fig.~\ref{fig:GNlargeN_PD} identical to the one presented in Fig.~1 of Ref.~\cite{Thies:2006ti}. By construction -- due to the renormalization condition of Eq.~\eqref{eq:gapeq} -- discrete chiral symmetry is broken with $\sigma=\sigma_0=1$ in the vacuum at $T=\mu=0$. At high chemical potentials and temperatures the discrete chiral symmetries is restored and therefore $\sigma=0$. At intermediate temperatures and chemical potentials one observes a first order phase transition line starting at zero temperature and non-zero chemical potential $\mu_1$ and ending in a critical point at $( \mu_{\mathrm{CP}}, T_{\mathrm{CP}} )$. Above $T_{\mathrm{CP}}$ discrete chiral symmetry is restored across a second order phase transition line starting at the critical point and ending on the $(\mu=0)$-axis at a non-zero temperature $T_\mathrm{C}$.
	
	The position of the critical point as well as $\mu_1$ and $T_\mathrm{C}$ can be computed with the help of known functions without the need of numerical minimization of the potential $V ( \sigma )$. We devote the remainder of this subsection to the derivation of the values for $T_\mathrm{C}$, $T_\mathrm{CP}$, $\mu_{\mathrm{CP}}$ and $\mu_1$ given in Refs.~\cite{Thies:2006ti,Wolff:1985av,Treml:1989} as well as expressions for specific lines in Fig.~\ref{fig:GNlargeN_PD} .\\
	
	For $T \rightarrow 0$ the integral $I_V ( 0, \mu )$ can be performed analytically,
		\begin{align}
			& I_V ( 0, \mu ) =	\vphantom{\bigg(\bigg)}	\label{eq:rgMFIT0}
			\\
			=\, & \tfrac{1}{2\pi} \, \Big( \mathrm{arsinh} \Big[ \sqrt{ \big( \tfrac{\mu}{\sigma} \big)^2 - 1 } \Big] - \mu \, \sqrt{\mu^2 - \sigma^2} \Big) \, \Theta \big[ \big( \tfrac{\mu}{\sigma} \big)^2 - 1 \big] \, ,	\vphantom{\bigg(\bigg)}	\nonumber
		\end{align}
	and we note $I_V ( 0, 0 ) = 0$ as all vacuum contributions are already integrated out and included in Eq.~\eqref{eq:rgMFV}. An analysis of Eq.~\eqref{eq:rgMFIT0} reveals, that the extremum at $\sigma = 0$ becomes a local minimum for $\mu > \mu_\mathrm{L} = \tfrac{1}{2}$. For ${\mu \in [ \mu_\mathrm{L}, \mu_\mathrm{R} ]}$ the potential has three minima (one at $\sigma = 0$ and two at $\sigma = \pm 1$) and at $\mu_1 = \tfrac{1}{\sqrt{2}} \simeq 0.707107$ all local minima become global minima signaling a first order phase transition at $\mu_1$. The notable chemical potentials $\mu_\mathrm{L}$, $\mu_1$ and $\mu_\mathrm{R}$ are marked on the $(T=0)$-axis in Fig.~\ref{fig:GNlargeN_PD}.

	For $T > 0$ the integral $I_V ( T, \mu )$ can be performed numerically or it can be rewritten in terms of known functions, see, \textit{e.g.}, Ref.~\cite{Actor:1986zf} Eqs.~(3.11) \textit{ff.} and Refs.~\cite{Actor:1985vh,Actor:1985xp},
		\begin{align}
			& I_V ( T, \mu ) =	\vphantom{\Bigg(\Bigg)}	\label{eq:rgMFIseries}
			\\
			= \, & - \tfrac{2}{\pi} \, T^2 \sum_{m=1}^{\infty}\tfrac{(-1)^{m + 1}}{m} \, K_1 \big( m \, \tfrac{\sigma}{T} \big) \, \cosh \big( m \, \tfrac{\mu}{T} \big) \, ,	\vphantom{\Bigg(\Bigg)}	\nonumber
		\end{align}
	with the modified Bessel function of second kind $K_n ( x )$ at $n = 1$. This representation is particularly useful when one is interested in a Ginzburg-Landau-type series expansion of $V ( \sigma )$, \textit{cf.} Refs.~\cite{Klevansky:1992qe,Buballa:2014tba,Buballa:2018hux}, around $\sigma = 0$ at $T > 0$, \textit{e.g.},\ to study second order phase transitions within a mean-field approach. Using Eqs.~\eqref{eq:rgMFV}, \eqref{eq:rgMFIT0} and \eqref{eq:rgMFIseries} one can show\footnote{The computation at $T > 0$ is based on the ascending series for $K_1 ( x )$, see, \textit{e.g.}, Ref.~\cite{abramowitz+stegun} Eq.~(9.6.11), and various identities for the polylogarithm and Riemann zeta function, see \textit{e.g.}, Refs.~\cite{abramowitz+stegun,Erdelyi1953}. For $T = 0$ the coefficients $\alpha_{2n}$ can be computed directly from Eq.~\eqref{eq:rgMFIT0} with the help of the series expansion for the $\mathrm{arcsinh} ( x )$ for $|x|>1$, see, \textit{e.g.}, Ref.~\cite{abramowitz+stegun} Eq.~(4.6.31).},
		\begin{align}
			V ( \sigma ) = \, & \sum_{n = 0}^{\infty} \alpha_{2n} \, \sigma^{2n} \, ,	\label{eq:GLpotential}
		\end{align}
	where the zeroth order coefficient reads
		\begin{align}
			\alpha_0 = \, &  - \tfrac{\pi}{6} \, T^2 - \tfrac{1}{2\pi} \, \mu^2 \, ,	\label{eq:alpha0}
		\end{align}
	and the quadratic order coefficient is given by
	\begin{widetext}
		\begin{align}
			\alpha_2 =
			\begin{cases}
				\tfrac{1}{2\pi} \, \ln ( 2 \mu ) \, ,	&	\text{for} \quad T = 0 \, ,  \mu > 0 \, ,	\vphantom{\bigg(\bigg)}
				\\
				- \tfrac{1}{2\pi} \, [ \gamma - \ln ( \pi T ) ] \, ,	&	\text{for} \quad T > 0 \, , \mu = 0 \, ,	\vphantom{\bigg(\bigg)}
				\\
				- \tfrac{1}{2\pi} \, \big[ \gamma - \ln ( 2 T ) + \mathrm{DLi}_{0} \big( \tfrac{\mu}{T} \big) \big] \, ,	&	\text{for} \quad T > 0 \, ,\mu > 0 \, .	\vphantom{\bigg(\bigg)}
			\end{cases}	\label{eq:alpha2}
		\end{align}
	The higher order coefficients  $n \geq 2$ can also be expressed in terms of known functions,
		\begin{align}
			\alpha_{2n} = \, &
			\begin{cases}
				- \tfrac{1}{\pi} \, \tfrac{( 2 n - 3 )!!}{2^{n + 1} ( n - 1)! n!} \, \mu^{2 - 2 n} \, ,	&	\text{for} \quad T = 0 \, , \mu >0 \, ,	\vphantom{\bigg(\bigg)}
				\\
				\tfrac{( - 1 )^n}{\pi^{2 n - 1}} \, \tfrac{4^{1 - 2 n} ( 4^n - 2 ) ( 2 n - 2)!}{( n - 1 )! n!} \, \zeta ( 2 n - 1 ) \, T^{2 - 2 n} \, ,	&	\text{for} \quad T > 0 \, , \mu = 0 \, ,	\vphantom{\bigg(\bigg)}
				\\
				- \tfrac{1}{\pi} \, \tfrac{2^{1 - 2 n}}{n! ( n - 1)!} \, \mathrm{DLi}_{2 - 2 n} \big( \tfrac{\mu}{T} \big) \, T^{2 - 2 n} \, ,	&	\text{for} \quad T > 0 \, , \mu > 0 \, .	\vphantom{\bigg(\bigg)}
			\end{cases}	\label{eq:alpha2n}
		\end{align}
	\end{widetext}
	Here, we defined
		\begin{align}
			\mathrm{DLi}_n ( z ) \equiv \big[ \tfrac{\partial}{\partial s} \mathrm{Li}_s ( -\mathrm{e}^z ) + \tfrac{\partial}{\partial s}\mathrm{Li}_s ( - \mathrm{e}^{-z} ) \big]_{s = n} \, ,
		\end{align}
	with the polylogarithm $\mathrm{Li}_m ( x )$\footnote{The numerical evaluation of $\mathrm{DLi}_n ( z )$ involves derivatives of the the polylogarithm $\mathrm{Li}_m ( x )$ \textit{w.r.t.} to $m$ evaluated at $m=n\leq0$. For the scope of this work we compute $\mathrm{DLi}_n ( z )$ numerically to 32 significant digits using \textit{Maple} \cite{Maple:2017.3}.}. We also used the Riemann zeta function $\zeta ( x )$ and the Euler-Mascheroni constant $\gamma \simeq 0.577216$.
	
	The expression \eqref{eq:alpha0} for $- \alpha_0$ is the Stefan-Boltzmann pressure of a massless free Fermi gas in $1 + 1$ dimensions, \textit{cf.} Ref.~\cite{Kleinert:2016}. The logarithmic divergent contribution $\log(\sigma^2)$ for $\sigma\rightarrow 0$ to $\alpha_2$ stemming from the vacuum term in Eq.~\eqref{eq:rgMFV} gets canceled exactly in Eq.~\eqref{eq:alpha2} by a corresponding contribution from the medium term.
	
	The expansion \eqref{eq:GLpotential} around $\sigma = 0$ can be used to compute the second-order phase-boundary in terms of known functions including the critical end point. The second order phase transition between the restored and a broken phase with small $\sigma > 0$ occurs at $\alpha_2 = 0$ while $\alpha_4 > 0$. In the vicinity of a second order phase transition one finds $\alpha_2 > 0$ and $\alpha_4 > 0$ in the restored phase while $\alpha_2 < 0$ and $\alpha_4 > 0$ holds in the broken phase in this context. Using Eq.~\eqref{eq:alpha2} we find the transition temperature at $\mu=0$
		\begin{align}
			T_\mathrm{C} = \tfrac{\mathrm{e}^\gamma}{\pi} \simeq 0.566933 \, .\label{eq:Tc_MF}
		\end{align}
	The curvature $\kappa$ of the second order phase boundary $T_\mathrm{C} \big( \tfrac{\mu}{T} \big) = T_\mathrm{C} \, \big[ 1 - \kappa \, \big( \tfrac{\mu}{T} \big)^2 + \ldots \big]$ at $\mu = 0$ can be computed using Eq.~\eqref{eq:alpha2} and is given by $\kappa \approx 0.241671$.

	The critical end point is located at the intersection of the $\alpha_2 = 0$ and $\alpha_4 = 0$ lines, \textit{cf.} Refs.~\cite{Buballa:2014tba,Buballa:2018hux}. Using Eq.~\eqref{eq:alpha2n} for $n = 2$ we determine $\tfrac{\mu_\mathrm{CP}}{T_\mathrm{CP}} \simeq 1.910669$ from the only root of $\mathrm{DLi}_{-2}$, which is $\mathrm{DLi}_{-2} ( \tfrac{\mu_\mathrm{CP}}{T_\mathrm{CP}} ) = 0$. Having the ratio $\tfrac{\mu_\mathrm{CP}}{T_\mathrm{CP}}$, we determine
		\begin{align}
			(\mu_\mathrm{CP}, T_\mathrm{CP} ) \simeq ( 0.608221, 0.318329 ) \, ,\label{eq:muCPTCP_MF}
		\end{align}
	using Eq.~\eqref{eq:alpha2}. Above $T_\mathrm{CP}$ we have the second order phase transition with $\alpha_4 > 0$, while below $T_\mathrm{CP}$ we have the spinodal region with $\alpha_4 < 0$ and $\alpha_6 > 0$ in the vicinity of the critical point. $\alpha_2 > 0$, $\alpha_4 < 0$ and $\alpha_6 > 0$ allows for three local minima in the Ginzburg-Landau potential \eqref{eq:GLpotential} when considered up to $n = 3$ which indicates a first order phase transition below $T_\mathrm{CP}$. At the critical point ($\alpha_2 = \alpha_4=0$ and $\alpha_6 > 0$) the non-trivial minima from the ($\alpha_2 > 0$, $\alpha_4 < 0$)-scenario and ($\alpha_2 < 0$, $\alpha_4 > 0$)-scenario merge in $\sigma = 0$. The $\alpha_2 = 0$ line below $T_\mathrm{CP}$ signals the appearance of a local minimum at $\sigma = 0$ and is called the left spinodal line ending in $\mu_\mathrm{L}$ at $T = 0$, in accordance with the results at $T = 0$ discussed earlier in this subsection. The location of the first order phase transition and the right spinodal line, indicating the disappearance of multiple local minima, are not accessible with Ginzburg-Landau expansion around $\sigma = 0$ and have to be determined by numerical minimization of renormalized grand canonical potential per spatial volume \eqref{eq:rgMFV}. Phase boundaries, the $\alpha_2=0$ and $\alpha_4=0$ lines and the spinodal region are displayed in Fig.~\ref{fig:GNlargeN_PD}.

\subsection{UV initial condition for the effective potential for RG flows with and without bosonic fluctuations}\label{subsec:UUV}

	Next, we again turn to the UV initial condition for the effective potential and continue the discussion of Sub.Sec.~\ref{subsec:comment_on_the_truncation}. For practical calculations within the FRG framework, where the flow is not integrable analytically, we can not initialize the RG flow directly at $\Lambda = \infty$. But rather we have to choose some very large and finite $\Lambda$\footnote{$\Lambda$ should be significantly larger than the internal scales (here mainly $h \sigma_0$ and $\partial_\sigma^2 U ( \sigma_0 )$) and external scales (here $\mu$ and $T$). For a detailed discussion, we refer to Refs.~\cite{Braun:2018svj,Koenigstein:2021syz}.}, where we specify the initial values for the flow via $\bar{\Gamma}_\Lambda [ \Phi ] = \mathcal{S} [ \Phi ]$.
	
	From the previous discussion, it is obvious (at least in the mean-field approximation), how to specify the initial condition for a numeric solution of Eq.~\eqref{eq:pdeq-U-MF} or rather Eq.~\eqref{eq:pdeq-u} with $Q ( t, \partial_\sigma u) = 0$. Via our discussion on asymptotic freedom, we were able to eliminate $g^2$ from the initial condition \eqref{eq:initial_potential} in favor of the UV cutoff $\Lambda$ and the combination $h \sigma_0$. Hence, we can simply use Eq.~\eqref{eq:initial-condition} as the initial potential at some large scale $\Lambda$. Initializing the (numeric) mean-field version of the RG flow \eqref{eq:pdeq-u} with the $\sigma$-derivative of \eqref{eq:initial-condition} and an arbitrary value for $h$ at some scale $\Lambda \ggg h$, one always finds that the IR minimum in vacuum is located at $\sigma_0$. Consequently and \textit{w.l.o.g.}\ we can rescale all dimensionful quantities in terms of $h$ and express the dimensionless field space variable $\sigma$ in multiples of $\sigma_0$. On the level of the equations, this amounts to setting $\sigma_0 = 1$ and $h = 1$. Other choices for $h$ and $\sigma_0$ correspond to different renormalization conditions, but all results are unique and can be transformed into each other via simple rescaling -- as already mentioned. However, we still have to ensure that other IR observables do not depend on the UV cutoff $\Lambda$, which is realized by choosing $\Lambda$ much larger than all internal and external model scales. This shown numerically in Sec.~\ref{subsec:numeric_consistency_check_mean-field} and discussed in the context of RG consistency.\\
	
	Including bosonic quantum fluctuations, it is less obvious, how to choose the initial condition for the RG flow, that means how to choose a meaningful value for $\frac{h^2}{g^2}$ in Eq.~\eqref{eq:uv-initial}. When performing calculations at finite $N$, each individual choice of $N$ represents a single model on its own. Hence, even if there is symmetry breaking for the vacuum flow equation \eqref{eq:vacuum_limit_flow_equation} including bosonic quantum fluctuations, the IR physics for different $N$ is not necessarily directly comparable. Thus, setting a unique renormalization condition for all $N$ in the IR, like fixing the position of the IR minimum and/or the IR curvature mass by tuning the various UV-initial conditions, is -- to the best of our knowledge -- not useful.
	
	We think that it is natural to use exactly the same initial condition for RG flows, namely Eq.~\eqref{eq:initial-condition}, with and without bosons and to fix the renormalization condition for the bosonic RG flows in the UV.	This might be counter-intuitive, because in a lot of RG studies for effective models of strongly correlated systems, the physics is fixed in the IR and the UV initial condition is tuned, in such a way that the RG flow ends up with an IR effective action having observables compatible with desired numerical values. Actually, our situation is much closer to what is done in FRG studies for QCD, see, \textit{e.g.}, Refs.~\cite{PhysRevD.69.025001,Braun:2003ii,PhysRevLett.106.022002,PhysRevD.87.076004,Herbst:2013ufa,PhysRevD.91.054035,PhysRevD.92.076012,PhysRevD.94.034016,Springer2017,Fu:2019hdw} as well as the lecture notes \cite{PawlowskiScript} and the recent review \cite{Dupuis:2020fhh} -- another asymptotically free QFT. Also there, the asymptotic freedom, which is derived from the one-loop beta function, is used to fix the gauge-coupling at some extremely large $\Lambda$ in the UV.
	
	Overall our formal arguments for choosing Eq.~\eqref{eq:initial-condition} are as follows. Using the same UV initial condition for all finite and infinite $N$ allows for a direct comparison of calculations at different $N$. This is the case, because the $\tfrac{1}{N}$-rescaled UV potential \eqref{eq:initial-condition} always describes a theory of $N$ asymptotically free fermions and a $\sigma$ mode that decouples from the system at the UV initial scale. Choosing $\Lambda \ggg h$ naturally leads to a large curvature mass $\partial_\sigma^2 U ( t, \sigma )$ and suppression of fluctuations of the $\sigma$ mode in the UV, which can be directly seen in Eq.~\eqref{eq:initial-condition}. This argument was already brought up in Sub.Sec.~\ref{subsec:comment_on_the_truncation}. On a formal level this can be seen by inspecting the fluid dynamic formulation of the RG flow equation \eqref{eq:pdeq-u} and especially its bosonic contribution in terms of a highly non-linear diffusion equation \eqref{eq:heat_equation_analogy}. One finds that the large curvature mass $\partial_\sigma^2 U ( t, \sigma )= \partial_\sigma u ( t, \sigma )$ in the propagators $\tfrac{1}{E_\sigma}$ yields a small diffusion coefficient $D ( t, \partial_\sigma u )$ and therefore a suppression of the diffusion along field space -- the bosonic contribution to the RG flow.
	
	Though, the more drastic argument, why bosonic fluctuations are actually totally absent in the UV, is a fundamental property of all diffusion equations of type $\partial_t u ( t, \sigma ) \propto \partial_\sigma^2 u ( t, \sigma )$. Independent of the finite diffusion coefficient the term $\partial_\sigma^2 u$ vanishes exactly for spatially linear $u ( t, \sigma ) \propto \sigma$ and the diffusion and dynamics stops. This is rather natural, if one considers diffusive processes from our everyday life, \textit{e.g.}, heat conduction. In the context of this work, it follows from the quadratic UV potential \eqref{eq:uv-initial} that $u ( 0, \sigma ) \propto \sigma$ and consequently the contribution from the $\sigma$ mode to the RG flow vanishes exactly. Bosonic fluctuations will be suppressed as long as the fermionic source/sink contributions to the RG flow do not alter the linear shape of $u ( t, \sigma )$, which is approximately the case until $k^2 ( t ) \approx ( h \sigma )^2$ for small $\sigma$. We conclude that the RG trajectories in theory space for RG flows including fermions and bosons at finite $N$ will approximately follow the mean-field RG trajectories for infinite $N$, as long as the UV initial potential is quadratic in $\sigma$. This behavior is indeed observed in our numeric computations in Sec.~\ref{sec:rg-flow-with-bosons}. It also justifies setting $Z_\varphi ( t ) = 1$ already in the UV and we expect to resemble the dynamics of the GN with the GNY model in LPA to a certain extend.
	
	Our last formal argument is based on the previous discussion and concerns the UV-cutoff independence of IR observables for RG flows at finite $N$. Because the RG trajectories at finite $N$ will approximately follow the mean-field RG trajectories at infinite $N$, we expect that choices of $\Lambda$, which are sufficiently large to ensure UV cutoff independence at infinite $N$, should also suffice to ensure RG consistency for RG flows at finite $N$, if the same UV initial potential is used. This is explicitly demonstrated in App.~\ref{app:uv-cutoff_independence}, where we numerically demonstrate UV cutoff independence of the calculations presented in Sec.~\ref{sec:rg-flow-with-bosons}.
	
	In summary, as a first approach to enable a comparison of RG flows within the GNY with bosonic quantum fluctuations at different $N$, we choose exactly the same UV initial condition \eqref{eq:initial-condition} with $h = 1$ and $\sigma_0 = 1$ for all RG flows. Hence, all dimensionful quantities are measured in terms of the UV value of the Yukawa coupling $h$ (which stays constant in our truncation anyhow), while field space is measured in multiples of the mean-field minimum $\sigma_0$, which has turned into a free additional parameter. Including bosonic fluctuations $\sigma_0$ is no longer the position of the vacuum IR minimum, but modifies the ratio of the Yukawa coupling $h$ and $\Lambda$ in the UV, compare Eq.~\eqref{eq:initial-condition}. Of course, this ratio will still influence the dimensionless position of the vacuum IR minimum even in the presence of bosons. Hence, it is most convenient for us to choose $\sigma_0 = 1$ in the initial condition \eqref{eq:initial-condition} to recover the infinite-$N$ results directly for $N \rightarrow \infty$, without trivial rescalings. Still, we also performed calculations for $\sigma_0 \neq 1$, which did not alter the qualitative results.\\
	
	Overall this implies that we do not perform computations for different $N$ on lines of ``constant IR physics''. We compute on ``constant UV physics'' for different $N$.

\subsection{(Numeric) consistency check with mean-field results}
\label{subsec:numeric_consistency_check_mean-field}

	This section is dedicated to a consistency check of our numeric implementation of the fermionic contribution to the RG flow. Furthermore, we comment on the concept of RG consistency \cite{Braun:2018svj} in the context of the discussed mean-field calculations.
	
	For the consistency checks in this section we use the numerical implementation presented in Sub.Sec.~\ref{subsec:numerical_implementation} and the corresponding App.~\ref{app:source_sink_implementation} and manually switch off the bosonic contribution (the diffusion) in the flow equation \eqref{eq:pdeq-u}. Since Eq.~\eqref{eq:pdeq-u} is formulated for $u ( t, \sigma ) = \partial_\sigma  U ( t, \sigma )$, we used the $\sigma$-derivative of Eq.~\eqref{eq:initial-condition} as the initial condition for $u ( t, \sigma )$. Furthermore, as already done in the previous Sub.Sec.~\ref{subsec:phase_diagram_mean_field}, we work in rescaled quantities. This means that $\sigma$ is measured in multiples of $\sigma_0$ and dimensionful quantities are measured in multiples of the Yukawa coupling $h$ -- implying \textit{w.l.o.g.} $h = 1$ and $\sigma_0 = 1$ on the level of the equations. Thus, the UV initial condition for the (numeric) RG flow in dimensionless variables explicitly reads
		\begin{align}
			& u ( t = 0, \sigma ) =	\vphantom{\Bigg(\Bigg)}	\label{eq:initial-condition_small_u}
			\\
			= \, & \tfrac{d_\gamma}{2 \pi} \, \sigma \, \bigg[ \mathrm{artanh} \bigg( \Big[ 1 + \big( \tfrac{1}{\Lambda} \big)^2 \Big]^{- \frac{1}{2}} \bigg) - \Big[ 1 + \big( \tfrac{1}{\Lambda} \big)^2 \Big]^{- \frac{1}{2}} \bigg] \, .	\vphantom{\Bigg(\Bigg)}	\nonumber
		\end{align}
	The explicit dependence on $\Lambda$ of $u ( t = 0, \sigma )$ realizes non-trivial IR minima at $\pm\sigma_0$ independent of the UV initial scale by construction. The notion of RG consistency requires the independence of all IR observables from the UV initial scale \cite{Braun:2018svj}. Translated to the UV this implies that the initial action has to be a classical action in the sense that all quantum fluctuations stemming from momenta $k>\Lambda$ have to be included in the initial action. This is implied in the derivation of the ERG equation \eqref{eq:wetterich} and a necessary condition for its exactness. Guaranteeing RG consistency -- or quantifying deviations stemming from its violation -- is a crucial step to get meaningful and predictive results in the IR, \textit{cf.}\ Refs.~\cite{Braun:2018svj,Koenigstein:2021syz} as well as Refs.~\cite{Braun:2003ii,Herbst:2013ufa,Springer2017,PhysRevD.87.076004} in the context of RG consistency and UV completions for low-energy effective theories of QCD.
	
	One way to realize RG consistency in practical computations is to choose $\Lambda$ significantly larger than all internal and external scales of the problem under consideration. In practical computations a sufficiently large $\Lambda$ can be found by studying the $\Lambda$-dependency of IR observables of interest. We will discuss this on mean-field level in this subsection while dedicating App.~\ref{app:uv-cutoff_independence} to a similar discussion including the effects of bosonic fluctuations.\\
	
	While the condensate is fixed in the IR in vacuum by construction, the corresponding sigma curvature mass, see Eq.~\eqref{eq:msigma_MF_Lambda}, is not. Comparing the expression \eqref{eq:msigma_MF_Lambda} at finite $\Lambda$ to the renormalized result of Eq.~\eqref{eq:msigma_MF} we conclude that a relative difference of for example $10^{-3}$ ($10^{-6}$) between $m_\sigma^2$ at finite and infinite $\Lambda$ requires a UV initial scale of around $40$ ($1200$). Considerations like this in vacuum give insight into internal model scales.
	
\begin{figure}
		\centering
		\includegraphics{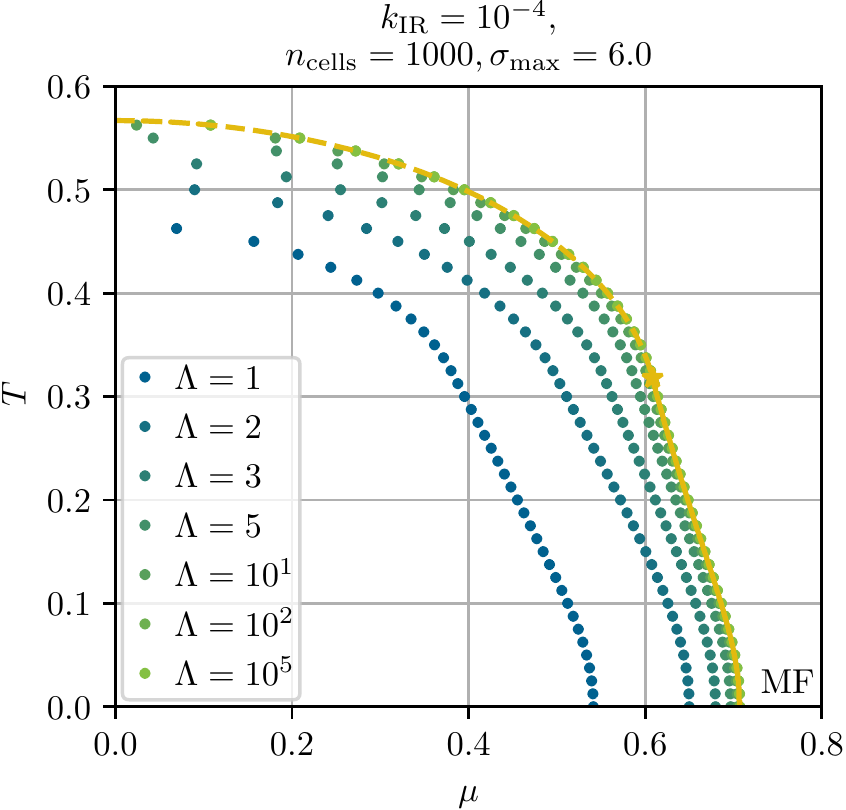}
		\caption{\label{fig:MF_consistency_check}
			Phase transition lines (equally colored dots) in the $\mu$-$T$-plane of the GNY model \eqref{eq:bgn-model} in the mean-field approximation for different values of the UV cutoff $\Lambda$. The curves are extracted from numerical solutions of the RG flows of $u ( t, \sigma )$ with the methods presented in Sub.Sec.~\ref{subsec:numerical_implementation}. As reference curve, we also plot the exact phase transition line (in yellow) below the data points, \textit{cf.}\ Fig.~\ref{fig:GNlargeN_PD}. RG flows for larger cutoffs $\Lambda$ are closer to the correct reference result.
		}
	\end{figure}

	Studying the $\Lambda$-dependence of observables at $\mu>0$ and/or $T>0$ we can asses the relation between $\Lambda$ and the external model scales. To this end we plot the phase transition lines in the $\mu$-$T$-plane, which we obtain via the numerical solution of the purely fermionic RG flow equation for various $\Lambda$ in Fig.~\ref{fig:MF_consistency_check}. The phase boundaries are extracted via the bisection method \cite{PresTeukVettFlan92,Press:1992zz} in the $\mu$-$T$-plane, by extracting the minimum of the IR potential from the cell-averages $\{ \bar{u}_i ( t_\mathrm{IR} ) \}$\footnote{We do not explicitly distinguish first- and second-order phase transitions in our numeric calculations and the bisection.
	}.
	As reference values for the mean-field phase boundaries in Fig.~\ref{fig:MF_consistency_check}, we use the exact results of the previous section, which are also plotted in Fig.~\ref{fig:GNlargeN_PD}. We observe a dependence of the phase boundary between the restored and broken phase on $\Lambda$. Increasing $\Lambda$ beyond $10^2$ we observe an apparent convergence of the numerically obtained phase boundaries, which eventually for $\Lambda \approx 10^5$ approaches the phase boundary of the renormalized mean-field computation. The general UV initial scale dependence at non-zero chemical potential and temperature is easily understood when considering the expression \eqref{eq:rg-U-MF} of the underlying IR potential. In RG based mean-field computations with the Limit regulator the UV initial scale acts as a sharp momentum cutoff. Using a small (relative to $\mu$ and $T$) UV initial scale limits the phase-space/loop-momentum in Eq.~\eqref{eq:rg-U-MF} significantly, which in turn leads to a cutoff of relevant thermal/density fluctuations. This manifests in the phase diagram in a shift of the phase transition line towards lower $\mu$ and $T$. The recovery of the renormalized mean-field results for the phase boundary was analytically discussed in the previous Sub.Sec.~\ref{subsec:phase_diagram_mean_field}. 
	
	We find that our numeric implementation of the source term in the RG flow equation \eqref{eq:pdeq-u} is capable of reproducing the conventional mean-field results. Furthermore, we already obtain some estimate for decent UV-cutoffs $\Lambda$ for computations involving bosonic quantum fluctuations.\\
	
	At this point we can close all our preliminary discussions and finally turn to the main results of our work -- bosonic quantum fluctuations.

\section{Bosonic quantum fluctuations in the Gross-Neveu-Yukawa model -- Results}
\label{sec:rg-flow-with-bosons}

	In this section, we present our (numeric) results for RG flows of the GNY model at non-zero $\mu$ and $T$ including bosonic quantum fluctuations in the LPA. Thereby we proceed as follows: First, we discuss the dynamics, which take place during a single RG flow at some fixed finite $N$, $\mu$, and $T$. This sets the stage for a detailed discussion of the effects that are induced by the chemical potential at very low temperatures. Afterward, we turn to our central result -- the absence of spontaneous $\mathbb{Z}_2$ symmetry breaking at finite $N$ and $T > 0$. We thereby present details on dependencies on $\mu$, $T$, and $N$ of the restoration scale $k_\mathrm{res}$, where the discrete chiral symmetry is restored, as well as a phase diagram during the RG flow.

\subsection{Symmetry breaking and restoration during the RG flow}

	This section is dedicated to an instructive discussion of symmetry breaking and restoration during RG flows. To get a better intuition on how this realizes during the RG flow and how the typical setup and dynamics looks like, we picked a single point in the $\mu$-$T$-plane, namely $\mu = 0.1$ and $T = 0.1$, where we at least expect some non-trivial condensation and vaporization phenomena and typical dynamics in the RG flow. We also fixed the number of fermions to $N = 2$.
	
	Furthermore, as discussed in Sub.Secs.~\ref{subsec:boundary_conditions} and \ref{subsec:numerical_implementation}, we have to specify the spatial extend of the computational domain $\sigma_\mathrm{max}$ as well as the number $n_\mathrm{cells}$ of computational cells, which are of size
		\begin{align}
			\Delta \sigma = \tfrac{\sigma_\mathrm{max}}{n_\mathrm{cells} - 1} \, .
		\end{align}
	We also have to choose the UV initial scale $\Lambda$ and the computational IR cutoff $k_\mathrm{IR}$. Hereby, one has to ensure that this choice of (computational) parameters does not influence our results, \textit{i.e.}, the IR physics. For detailed checks and numerical tests, we refer to App.~\ref{app:numerical_tests} and fix the parameters for all upcoming calculations in this section as presented in Table \ref{tab:numerical_parameters}.
		\begin{table}[b]
			\caption{\label{tab:numerical_parameters}%
				Numerical parameters used for all calculations of this section. The numerical parameters have been chosen according to the tests in App.~\ref{app:numerical_tests} and the discussions of the previous sections.
			}
			\begin{ruledtabular}
				\setlength\extrarowheight{2pt}
				\begin{tabular}{l c c c c c c}
					parameter	&	$\sigma_\mathrm{max}$	&	$n_\mathrm{cells}$	&	$\Lambda$	&	$k_\mathrm{IR}$	&	$h$	&	$\sigma_0$
					\\
					value		&	$6$						&	$1000$				&	$10^{5}$	&	$10^{-4}$		&	$1$	&	$1$
				\end{tabular}
			\end{ruledtabular}
		\end{table}
	As UV initial condition, we directly use Eq.~\eqref{eq:initial-condition_small_u}, where we already set $h = 1$ and $\sigma_0 = 1$.\\
	
	Using this setup, we obtain the numeric results for the RG flow of $U ( t, \sigma )$\footnote{The scale dependent potential $U ( t, \sigma )$ is calculated from the cell averages $\{ \bar{u}_i ( t ) \}$ by direct Riemann summation, compensating for the half-sized cells next to the computational interval boundaries \cite{Koenigstein:2021syz}, \textit{i.e.}, $ U ( t, \sigma_i ) = \Delta \sigma \sum_{m = 0}^{i} \frac{\bar{u}_m ( t )}{( 1 + \delta_{m 0} + \delta_{m i})}$.} and $u ( t, \sigma )$ depicted in Fig.~\ref{fig:flow_N=2,T=0.1,mu=0.1}.
		\begin{figure}
			\centering
			\includegraphics{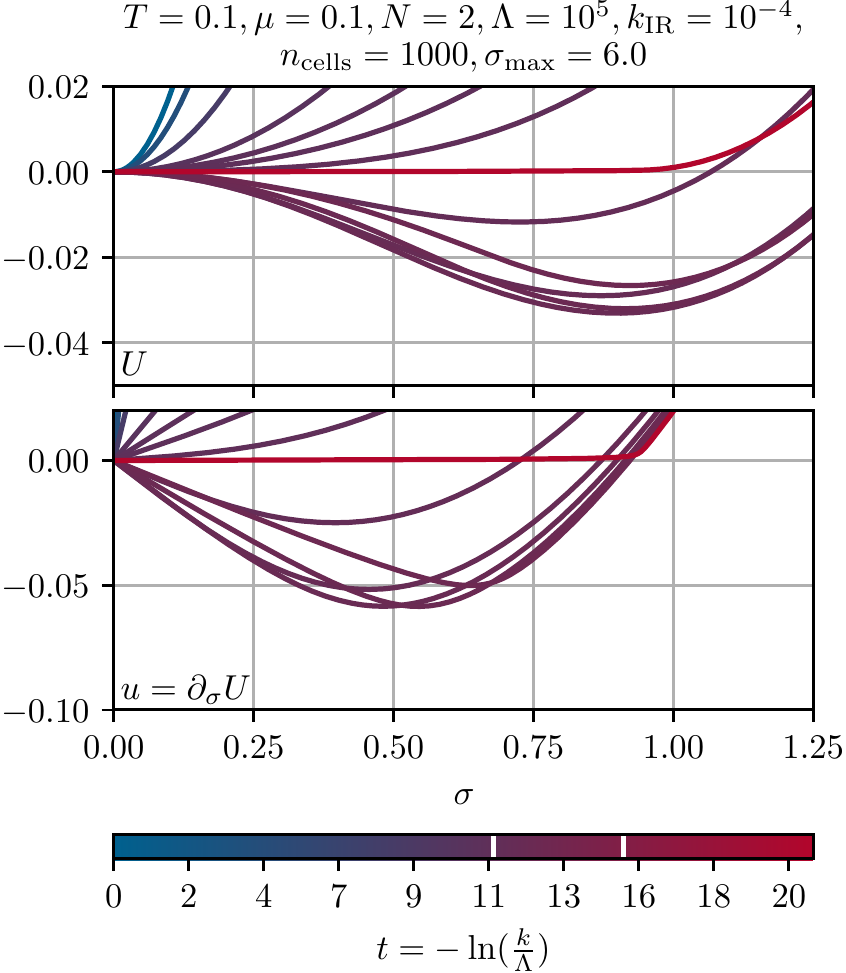}
			\caption{\label{fig:flow_N=2,T=0.1,mu=0.1}
				RG flow of the scale dependent effective potential $U ( t, \sigma )$ (upper panel) and its $\sigma$-derivative (the fluid) $u ( t, \sigma ) = \partial_\sigma U ( t, \sigma )$ (lower panel) from the UV ({blue}) to the IR ({red}) at $T = 0.1$ and $\mu = 0.1$. For the sake of simplicity (and using the (anti-)symmetry in $\sigma$) the functions $u ( t, \sigma )$ and $U ( t, \sigma )$ are plotted for positive $\sigma$ only. The different RG-times are encoded via the colored bar-legend below the plots. The white vertical lines in the colored bar-legend denote the RG times (scales) when the $\mathbb{Z}_2$ symmetry is broken (condensation) and restored (vaporization). All other parameters are stated above the panels.
			}
		\end{figure}
	Additionally, in Fig.~\ref{fig:k_N=2, T=0.1, mu=0.1} we show the RG flow of the scale-dependent minimum $\sigma_\mathrm{min} ( t )$ and the scale-dependent curvature mass, which is evaluated at the moving scale-dependent minimum\footnote{We determine the position of the scale dependent minimum ${\sigma_\mathrm{min} ( t ) = \Delta \sigma \cdot i_\mathrm{min} ( t )}$, by searching for the position $i_\mathrm{min} ( t )$ of cells next to zero-crossings in the list $\{ u ( t, \sigma_i ) \}$ combined with a check of the list $\{ U ( t, \sigma_i ) \}$ for its smallest entry. The curvature mass is calculated at this minimum via a simple right-derivative stencil, hence $\partial_\sigma u ( t, \sigma ) \big|_{\sigma_\mathrm{min} ( t )} = \frac{1}{\Delta \sigma} \, [ \bar{u}_{i_\mathrm{min} + 1} ( t )- \bar{u}_{i_\mathrm{min}} ( t ) ]$.},
		\begin{align}
			m^2_{\sigma} ( t ) = \, & \partial_\sigma^2 U ( t, \sigma )\big|_{\sigma_{\min}(t)} =	\vphantom{\bigg(\bigg)}
			\\
			= \, & \partial_\sigma u ( t, \sigma )\big|_{\sigma_{\min}(t)} \, .	\vphantom{\bigg(\bigg)}	\nonumber
		\end{align}
	
	For optical guidance and better detection, when the plateau in IR is reached, we also introduce the changing rate of the curvature mass
		\begin{align}
			\partial_t m_\sigma^2 ( k_j ) \equiv - k_j \, \frac{m^2_\sigma( k_{j + 1} ) - m^2_\sigma ( k_j )}{  k_{j + 1} - k_j} \, ,	\label{eq:changing_rate_mass}
		\end{align}
	which we evaluated for the plots at $j \in \{ 0, 1, \ldots, 998 \}$ intermediate RG scales $k_j$ between $k = \Lambda$ and $k = k_\mathrm{IR}$, such that the belonging RG-times are equidistantly distributed.
		\begin{figure}
			\centering
			\includegraphics{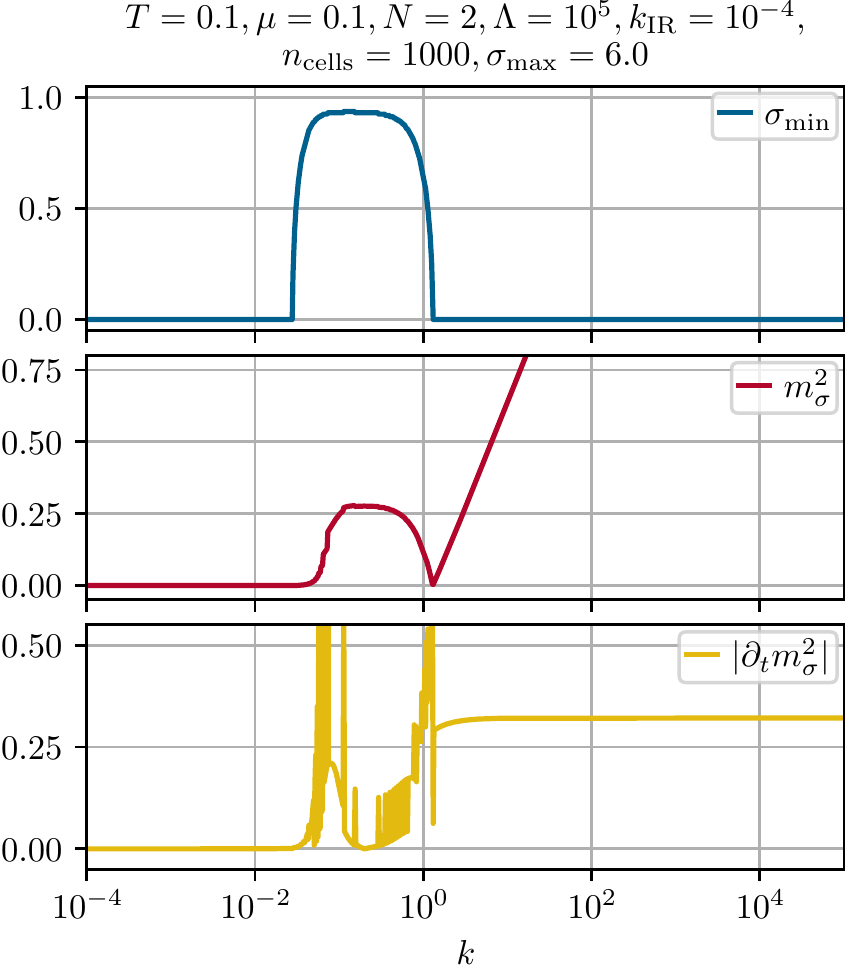}
			\caption{\label{fig:k_N=2, T=0.1, mu=0.1}
				RG flow of the minimum $\sigma_{\mathrm{min}} ( k )$ of the scale dependent effective potential $U ( k, \sigma )$ (upper panel), the squared curvature mass $m_\sigma^2 ( k ) = \partial_\sigma u ( t, \sigma ) \big|_{\sigma_\mathrm{min} ( t )}$ at the moving minimum $\sigma_\mathrm{min} (t)$ (middle panel), and the relative change of the squared curvature mass $| \partial_t m_\sigma^2 ( k ) |$ (lower panel) according to Eq.~\eqref{eq:changing_rate_mass} plotted as functions of the RG scale $k ( t )$ for $T = 0.1$ and $\mu = 0.1$. The plot corresponds to the RG flows of $u ( t, \sigma )$ and $U ( t, \sigma )$ of Fig.~\ref{fig:flow_N=2,T=0.1,mu=0.1}. Note, that the curvature mass at the moving minimum in the IR is non-zero, $m_\sigma^2 ( t_\mathrm{IR} ) \approx 4.60\cdot10^{-4}$.
			}
		\end{figure}
			
	From Figs.~\ref{fig:flow_N=2,T=0.1,mu=0.1} and \ref{fig:k_N=2, T=0.1, mu=0.1} we observe the following dynamics: The flow for $u ( t, \sigma )$ starts with the UV initial condition that is linear in $\sigma$. At the beginning of the flow the fermions are active and clearly dominate the dynamics, according to the sink (source) contribution in the fluid dynamic language. We find that this sink (source) contribution causes the $\mathbb{Z}_2$ symmetry breaking and generation of a non-trivial minimum at $k(t \approx  11.2) = 1.31$, which is roughly at the order of the model scales, which are of order $1$ (\textit{cf.} position of the intermediate minimum or value of $h$). Shortly after the non-trivial minimum has formed, the sink (source) is still active, but the diffusion caused by the bosonic contributions sets in, due to the negative gradients $\partial_\sigma u ( t, \sigma )$ close to $\sigma = 0$, which enhance the diffusion coefficient \eqref{eq:heat_equation_analogy}. Interestingly, when the position of the minimum (the value of the condensate) has settled, it is approximately of the same order of magnitude as for the mean-field calculations, even though $N = 2$ is everything but close to $N \rightarrow \infty$. As a matter of fact, this signals that the diffusion is weak at $\sigma > \sigma_\mathrm{min} ( t )$. Subsequently, for approximately another two orders of magnitude in RG scale $k ( t )$, the minimum keeps its position $\sigma_\mathrm{min} \approx 0.93$. Also the bosonic curvature mass seems to freeze at $m_\sigma^2 \approx 0.28$ and the potential $U ( t, \sigma )$ is not changing much. However, having a closer look at $u ( t, \sigma )$, we find that the diffusion in field space direction $\sigma$ causes some highly non-linear dynamics, especially close to the point, where the gradient $\partial_\sigma u ( t, \sigma )$ changes its sign. Suddenly, at $k ( t \approx 15.1) = 2.76\cdot10^{-2}$, we observe a destabilization of the condensate $\sigma_\mathrm{min} ( t )$ and also $m_\sigma^2 ( t )$ starts changing drastically. The inclusion of IR modes in a small momentum range leads to a complete vaporization of the condensate. Additionally, inspecting $U ( t, \sigma )$, we find that meanwhile the potential turned convex -- as it should be the case in the IR. This flattening of the potential is completely driven by the highly non-linear diffusion, compare our discussion in Ref.~\cite{Steil:2021cbu}. Finally, we find that the dynamics completely freezes and that we indeed integrated out all fermionic and bosonic quantum fluctuations. This can be seen best by looking at the absolute value of the changing rate of the squared curvature mass $| \partial_t m_\sigma^2 ( t ) |$, but also directly from $m_\sigma^2 ( t )$ or $\sigma_\mathrm{min} ( t )$. Note, that $m_\sigma^2 ( t_\mathrm{IR} ) \approx 4.60\cdot10^{-4}$, which is not visible from the plot, while $\sigma_\mathrm{min} ( t ) = 0$ already shortly after $k ( t \approx 15.1) = 2.76\cdot10^{-2}$.

	Overall, we observed that the fermions were indeed able to form a condensate, which however does not survive the long-range bosonic quantum fluctuations in the deep IR. This dynamics might also be referred to as precondensation \cite{Boettcher:2012cm,Boettcher:2012dh,Boettcher:2013kia,Boettcher:2014tfa,Roscher:2015xha,Khan:2015puu}.

\subsection{The role of the chemical potential in the fluid dynamic setup}
\label{subsec:chemical_potential_mf_vs_with_bosons}

	Before we come to our discussion on symmetry breaking an restoration in RG flows for arbitrary $N$ and arbitrary points in the $\mu$-$T$ plane, we briefly return to the discussion in Sub.Sec.~\ref{subsec:chemical_potential_shock_wave} on the role of the chemical potential in the fluid dynamic formulation of the RG flow equation \eqref{eq:pdeq-u}. We therefore accentuate our discussion from Sub.Sec.~\ref{subsec:chemical_potential_shock_wave} with explicit calculations and plots of RG flows at very low as well as vanishing temperature and moderate chemical potential. \textit{W.l.o.g.}\ we choose $\mu = 0.6$ and $T = 0.00625$ or $T = 0$.\\
	
	Therefore we present Fig.~\ref{fig:flow_MF_T=0.0,mu=0.6} as the reference plot of our discussion. It shows the mean-field RG flow for $u ( t, \sigma )$ and $U ( t, \sigma )$ at $T = 0$ and $\mu = 0.6$.
		\begin{figure}
			\centering
			\includegraphics{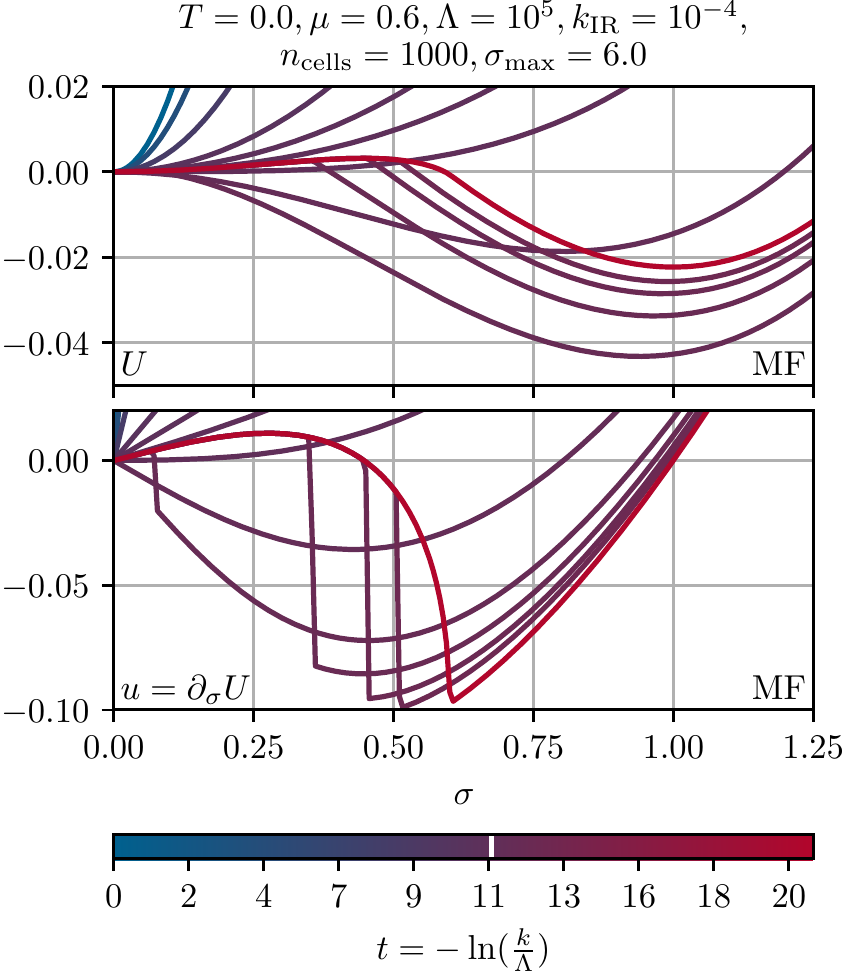}
			\caption{\label{fig:flow_MF_T=0.0,mu=0.6}
				Mean-field (infinite-$N$) RG flow (sink term only) of the scale dependent effective potential $U ( t, \sigma )$ (upper panel) and its $\sigma$-derivative (the fluid) $u ( t, \sigma ) = \partial_\sigma U ( t, \sigma )$ (lower panel) from the UV ({blue}) to the IR ({red}). For the sake of simplicity (and using the (anti-)symmetry in $\sigma$) the functions $u ( t, \sigma )$ and $U ( t, \sigma )$ are plotted for positive $\sigma$ only. The different RG-times are encoded via the colored bar-legend below the plots. The white vertical line in the colored bar-legend denotes the RG time (scales) when the $\mathbb{Z}_2$ symmetry is broken (condensation). There is no symmetry restoration at $T = 0$ and $\mu = 0.6$ in mean field. All other parameters are stated above the panels.
			}
		\end{figure}
	As described in Sub.Sec.~\ref{subsec:chemical_potential_shock_wave} the chemical potential enters the RG flow of $u ( t, \sigma )$, which is entirely described via the source/sink equation, suddenly during the RG flow in terms of a discontinuity of $u ( t, \sigma )$ in field space. This discontinuity appears at $\sigma = 0$ when $k^2 ( t ) \approx \mu^2$ and moves towards larger $| \sigma |$ until $\sigma^2 \approx \tfrac{\mu^2}{h^2}$ ($h = 1$). Formally, this discontinuity leads to infinite negative gradients $\partial_\sigma u ( t, \sigma )$ and impedes the study of bosonic quantum fluctuations at $T = 0$ and $\mu \neq 0$ within our current setup. This is, because $E^2_\mathrm{b} ( t, \partial_\sigma u ) = k^2 ( t ) + \partial_\sigma u ( t, \sigma ) \approx \mu^2 + \partial_\sigma u ( t, \sigma ) < 0$, which leads to an abrupt overshooting over the poles of the bosonic propagators $\tfrac{1}{E_\mathrm{b}}$ and drives the diffusion coefficient $D ( t, \partial_\sigma u )$ from Eq.~\eqref{eq:heat_equation_analogy} negative.
	
	After spatial integration of $u ( t, \sigma )$ the remnant of the discontinuity is clearly visible in $U ( t, \sigma )$ in terms of a moving cusp. Additionally, it is worth mentioning that the potential $U ( t, \sigma )$ is in the symmetry broken phase and everything but flat for small $|\sigma|$ in the IR, which violates a fundamental property of thermodynamic potentials and $\Gamma [ \bar{\psi}, \psi, \varphi ]$, namely convexity. This is a typical mean-field/infinite-$N$ artifact.\\
	
	In order to study, how the chemical potential influences the bosonic RG flows, hence the diffusion in field space, we first slightly increase the temperature of the mean-field calculation to $T = 0.00625$, while keeping the same chemical potential $\mu = 0.6$. In Fig.~\ref{fig:flow_MF_T=0.00625,mu=0.6} it is clearly visible that the huge negative gradients $\partial_\sigma u ( t, \sigma )$ are still present and the overall shape and flows of $u ( t, \sigma )$ and $U ( t, \sigma )$ do not change much compared to Fig.~\ref{fig:flow_MF_T=0.0,mu=0.6}.
		\begin{figure}
			\centering
			\includegraphics{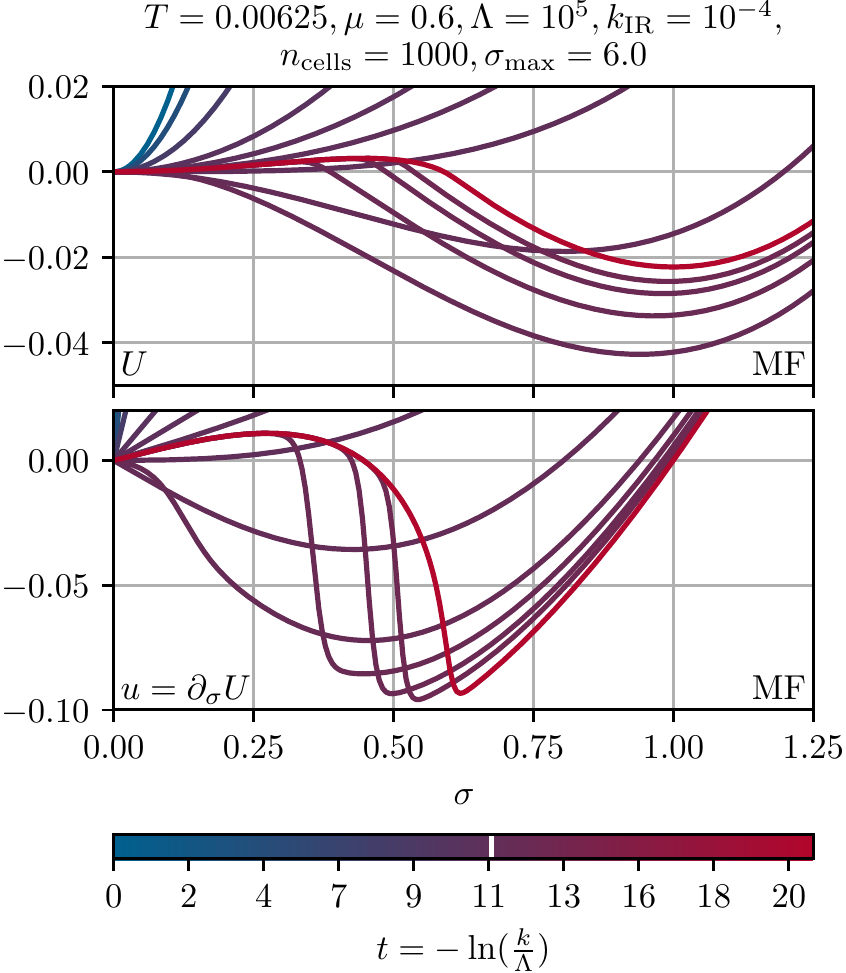}
			\caption{\label{fig:flow_MF_T=0.00625,mu=0.6}
				Same mean-field RG flow as in Fig.~\ref{fig:flow_MF_T=0.0,mu=0.6} but for $T = 0.00625$ instead of $T = 0$.
			}
		\end{figure}
	Though, already very small temperatures are able to smear out the sharp edges of the jumps, which smoothens $u ( t, \sigma )$ significantly already without any diffusive contributions from the bosons. This effect stemming from the Fermi-Dirac distribution function \eqref{eq:distribution_function} enables the inclusion of bosons, since gradients are still large, but finite.\\
	
	The next question is, what happens, if the number of fermions is finite and bosonic fluctuations enter the RG flow as non-linear diffusion on the level of $u ( t, \sigma )$. To this end, we plot the same RG flows as before at $T = 0.00625$ and $\mu = 0.6$ for $N = 16$ fermions in Fig.~\ref{fig:flow_N=16,T=0.00625,mu=0.6}.
		\begin{figure}
			\centering
			\includegraphics{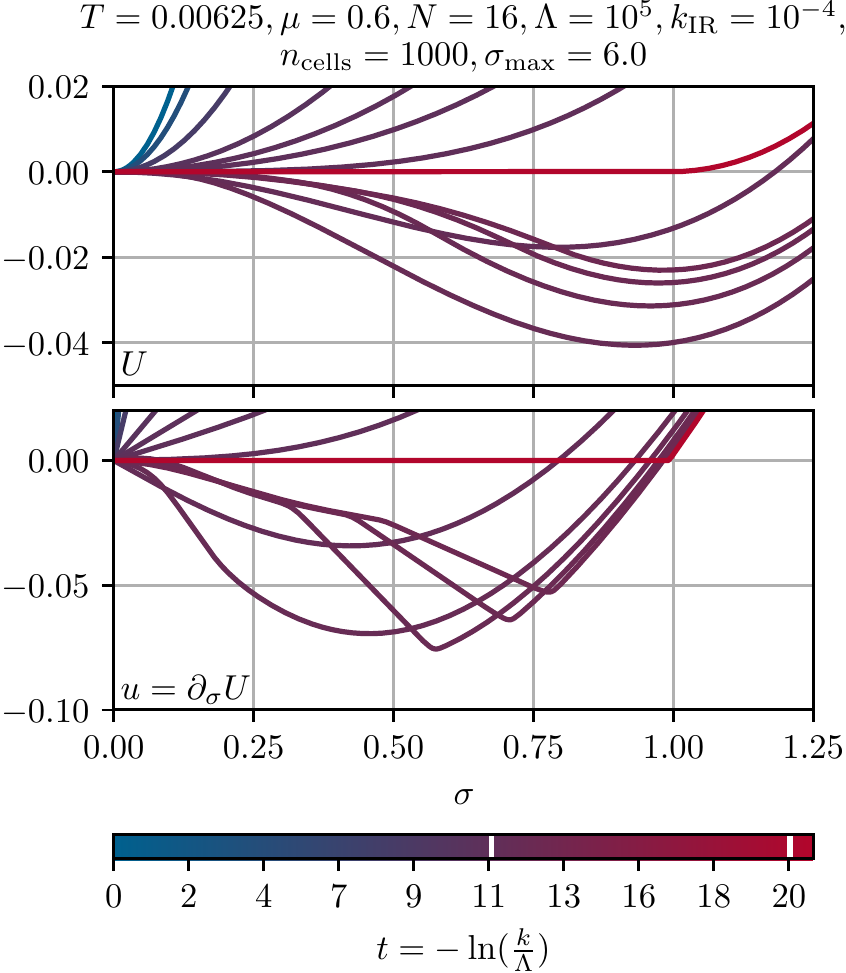}
			\caption{\label{fig:flow_N=16,T=0.00625,mu=0.6}
				Same RG flow as in Fig.~\ref{fig:flow_MF_T=0.0,mu=0.6} but for $T = 0.00625$ and $N = 16$ instead of $T = 0$ and $N\rightarrow\infty$, thus involving the effects of bosonic quantum fluctuations (diffusion).
			}
		\end{figure}
	Even though the number of fermions seems to be rather large, the overall picture changes drastically when compared to the situation in the limit $N\rightarrow\infty$. We observe that the chemical potential is still clearly visible on the level of $u ( t, \sigma )$ in terms of rather large gradients and cusps. But it is hardly visible in $U ( t, \sigma )$. The highly non-linear character of the diffusion does not really smear out the cusps, but results in the somewhat strange movement of the straight part of $u ( t, \sigma )$ between the two pronounced edges. Finally, the greatest difference to the mean-field calculations is that the diffusion vaporizes the condensate and fully restores the $\mathbb{Z}_2$ symmetry in the IR\footnote{We have checked numerically that there is indeed only a trivial minimum at $\sigma = 0$, which is hardly visible, because $u ( t_\mathrm{IR}, \sigma )$ and $U ( t_\mathrm{IR}, \sigma )$ are extremely close to the $\sigma$-axis in the relevant region. For details, we refer to the discussions within the next subsections.}. Additionally, the potential $U ( t, \sigma )$ turns convex in the IR, as expected in the case of finite $N$. Already from this calculation it is obvious that large but finite numbers of $N$ yields entirely different results than the $N \rightarrow \infty$ limit, as prominently stated in various publications \cite{Witten:1978qu,ZinnJustin:2002ru,Rosenstein:1990nm} and exemplified in great detail in part III of our parallel series of publications on zero-dimensional $O(N)$-symmetric toy models \cite{Steil:2021cbu}.\\
	
	Finally, we further decrease the number of fermions to $N = 2$ and again study the RG flows at $T = 0.00625$ and $\mu = 0.6$. In the corresponding Fig.~\ref{fig:flow_N=2,T=0.00625,mu=0.6} we observe that the diffusion via the sigma mode sets in much earlier during the RG flow and the intermediate symmetry breaking is less drastic. The reason is rather obvious: Changing $N$ in Eq.~\eqref{eq:zero_t_flow_equation} changes the relative strength between bosonic and fermionic interactions (fluctuations). On the level of the fluid dynamic equation \eqref{eq:pdeq-u} this implies that the flow is either more diffusion (for small $N$) or more source/sink (for large $N$) dominated. Still, even for $N = 2$ the chemical potential is clearly visible in form of a slightly smeared and moving cusp in $u ( t, \sigma )$. Apart from this, the qualitative picture is similar to the $N = 16$ scenario.
	\begin{figure}
		\centering
		\includegraphics{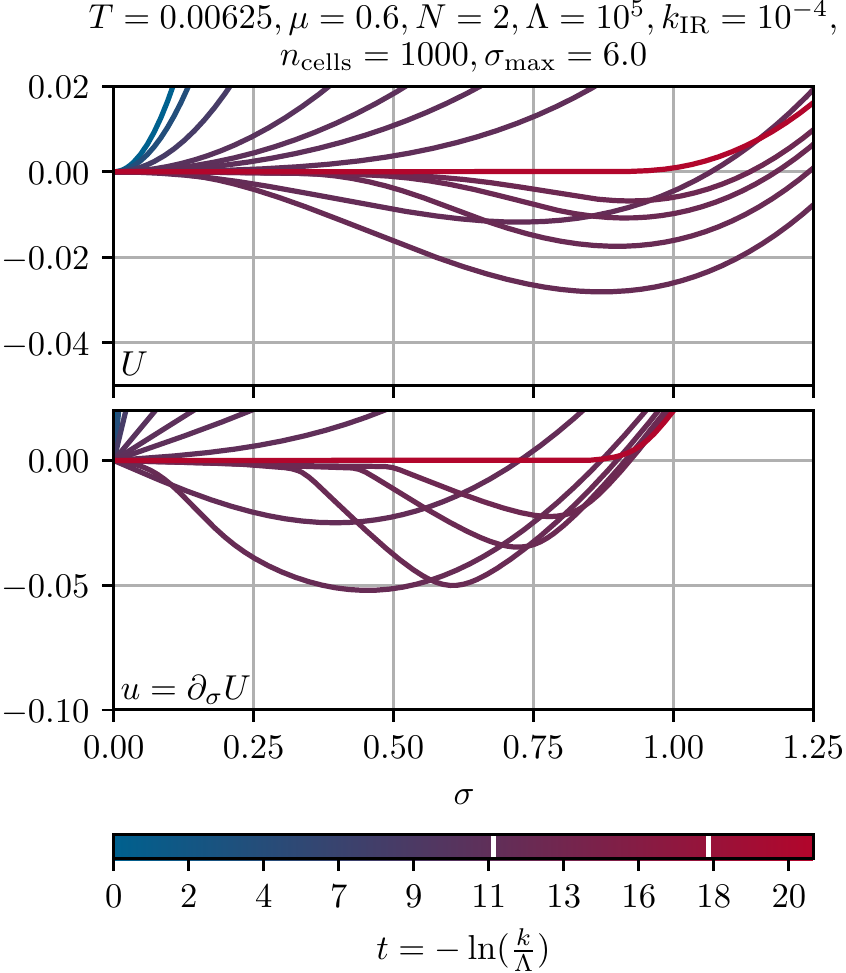}
		\caption{\label{fig:flow_N=2,T=0.00625,mu=0.6}
			Same RG flow as Fig.~\ref{fig:flow_MF_T=0.0,mu=0.6} but for $T = 0.00625$ and $N=2$ instead of $T = 0$ and $N\rightarrow\infty$, thus involving the effects of bosonic quantum fluctuations (diffusion).
		}
	\end{figure}

\subsection{Varying the number of flavors \texorpdfstring{$N$}{N}}
\label{subsec:variableN}

	Next, we turn to a more systematic analysis of our previous findings. We start by analyzing the relation between the RG scale $k_\mathrm{res}$, where the $\mathbb{Z}_2$ symmetry is restored (if it was initially broken by the fermions) and the number of fermionic flavors $N$.
	
\begin{figure}
			\centering
			\includegraphics{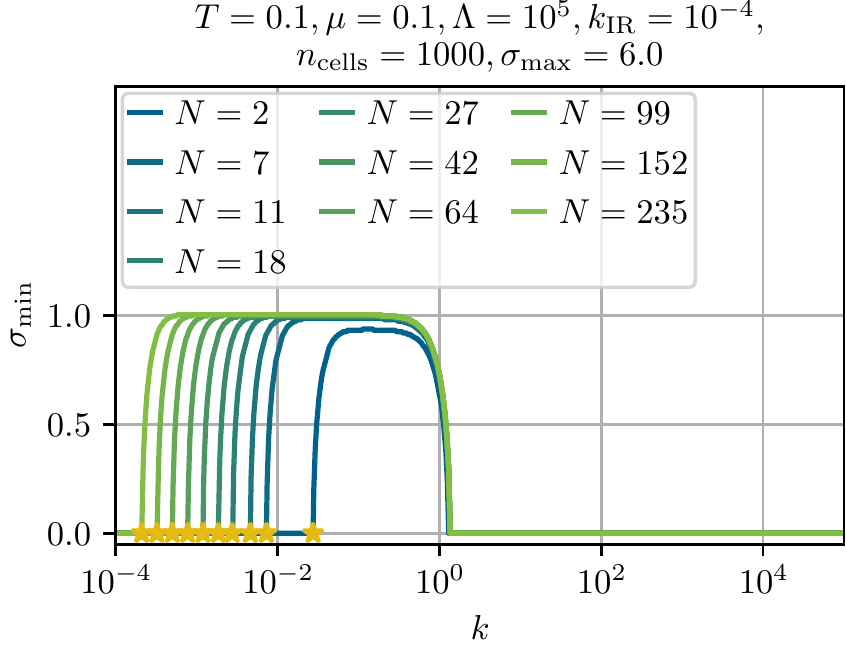}
			\caption{\label{fig:Nf_k_sigma_min}
				RG-flows of the value of the condensate (the position of the minimum) $\sigma_{\mathrm{min}} ( k )$ for various $N$ as functions of the RG scale $k$ at $T = 0.1$ and $\mu = 0.1$. The yellow stars mark the RG scales $k_\mathrm{res}$, where the $\mathbb{Z}_2$ symmetry is restored, due to a vanishing condensate.
			}
		\end{figure}

	As a first step to get an overall impression, we again fix $\mu = 0.1$ and $T = 0.1$ and look at the condensate $\sigma_\mathrm{min} ( t )$ as a function of the RG scale $k ( t )$ for various selected values of $N$. The numeric results are depicted in Fig.~\ref{fig:Nf_k_sigma_min}.
	The results are rather intuitive. The RG time period, in which we find a non-zero condensate ${\sigma_\mathrm{min} ( t ) \neq 0}$, strongly depends on $N$. For small $N$, the $\mathbb{Z}_2$ symmetry restores almost at model scales, which are set by $h = 1$, while for larger $N$ one finds that the restoration scale $k_\mathrm{res}$ moves several orders of magnitude on the RG scale towards the IR\footnote{This is the reason, why we cannot choose arbitrarily large $N$ for numeric calculations. If $N$ is too large, we have to integrate to too small RG scales to find symmetry restoration. However, this is numerically not possible, due to general limitations of the numerical precision during numerical RG time evolution.}. Furthermore, we observe that in the time periods with broken $\mathbb{Z}_2$ symmetry, the position of the minimum $\sigma_\mathrm{min} ( t )$ is approaching its mean-field value ${\sigma_\mathrm{min} = 1}$ rapidly while increasing $N$ in Fig.~\ref{fig:Nf_k_sigma_min}. The reason for this behavior is that the precondensation -- the formation of a non-persistent non-trivial minimum during the RG flow -- due to fermionic quantum fluctuations does not depend on $N$.
	
\begin{figure}
			\centering
			\includegraphics{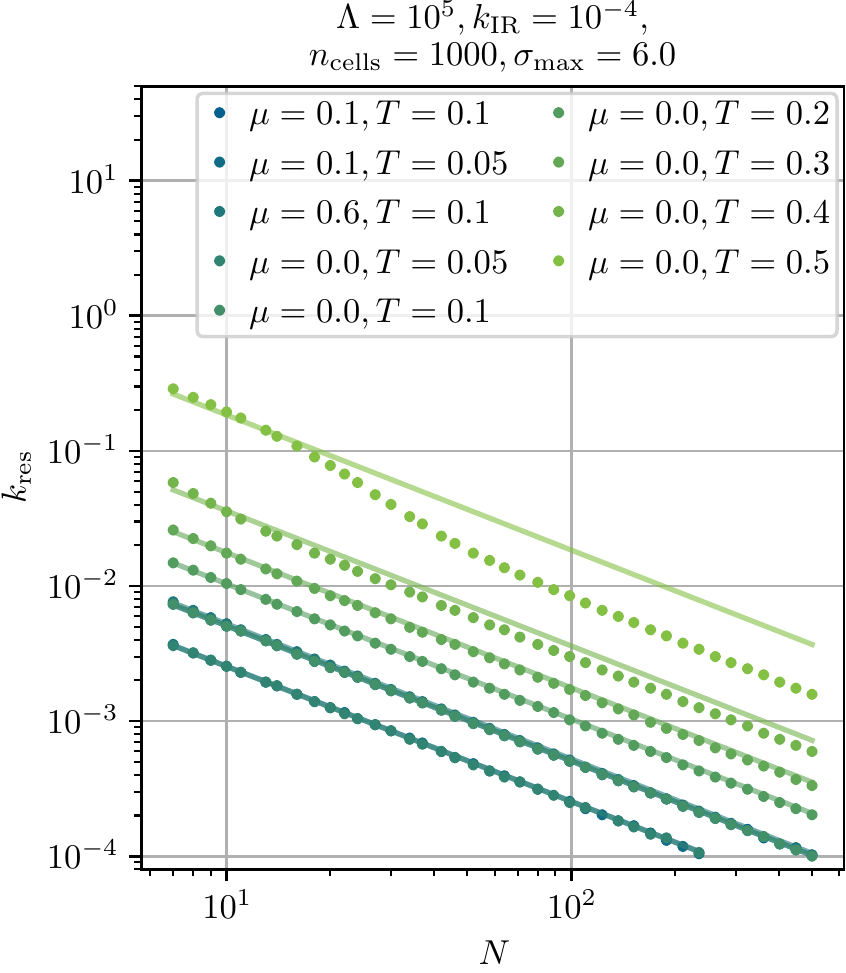}
  			\caption{\label{fig:Nf_k_restored}
				Restoration scale $k_\mathrm{res}$, where the $\mathbb{Z}_2$ symmetry is restored during the RG flow, as a function of the number of fermions $N$ at various points in the $\mu$-$T$-plane, where the $\mathbb{Z}_2$ symmetry was dynamically broken at $k > k_\mathrm{res}$ by the fermions. For $\mu = 0.1$ and $T = 0.1$ the data points correspond to the yellow stars in Fig.~\ref{fig:Nf_k_sigma_min}. For the various ($\mu$,$T$)-tuples we include $k_\mathrm{res} ( N ) \propto N^{-1}$ fits ought to be used for optical guidance. Note, that the points for equal temperatures but different chemical potentials are lying on top of each other. In Sub.Sec.~\ref{subsec:variablemu} we will discuss this in detail.
			}
		\end{figure}

	The direct follow up question is, if there is some fixed relation between $k_\mathrm{res}$ and $N$ and if we can expect to recover the mean field result for $N \rightarrow \infty$, where $k_\mathrm{res} = 0$, which is not reachable in practical computations at finite $N$ involving bosonic fluctuations. Therefore, we calculate and plot $k_\mathrm{res}$ as a function of $N$ for different combinations of $\mu$ and $T$ in Fig.~\ref{fig:Nf_k_restored} in a $\log$-$\log$ plot. The values for $\mu$ and $T$ in Fig.~\ref{fig:Nf_k_restored} lie in the symmetry broken phase of the MF phase diagram and consequently precondensation occurs during the RG flow, which is necessary to define and discuss the restoration scale $k_\mathrm{res}$.
	Already from the numeric data points, it seemed as if we found some power law behavior for $k_\mathrm{res} ( N )$, which was confirmed for all combinations of $\mu$ and $T$, as long as $T$ is sufficiently small\footnote{For $\mu = 0.0$ and $T = 0.5$ the relation $k_\mathrm{res} ( N )$ does not obey a strict power law \eqref{eq:N-k} but slightly deviates. We believe that the reason for this behavior is the fact that for sufficiently large $T$ one is already close to the second order phase transition, where the $\mathbb{Z}_2$ symmetry is restored by thermal fluctuations (also in mean-field) and not primary by bosonic quantum fluctuations.}. The straight lines are fits of the function
		\begin{align}
			k_{\mathrm{res}} ( N ) \propto N^{-1}	\label{eq:N-k}
		\end{align}
	to our data points. This strongly supports the hypothesis that for all finite $N$ the discrete chiral symmetry is never broken in the IR for $T>0$, while for $N \rightarrow \infty$, the mean field result is recovered. In the following discussions we will mainly focus on $N = 2$. Nevertheless, we checked for various other values of $N$ that our overall findings are similar for general finite $N > 1$.

\subsection{Varying the temperature \texorpdfstring{$T$}{T}}\label{subsec:variableT}

	In this subsection we focus on the relation between the restoration scale $k_\mathrm{res}$ and the temperature $T$. This relation is exemplified for $N = 2$.\\
	
\begin{figure}
			\centering
			\includegraphics{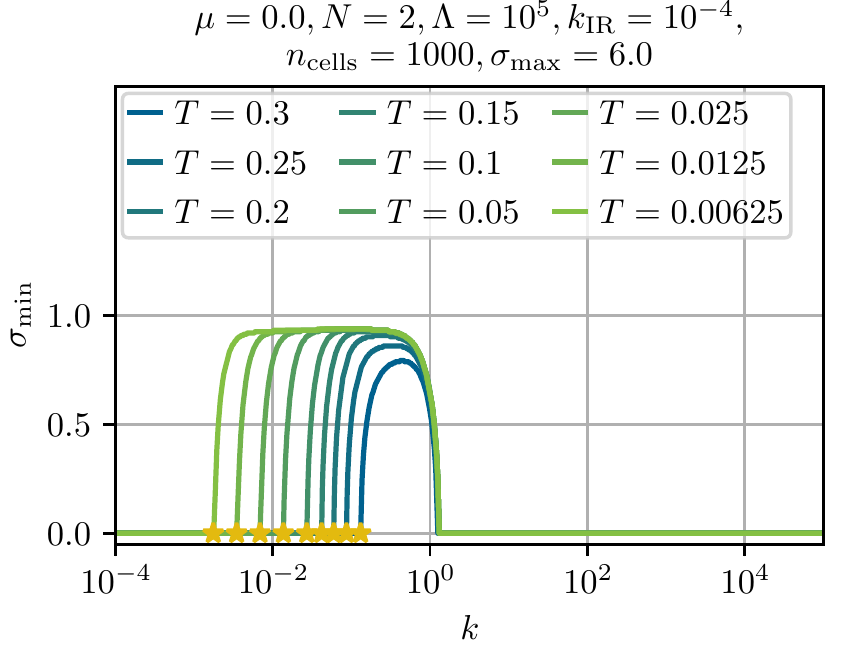}
			\caption{\label{fig:T_k_sigma_min}
				Value of the condensate (the position of the minimum) $\sigma_{\mathrm{min}} ( k )$ for various $T$ as a function of the RG scale $k ( t )$. The yellow stars mark the RG scales $k_\mathrm{res}$, where the $\mathbb{Z}_2$ symmetry is restored, due to a vanishing condensate.
			}
		\end{figure}

	We start by setting $\mu = 0$ and solely focusing on the evolution of $\sigma_\mathrm{min} ( t )$ along the RG scale $k ( t )$ for different temperatures $T$. This is plotted in Fig.~\ref{fig:T_k_sigma_min}.
	We find, that by decreasing the temperature, the RG time period of broken $\mathbb{Z}_2$ symmetry becomes longer and one has to go deeper into the IR to find symmetry restoration for smaller temperatures. Additionally, we observe remnants of the mean-field second order phase transition, because for larger temperatures the value of the intermediate condensate $\sigma_\mathrm{min} ( t )$ is smaller than for smaller temperatures. In general we observe no SSB in the IR for all $T>0$ and $\mu=0$. For temperatures above the mean-field critical temperature $T_\mathrm{C}\simeq  0.567$, see Eq.~\eqref{eq:Tc_MF}, symmetry restoration is driven by thermal fluctuations while for $0 < T < T_\mathrm{C}$ bosonic quantum fluctuations seem to drive symmetry restoration at finite $N$ and $\mu = 0$.\\	
		\begin{figure}
			\centering
			\includegraphics{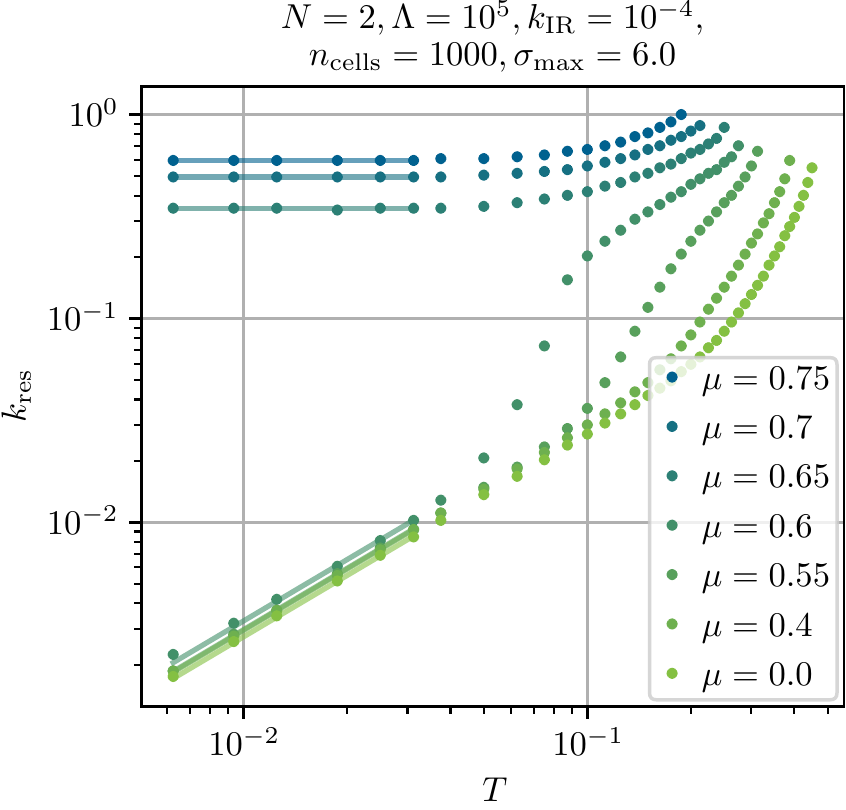}
			\caption{\label{fig:T_k_restored}
				Restoration scale $k_\mathrm{res}$, where the $\mathbb{Z}_2$ symmetry is restored during the RG flow, as a function of temperature $T$ for a constant number of fermionic flavors $N = 2$ and various values of $\mu$. The points corresponding to ${\mu = 0.65}$, ${\mu = 0.7}$, and ${\mu = 0.75}$ correspond to regions in the $\mu$-$T$ plane, where the $\mathbb{Z}_2$ symmetry is restored by the chemical potential, already before bosonic fluctuations are active, \textit{cf.}\ Fig.~\ref{fig:phase_boundaries}. For temperatures larger than the points, which are shown in the plot, there is never symmetry breaking throughout the entire RG flow. For $T\lesssim 0.03$ we plot $k_{\mathrm{res}}(T) \propto T^{0}$ for $\mu>0.6$ and $k_{\mathrm{res}}(T) \propto T$ for $\mu<0.6$ ought to be used for optical guidance.
			}
		\end{figure}
	In Fig.~\ref{fig:T_k_restored} we present results for the temperature dependence of the symmetry restoration scale $k_\mathrm{res} ( T )$ for various chemical potentials.  We observe that for large temperatures, which are already close to the mean-field critical temperature, the influence of the thermal fluctuations, also in the fermionic loop contribution is too large to have a simple relation between $k_\mathrm{res}$ and $T$. For $T\gtrsim 0.03$ we find that $k_\mathrm{res} ( T )$ depends non-trivially on $\mu$ indicating a complicated interplay of thermal, density and quantum fluctuations that leads to symmetry restoration.
	
	For low temperatures we can identify two distinct trends in Fig.~\ref{fig:T_k_restored}. On the one hand, the points $k_\mathrm{res} ( T )$ for $\mu>0.6$ become insensitive to $T$ for $T\lesssim 0.03$ while showing a clear dependence on $\mu$: $k_\mathrm{res} ( T ) = c ( \mu ) \, T^0$. In this regime it is not the bosonic quantum fluctuations that restore the symmetry, but rather density fluctuations related to the chemical potential. This is already the case in the limit $N \rightarrow \infty$ (mean-field). In the fluid dynamical interpretation of fermions as a source/sink term, symmetry gets restored at large chemical potentials due to the manifestation as a source in this scenario, \textit{cf.}\ Sub.Sec.~\ref{subsec:chemical_potential_shock_wave}.
	
	On the other hand, the points $k_\mathrm{res} ( T )$ for $\mu < 0.6$ become rather insensitive to $\mu$ for $T \lesssim 0.03$ while showing a clear linear dependence on $T$, \textit{viz.} $k_\mathrm{res} ( T ) = c ( \mu ) \, T$, where the constant $c(\mu)$ depends only very weakly on $\mu$. This implies that for small $T$ the fermionic contributions to the flow are almost negligible and predominantly the first non-zero Matsubara mode $2 \pi  T$ controls $k_\mathrm{res}$, which also explains the linear relation between $k_\mathrm{res}$ and $T$.
	
	For $T<0.3$ and $\mu<0.6$ we find symmetry restoration at a finite $k_\mathrm{res}$ -- consequently no SSB in the IR -- at finite $N = 2$, which is in direct contrast to the $N \rightarrow \infty$ (mean-field) results, where symmetry is still broken in this regime. Assuming that the functional trends identified in Fig.~\ref{fig:T_k_restored} at low $T$ hold in the limit $T \longrightarrow 0$, the linear relations $k_\mathrm{res} ( T ) \propto T^1$ suggest SSB in the IR at $T = 0$ even at finite $N$. We will come back to this possibility in Sub.Sec.~\ref{subsec:PDfiniteN} when discussing the phase diagram after we explore the situation in vacuum in Sub.Sec.~\ref{subsec:vacFiniteN_LD1}.\\

\begin{figure} %TODO: Niklas: bitte den plot neu beschriften mit T_\mathrm{pc}
			\centering
			\includegraphics{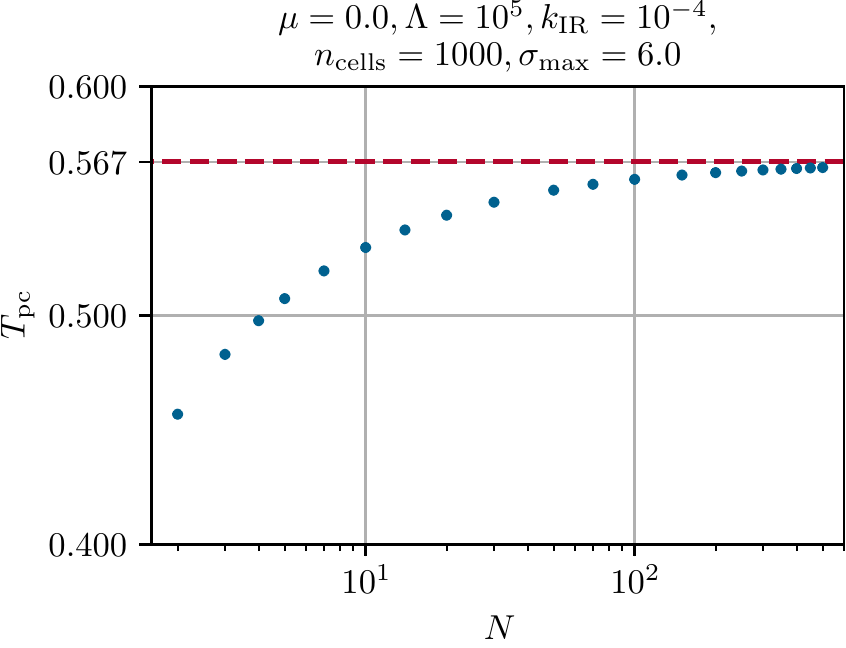}
			\caption{\label{fig:precondensation_T}
				Precondensation temperature $T_\mathrm{pc} ( N )$ as a function of the number of fermions $N$ at $\mu = 0$. Hereby $T_\mathrm{pc} ( N )$ is defined as the temperature that is needed to keep the system in the $\mathbb{Z}_2$ symmetric phase over the entire RG flow, meaning that $\sigma_\mathrm{min} ( k ) = 0$ at all scales $k$. The {red-dashed} line marks the critical temperature $T_\mathrm{C}\simeq 0.567$ from mean-field calculations, see Eq.~\eqref{eq:Tc_MF}.
			}
		\end{figure}

	Before we turn to further discussion concerning the chemical potential, we conclude this subsection on temperature dependencies with another plot, namely Fig.~\ref{fig:precondensation_T}. With this figure we study the dependence of the temperature $T_\mathrm{pc}$ on $N$. $T_\mathrm{pc}$ is the precondensation temperature \cite{Boettcher:2012cm,Boettcher:2012dh,Boettcher:2013kia,Boettcher:2014tfa,Roscher:2015xha,Khan:2015puu}, which is defined in our work as the threshold temperature above which the system is always in the symmetric phase for all $k(t)$ at $\mu = 0$ and $\mathbb{Z}_2$ symmetry is never broken during the RG flow. We observe that 
	$T_\mathrm{pc}$ approaches the mean-field value for the critical temperature $T_\mathrm{C} \simeq 0.567$ while increasing $N$. It should however be stressed, that for finite $N$ $T_\mathrm{pc}$ is not a critical temperature associated with a second order phase transition to a symmetry broken phase in the IR. While symmetry breaking occurs for $T < T_\mathrm{pc} ( N )$ during the RG flow, bosonic fluctuations restore symmetry in the IR for all finite $N$. Only in the limit $N \rightarrow \infty$ the mean-field result of a second order phase transition at $\lim_{N \rightarrow \infty} T_\mathrm{pc} ( N ) = T_\mathrm{C}$ is recovered, which again qualitatively confirms the consistency of our numeric results with the mean-field calculations.
	
\subsection{Varying the chemical potential \texorpdfstring{$\mu$}{mu}}\label{subsec:variablemu}

\begin{figure}
			\centering
			\includegraphics{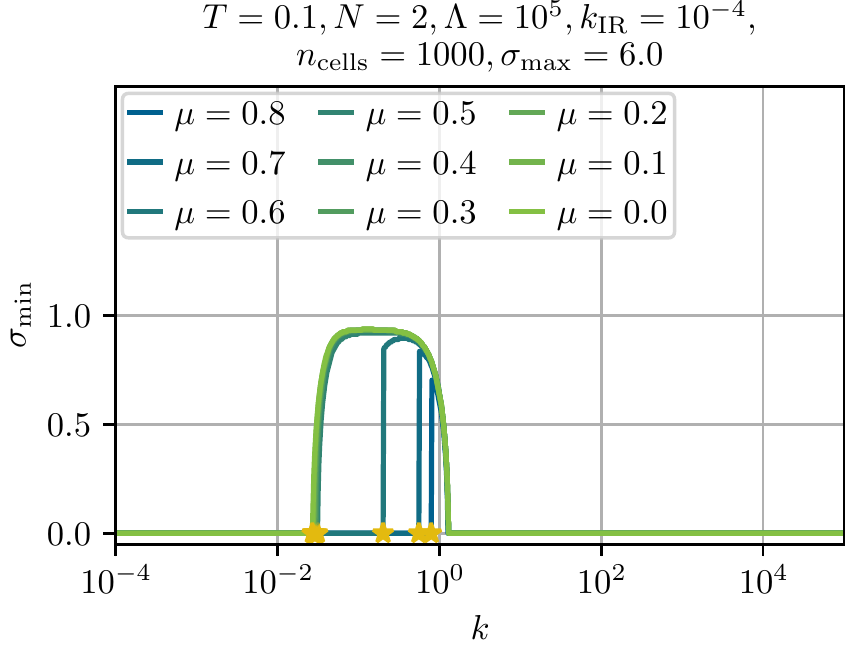}
			\caption{\label{fig:mu_k_sigma_min}
				Value of the condensate (position of the minimum) $\sigma_\mathrm{min} ( k )$ for various $\mu$ as a function of the RG scale $k ( t )$ at constant temperature $T = 0.1$ and constant $N = 2$. The yellow stars mark the RG scales, where the $\mathbb{Z}_2$ symmetry is restored.
			}
		\end{figure}

	To further study the relation between $k_\mathrm{res}$ and $\mu$, we proceed as follows. First, we fix $N = 2$ and $T = 0.1$ and again look at $\sigma_\mathrm{min} ( t )$ plotted over the RG scale $k ( t )$ in Fig.~\ref{fig:mu_k_sigma_min}.
	Here, we observe that for large $\mu>0.6$ the fermionic density fluctuations restore the symmetry during the RG flow, signaled by a $k_\mathrm{res}$ which is slightly smaller than $\mu$ but of the same order of magnitude. The strip at $T=0.1$ with $\mu>0.6$ is in the restored phase of the mean-field phase diagram and the dynamics at finite and infinite $N$ are dominated by fermionic density fluctuations mediated by the chemical potential. Small values of $\mu<0.6$ cannot influence a large region in field space. The source contributions at small $\mu$ are insufficient to restore the symmetry (compare with our previous discussions). For $\mu<0.6$ the restoration scale $k_\mathrm{res}$ is always the same and is set by the temperature.
		\begin{figure}
			\centering
			\includegraphics{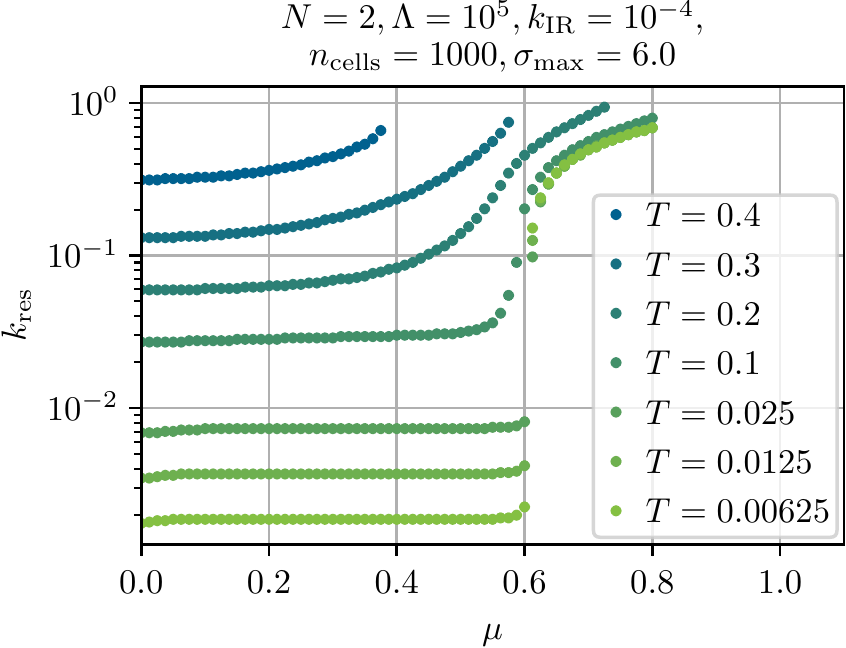}
			\caption{\label{fig:mu_k_restored}
				Restoration scale $k_\mathrm{res}$, where the $\mathbb{Z}_2$ symmetry gets restored during the RG flow, as a function of the chemical potential $\mu$ for different temperatures for constant $N = 2$.
			}
		\end{figure}
	
	This behavior is visualized even better in Fig.~\ref{fig:mu_k_restored}, where we plot $k_\mathrm{res} ( \mu )$ for various $T$ and note that $k_\mathrm{res} ( \mu )$ becomes insensitive to $\mu$ for small chemical potentials. We observe remnants -- the jump/large gradient in $k_\mathrm{res} ( \mu )$ at $\mu\approx 0.6$ around $T\approx0.3$ --  of the mean-field first order phase transition below the mean-field critical point $(\mu_\mathrm{CP},T_\mathrm{CP})\simeq(0.6,0.3)$.

\subsection{Computations in vacuum}\label{subsec:vacFiniteN_LD1}
	Before we conclude this section with the discussion of the phase diagram at finite $N$ we turn to selected results in vacuum at vanishing temperature and chemical potential. Direct numerical computations at $T=0$ and $\mu>0$ and finite $N$ were not possible within this work as discussed at length in the previous subsections. Computations at $T=0$ and vanishing chemical potential $\mu=0$ are however possible at finite $N$. In this subsection we discuss a specific vacuum flow at $N=2$ obtained by numerical solution of the vacuum flow Eq.~\eqref{eq:vacuum_limit_flow_equation} (strictly speaking the $\sigma$-derivative of Eq.~\eqref{eq:vacuum_limit_flow_equation}) with the one-dimensional Litim regulator also used for the previous computations at $T>0$.\\

	The vacuum RG flow for $N=2$ is displayed in Fig.~\ref{fig:flow_L1D_N=2,T=0.0,mu=0.0}, showing the scale evolution from the UV and the initial condition~\eqref{eq:initial-condition_small_u} towards the IR. The corresponding flows of the running minimum $\sigma_{\mathrm{min}} ( k )$, the squared curvature mass $m_\sigma^2 ( k ) = \partial_\sigma u ( t, \sigma )$ at the IR minimum $\sigma_\mathrm{min}>0$ with the corresponding changing rate  $| \partial_t m_\sigma^2 ( k ) |$ according to Eq.~\eqref{eq:changing_rate_mass} are plotted in Fig.~\ref{fig:k_L1D_N=2,T=0.0,mu=0.0}. We observe SSB in the IR indicated by the non-zero minimum $\sigma_\mathrm{min} > 0$. The value of $\sigma_\mathrm{min}\approx 0.907$ for $N = 2$ is slightly smaller than the mean-field value in the limit $N \rightarrow \infty$ of $\sigma_\mathrm{min} = \sigma_0 = 1$. The curvature mass squared for $N = 2$ is extremely small with $m_\sigma^2 \approx 1.04 \cdot 10^{-5}$ , significantly smaller than the mean-field value of $m_\sigma^2 =\tfrac{1}{\pi}\simeq 0.318$, see Eq.~\eqref{eq:msigma_MF}, in the limit $N \rightarrow \infty$. The changing rate  $| \partial_t m_\sigma^2 ( k ) |$ indicates an extremely long dynamical range in RG scale $k$. The IR cutoff of $k_\mathrm{IR}=10^{-4}$ is arguably not low enough and integration deeper into the IR should be performed to ensure that all relevant long-range bosonic vacuum fluctuations are included. However the computations in vacuum are numerically extremely demanding. Lower IR cutoff would require better spatial resolution and potentially even higher numerical precession for numerical RG time evolution. Both increase the computational time significantly. Computations with lower IR cutoffs were infeasible within the scope of this work. This limitation also implicitly excludes studies at significantly higher finite $N$ since the dynamics get shifted to even lower RG scales for $N>2$ as discussed in Sub.Sec.~\ref{subsec:variableN}.

	The result of SSB in the IR for finite $N=2$ at $T=\mu=0$ is supported by the results discussed in App.~\ref{app:vac_flow_L2D} obtained from vacuum flows using two-dimensional Litim regulators. A non-vanishing $\sigma_\mathrm{min}$ in vacuum is also supported by the results at finite but low temperature of Sub.Sec.~\ref{subsec:variableT} and especially the results presented in Fig.~\ref{fig:T_k_restored}.

	\begin{figure}
		\centering
		\includegraphics{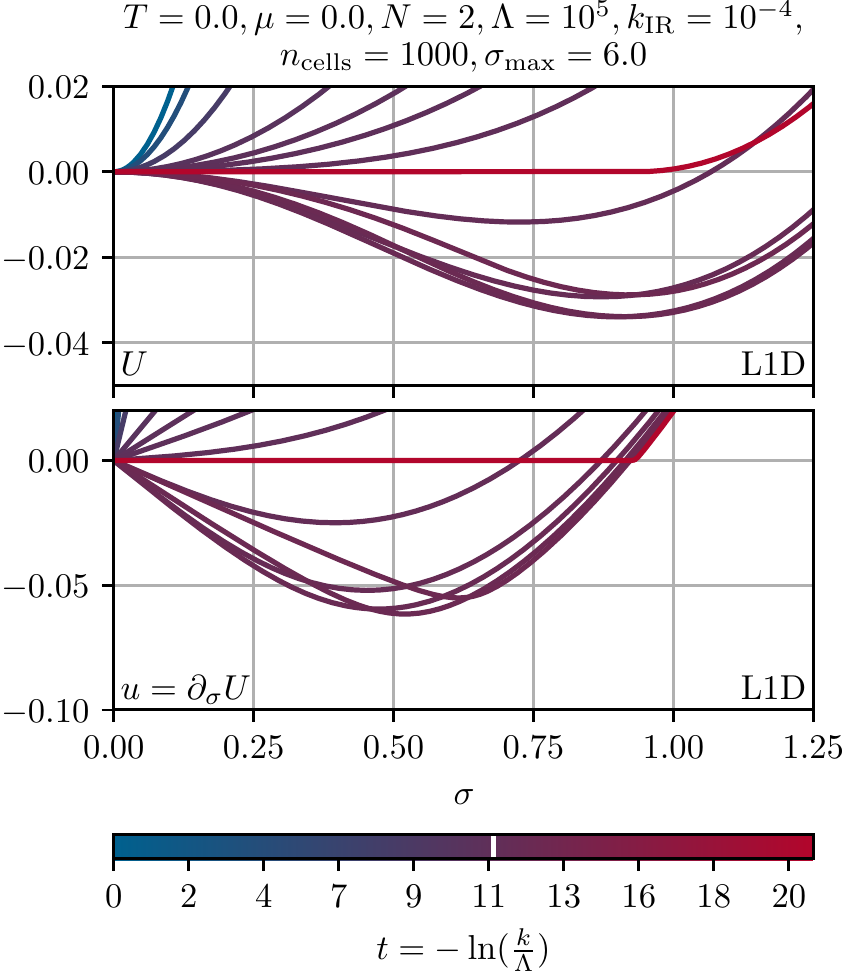}
		\caption{\label{fig:flow_L1D_N=2,T=0.0,mu=0.0} 
			RG flow of the scale dependent effective potential $U ( t, \sigma )$ (upper panel) and its $\sigma$-derivative (the fluid) $u ( t, \sigma ) = \partial_\sigma U ( t, \sigma )$ (lower panel) from the UV ({blue}) to the IR ({red}) in vacuum ($T = \mu =0$). For the sake of simplicity (and using the (anti-)symmetry in $\sigma$) the functions $u ( t, \sigma )$ and $U ( t, \sigma )$ are plotted for positive $\sigma$ only. The different RG-times are encoded via the colored bar-legend below the plots. The white vertical line in the colored bar-legend denotes the RG time (scales) when the $\mathbb{Z}_2$ symmetry is broken (condensation). We do not find symmetry restoration in vacuum for finite $N=2$ within the RG flow for $k\geq k_\mathrm{IR}=10^{-4}$. Results were obtained using the conventional one-dimensional, spatial LPA-optimized (Litim) regulator. All other parameters are stated in the figure.
		}
	\end{figure}
	\begin{figure}
		\centering
		\includegraphics{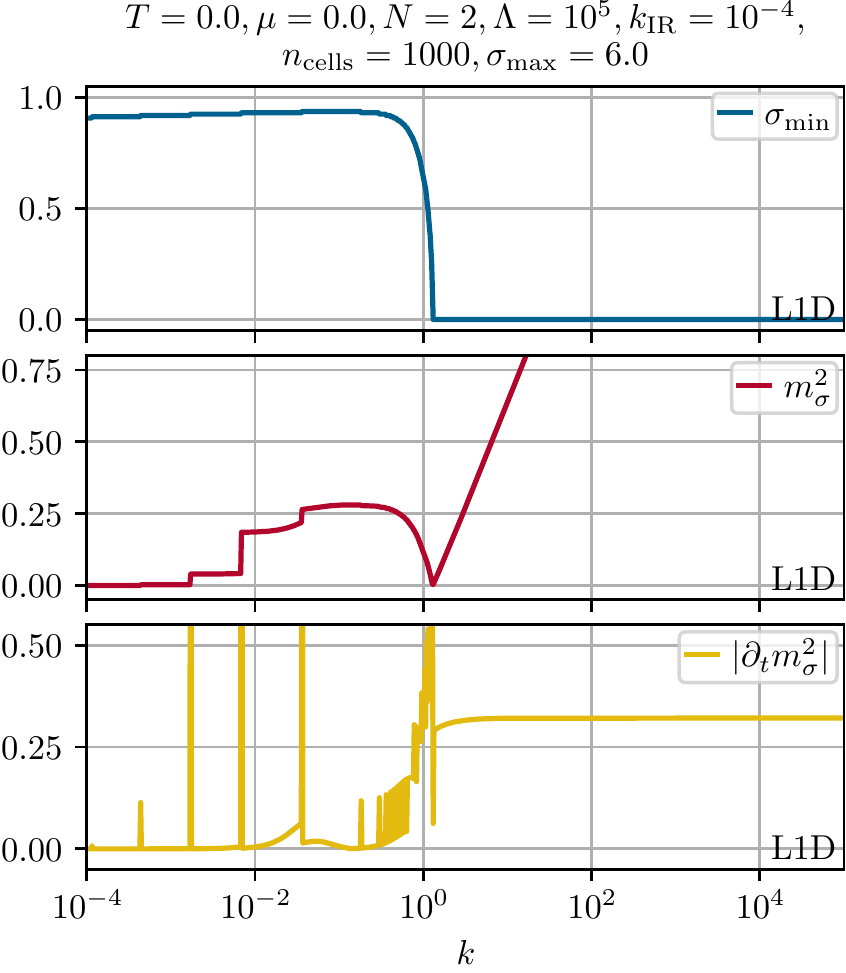}
		\caption{\label{fig:k_L1D_N=2,T=0.0,mu=0.0}
			RG flow of the minimum $\sigma_{\mathrm{min}} ( k )$ of the scale dependent effective potential $U ( k, \sigma )$ (upper panel), the squared curvature mass $m_\sigma^2 ( k ) = \partial_\sigma u ( t, \sigma )$ (middle panel), and the relative change of the squared curvature mass $| \partial_t m_\sigma^2 ( k ) |$ (lower panel) according to Eq.~\eqref{eq:changing_rate_mass} plotted as functions of the RG scale $k ( t )$ in vacuum ($T = \mu =0$). The plot corresponds to the RG flows of $u ( t, \sigma )$ and $U ( t, \sigma )$ of Fig.~\ref{fig:flow_L1D_N=2,T=0.0,mu=0.0}.
		}
	\end{figure}
	
\subsection{The phase diagram}\label{subsec:PDfiniteN}
	With this subsection we finally turn to a discussion of the phase diagram in the $\mu$-$T$-plane at finite $N$. We focus explicitly on $N = 2$ but the qualitative statements should, following Sub.Sec.~\ref{subsec:variableN}, generalize to finite $N > 2$.
	
\begin{figure}
			\centering
			\includegraphics{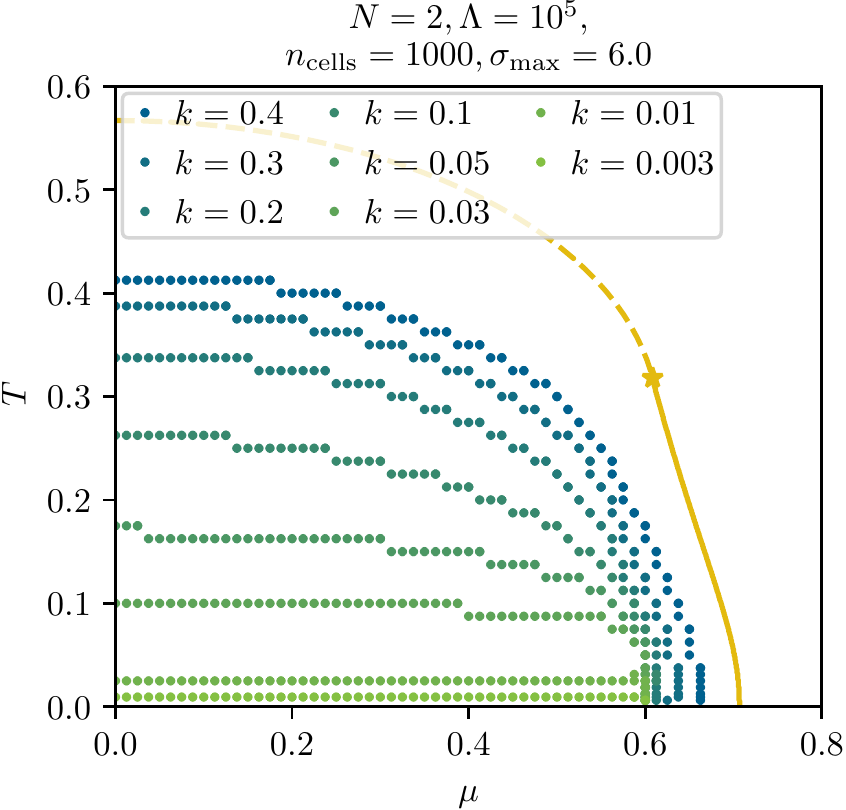}
			\caption{\label{fig:phase_boundaries}
				Phase transition lines (equally colored dots) for the GNY model with $N = 2$ in the $\mu$-$T$ plane depending on the RG scale $k ( t )$. For values $k ( t )$ that are slightly larger or smaller than the values, which are explicitly stated in the figure legend, there is no symmetry broken regime at non-zero temperatures. The plot is extracted from independent RG flows at points $( \mu_i, T_j )$ with $\mu_i = 0.0125 \cdot i$, $i \in \{ 0, \ldots, 65 \}$ and $T_j = 0.0125 \cdot j$, $j \in \{ 1, \ldots, 48 \}$. For better resolution at small temperatures, we also included calculations at points with $T_j = 0.00625 \cdot j$, $j \in \{ 1, \ldots, 4 \}$ and the same $\mu_i$ as before. The yellow solid/dashed line is the mean-field result for the phase transition line and taken from Fig.~\ref{fig:GNlargeN_PD}. It is only plotted as reference and serves as optical guidance.
			}
		\end{figure}

\begin{figure}
			\centering
			\includegraphics{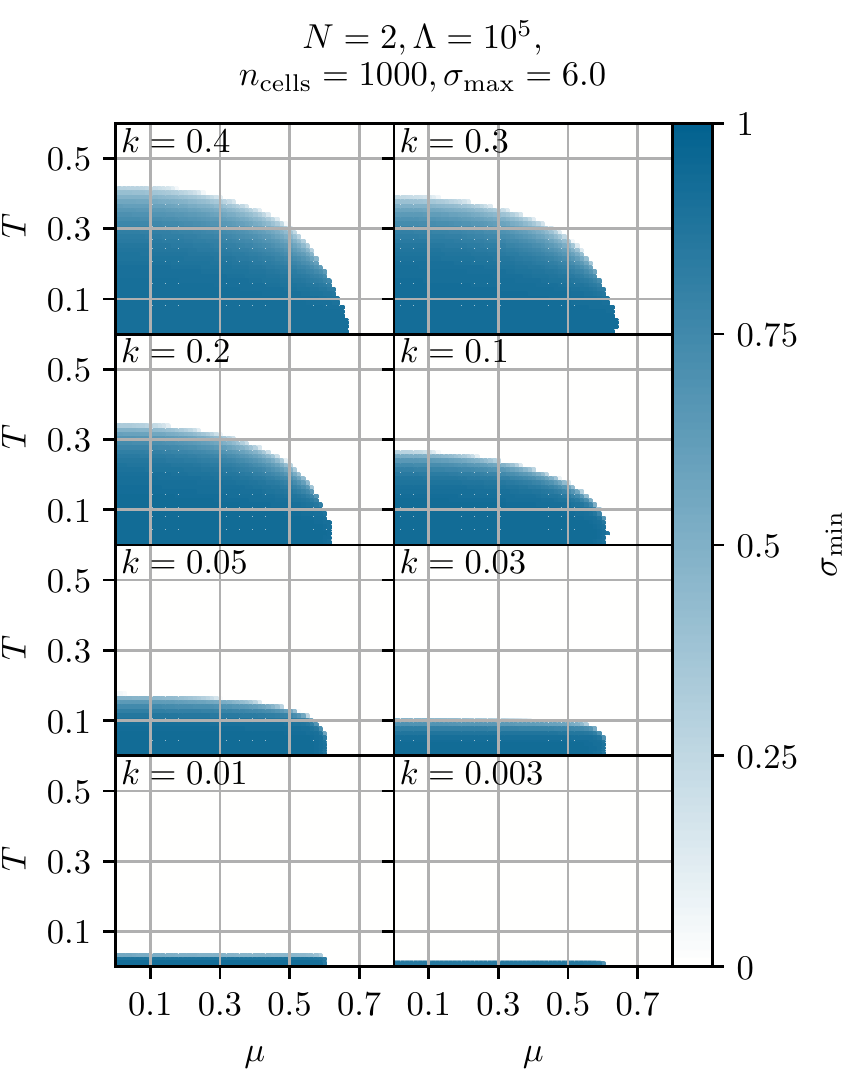}
			\caption{\label{fig:phase_diagram}
				Phase diagram of the GNY model in the $\mu$-$T$-plane with $N = 2$ at selected RG scales $k ( t )$ during the RG flow, where condensation is present. The single plots show the value of the minimum $\sigma_\mathrm{min}$ over $\mu$ and $T$, such that the type of phase transition is roughly recognizable. The figure complements Fig.~\ref{fig:phase_boundaries} and is based on the same RG flows.
			}
		\end{figure}

	In Fig.~\ref{fig:phase_boundaries} we plot the phase transition lines in the $\mu$-$T$ plane for different $k ( t )$. For slightly larger and smaller values of $k ( t )$ -- including the physical point in the IR -- than those that are presented in the plot legend, there is no phase with $\mathbb{Z}_2$ symmetry breaking at finite temperature. For larger $k(t)$ not enough momentum modes are included to allow for the formation of a non-trivial minimum, while for lower scales $k(t)$ bosonic long-range fluctuations already vaporized the condensate. With Fig.~\ref{fig:phase_diagram}, we present complementary density plots for the condensate at the selected $k(t)$ of Fig.~\ref{fig:phase_boundaries}.

	We find that when symmetry breaking sets in, the phase transition line looks similar to its infinite-$N$ counter part (yellow line). However, the region of broken $\mathbb{Z}_2$ symmetry is smaller at its formation at $k(t)\approx0.4$, since thermal bosonic fluctuations work against the symmetry breaking induced by the fermions. As soon as one further decreases $k ( t )$, the symmetry broken regime shrinks drastically. Interestingly, we can observe directly, that for small $T$ and late RG times the phase boundary is almost independent of $\mu$. This was already discussed in the previous subsections. Ultimately, the entire $\mathbb{Z}_2$ symmetry broken phase vanishes for $T > 0$, such that plotting a ``phase diagram'' in the $\mu$-$T$-plane at the physical point in the IR is kind of pointless. Still, Figs.~\ref{fig:phase_boundaries} and \ref{fig:phase_diagram} clearly show the region in the $\mu$-$T$-plane, where the precondensation phenomenon \cite{Boettcher:2012cm,Boettcher:2012dh,Boettcher:2013kia,Boettcher:2014tfa,Roscher:2015xha,Khan:2015puu} takes place.

	It is also noteworthy, that the time period during the RG flow, hence the range of RG scales, where we find $\mathbb{Z}_2$ symmetry breaking, is rather small (approximately $k \in [ 0.4, 0.003 ]$ spanning over roughly two orders of magnitude), if compared to the total flow-time, respectively the nine-orders of magnitude between the UV scale $k_\mathrm{UV} = \Lambda = 10^5$ and the IR scale $k_\mathrm{IR} = 10^{-4}$. A dynamical range of roughly two orders of magnitude starting at around $k(t)\approx 1$ was also observed in the previous subsections for computations at $T>0$ and $N=2$.\\
	
	We conclude this subsection with a few remarks regarding the situation at zero temperature and finite $N$. From the $N=2$ computations in vacuum, presented in the previous Sub.Sec.~\ref{subsec:vacFiniteN_LD1}, we have strong reasons to believe that $\mathbb{Z}_2$ symmetry breaking persists in vacuum for finite $N$ at low RG scales in the IR. The results obtained by direct computations at low temperatures and $\mu\geq 0$ discussed in Sub.Sec.~\ref{subsec:variableT} support this notion and suggest that $\mathbb{Z}_2$ symmetry breaking persists at non-zero $\mu$ until a chemical potential of $\mu\approx 0.6$ is reached. For $\mu\gtrsim 0.6$ at $T=0$ and $N=2$ we do not expect symmetry breaking in the IR. In order to give a more definite and refined picture of the situation at $T=0$ and $\mu\geq 0$ further computations as well as research and development are required including computations at even lower RG scales as well as direct computations at $T=0$ and $\mu>0$ for finite $N$.
	
	A phase transition at zero temperature driven by an external parameter (or field) rather than thermal fluctuation is called a quantum phase transition. In the context of the GN(Y) model at zero temperature the chemical potential acts as such an external parameter. Fermionic vacuum fluctuations are responsible for symmetry breaking in vacuum and low $\mu$, while density fluctuations induced by a non-zero chemical potential (at finite and infinite $N$) as well as bosonic quantum fluctuations (at finite $N$ only) drive the system towards symmetry restoration. For a general pedagogic discussion of quantum phase transitions we refer to the textbook~\cite{Sachdev:2011} as well as the review article~\cite{Dutta:2010}. There are multiple systems known to exhibit a quantum phase transition, see, \textit{e.g.}, Refs.~\cite{Sachdev:2011,Dutta:2010} and references therein.
	
	In the context of the GN model with its various connections to solid-state and/or spin systems, see Sec.~\ref{sec:introduction}, the so called transverse quantum Ising model, see, \textit{e.g.} Refs.~\cite{Fradkin:2013,Dutta:2010}, is particularly interesting. The path integral in discretized imaginary time of the one-dimensional quantum Ising model is equivalent to the partition function of the two-dimensional Ising model from classical statistical physics \cite{Fradkin:2013}. The one-dimensional transverse quantum Ising model in an external field $h$ with ferromagnetic interactions $J_x\geq0$, \textit{cf.} Ref.~\cite{Dutta:2010} Eqs.~(7) and (9), is a seminal example of a system exhibiting a quantum phase transition at vanishing temperature. The external field introduces quantum fluctuations in the model causing a quantum phase transition at a critical value $h_c$ from an ordered, ferromagnetic phase with broken $\mathbb{Z}_2$ symmetry to a disordered, paramagnetic phase with restored $\mathbb{Z}_2$ symmetry \cite{Fradkin:2013,Dutta:2010}. This situation is very reminiscent of the proposed scenario in the GNY model, which is put forward in the previous paragraphs of this work, again underlying the intricate connection between the GN model and solid-state spin systems in continuum. Understanding and formalizing this connection further in studies of the GN(Y) model from a high-energy physics perspective seems extremely promising especially in the context of the finite-$N$ phenomenology of the GN(Y) model.
	
\section{Conclusion and outlook}
\label{sec:conclusion_outlook}

	Using the FRG framework we have performed computations for the Gross-Neveu-Yukawa (GNY) model at finite and infinite number of flavor $N$ at finite temperature $T\geq0$ and quark chemical potential $\mu\geq0$.\\

	For infinite $N$ the bosonic fluctuations are completely suppressed within the GNY model, which is then equivalent to the Gross-Neveu (GN) model. Within the FRG framework in the limit $N \rightarrow \infty$ we recover well-known infinite-$N$ (mean-field) results for the homogeneous phase diagram of the GN model. The discrete chiral $\mathbb{Z}_2$ symmetry is spontaneously broken at small chemical potentials and temperatures and gets restored at high temperatures and chemical potentials across a second-order (first-order) phase transition for high temperatures (low temperatures). The restoration of $\mathbb{Z}_2$ symmetry in the infinite-$N$ limit is purely driven by fermionic thermal and density fluctuations.\\

	Computations at finite $N$ have been performed with the GNY model in the FRG framework using the local potential approximation (LPA) with one-dimensional Litim regulators. At finite $N$ the GNY model and GN model are not equivalent in LPA truncation. We argue however that, due to our specific choice for the classical action of the GNY model, the phenomenology of GNY model in LPA should be qualitatively similar to the one of the GN model. Direct computations with the GN model in LPA truncation within the FRG framework are not feasible. A proper resolution of the differences between the GNY and GN model at finite $N$ would be possible by improving the truncation scheme. Especially the addition of a scale dependent wave function renormalization for the scalar $\sigma$-channel is a natural next step in this direction. The improvement of the truncation scheme (including also additional scale and possibly field dependent terms like a Yukawa coupling $h_t(\sigma)$ \textit{etc.}) is in general a necessary step to assess the quality and predictive power of the presented LPA results and are subject of future works. 

	Numerical computations with the LPA flow equation within this work are based on a reformulation of the corresponding FRG flow equation as a conservation law put forward by some authors and collaborators in Refs.~\cite{Grossi:2019urj,Grossi:2021ksl,Koenigstein:2021syz,Koenigstein:2021rxj,Steil:2021cbu,Steil:2021partIV}. The flow equation in this setup manifests as a non-linear diffusion equation with a sink/source term, where the diffusive contributions can be clearly attributed to bosonic quantum fluctuations, while fermionic fluctuations enter the flow via a sink/source term. Numerical FRG scale evolution from the UV to the physical point in the IR is performed using a potent semi-discrete finite volume discretization of the flow equation from the field of computational fluid dynamics.\\ 

	Numerical results for various finite $N$ and especially $N=2$ at non-zero temperature and/or chemical potential were presented and discussed in Sec.~\ref{sec:rg-flow-with-bosons} and have revealed that there is no spontaneous $\mathbb{Z}_2$ symmetry breaking at non-zero temperature for finite $N$. This binary result is in agreement with heuristic arguments of L.~D.~Landau \textit{et al.} \cite{Landau:1980mil} and in the context of the GN model of R.~F.~Dashen \textit{et al.} \cite{Dashen:1974xz}, which were summarized and discussed in Sub.Sec.~\ref{subsec:z2breaking}. The situation at vanishing temperature is not completely settled yet. 

	Direct computations in vacuum are numerically challenging but possible and suggest spontaneous $\mathbb{Z}_2$ symmetry breaking even for finite $N$ at $T=\mu=0$, see Sub.Secs.~\ref{subsec:vacFiniteN_LD1} and \ref{subsec:PDfiniteN} as well as App.~\ref{app:vac_flow_L2D}. Long-range bosonic quantum fluctuations do not appear to be strong enough to restore the $\mathbb{Z}_2$ symmetry initially spontaneously broken by the fermionic interactions at finite $N$ in vacuum. Further computations including flows to even lower RG scales should be performed to further understand our findings. This might however require significant computational resources or further research and development.
	
	Computations at zero and very low temperatures and non-zero chemical potentials within the used LPA flow equation are also challenging and arguably, see Sub.Sec.~\ref{subsec:chemical_potential_shock_wave}, impossible at zero temperature. This is a novel aspect of the LPA flow equation, which has not been discussed at this level in literature to the best of our knowledge. It might even point to general conceptual problems with the LPA flow equation. Further research and development will be necessary to better understand this aspect, observed here in the LPA truncation. Direct computations beyond very low temperatures are however possible without conceptual or numerical challenges even at non-zero chemical potentials, see Sub.Secs.~\ref{subsec:variableT} and \ref{subsec:vacFiniteN_LD1} for results and discussions. Considering the vacuum results and an extrapolation from results at low temperatures we have strong reasons to believe that a quantum phase transition between a phase of broken $\mathbb{Z}_2$ symmetry at low chemical potentials and a restored phase at higher chemical potentials is a highly likely scenario at $T = 0$ and finite $N$. This supports also the rather general narrative in this paper that the phenomenology of the GN(Y) model is very similar to the phenomenology of solid-state spin systems. A summary of some known analogies was given in Sub.Sec~\ref{subsec:GN_pheno}, while in the context of a possible quantum phase transition the transverse quantum Ising model might be of particular importance, see last Sub.Sec.~\ref{subsec:PDfiniteN}. Further research especially regarding the role of the GN model in solid-state physics could be very rewarding.\\
	
	Repeating the study presented here for the GN(Y) model for other $1 + 1$ dimensional models including four fermi-interactions is a very interesting prospect for future works and partially in preparation. A FRG study at non-zero $T$ and $\mu$ for the Thirring model falls into this category. For an overview of established models and results in the infinite-$N$ limit see Tab.~1 of Ref.~\cite{Thies:2020ofv} and references therein. Many model extensions of the GN model include a continuous chiral symmetry and as such SSB leads to the presence of massless Nambu-Goldstone bosons and thus such models would allow explicit studies of the Coleman-Mermin-Wagner-Hohenberg theorem, see Sub.Sec.~\ref{subsec:z2breaking}. Apart from studies with selected four Fermi-interaction channels a systematic study using a Fierz-complete basis of four Fermi-interactions, \textit{cf.} Refs.~\cite{Braun:2017srn,Braun:2018bik,Braun:2019aow}, would also be a very promising direction to gain a detailed understanding of $1+1$ dimensional four-Fermi theories and SSB in them. In this context an investigation of the competition between the $U ( 1 )_\mathrm{A}$ anomaly and the Coleman-Mermin-Wagner-Hohenberg theorem would be interesting.
	
	Another extremely interesting research direction is the study of the GN(Y) in finite volumes, hence in a one-dimensional spatial box. We plan to repeat the analysis of this work for the GNY in a finite spatial volume along the lines of Refs.~\cite{Braun:2004yk,Braun:2005gy,Braun:2005fj,Braun:2011iz,Braun:2011uq,Klein:2017shl} elsewhere, in order to directly analyze the effects of a finite sized spatial volume and to compare our results to the ones obtained with lattice Monte-Carlo simulations \cite{Cohen:1981qz,Cohen:1983nr,Karsch:1986hm,Lenz:2020bxk,Lenz:2020cuv,Pannullo:2019bfn,Pannullo:2019prx}. A study in finite volumes at infinite and finite number of flavors $N$ is also of conceptual and pedagogical interests when it comes to the general question of symmetry breaking (or remnants/signatures similar to it) in finite spatial volumes $L$. In this context the specific order of limits like $N \rightarrow \infty$, $L \rightarrow \infty$ and potentially $\tfrac{1}{T} \rightarrow \infty$ is expected to be extremely relevant.	The FRG framework allows for a study of all three aforementioned limits in a systematic fashion.

\begin{acknowledgments}
	A.K., S.R., M.J.S., J.S., and N.Z.\ acknowledge the support of the \textit{Deutsche Forschungsgemeinschaft} (DFG, German Research Foundation) through the collaborative research center trans-regio  CRC-TR 211 ``Strong-interaction matter under extreme conditions''-- project number 315477589 -- TRR 211.
		
	A.K.\ acknowledges the support of the \textit{Friedrich-Naumann-Foundation for Freedom}.
		
	A.K.\ and M.J.S.\ acknowledge the support of the \textit{Giersch Foundation} and the \textit{Helmholtz Graduate School for Hadron and Ion Research}.
	
	N.~Z.\ acknowledges the support of the \textit{Federal Ministry of Education and Research (Germany)} (BMBF) via the \textit{Deutschlandstipendium} scholarship program.
	
	We thank J.~Braun, M. Buballa, and D.~H. Rischke for valuable comments and enlightening discussions on the content and the first drafts of the manuscript. Furthermore, we are grateful for their encouragement to realize this project as well as their supervision of PhD and Master theses, which are associated to this project.
	
	We thank E.~Grossi and N.~Wink for their collaboration in related projects \cite{Koenigstein:2021syz,Koenigstein:2021rxj}, whose formal developments were important for this project.
	
	We further thank F.~Divotgey, J.~Eser, F.~Giacosa, L.~Kurth, L. Pannullo, A.~Sciarra, M.~Wagner, M.~Winstel for valuable discussions.

	We thank J.~Braun, L. Pannullo, M.~Wagner, N.~Wink for useful comments on the manuscript.

	We thank L. Pannullo for bringing Refs.~\cite{Mandl2021,Nonaka2021} to our attention.

	We thank the organizers of the \textit{1st workshop on Low Dimensional Quantum Many Body Systems} for the opportunity to present this project and we thank all participants for a stimulating environment and comments and discussions regarding this project.\\
	
	All numerical results as well as all figures in this work were obtained using \textit{Python 3} \cite{10.5555/1593511} with various libraries \cite{2020SciPy-NMeth,Hunter:2007,2020NumPy-Array}, if not explicitly stated otherwise. Some of the results were cross-checked with the \texttt{Mathematica} \cite{Mathematica:12.1} code basis of Refs.~\cite{Koenigstein:2021syz,Koenigstein:2021rxj,Steil:2021cbu,Steil:2021partIV}. The ``Feynman'' diagrams in Eqs.~\eqref{eq:wetterich} and \eqref{eq:pdeq-U} were generated using \texttt{Axodraw Version 2} \cite{Collins:2016aya}.
\end{acknowledgments}

\appendix

\section{Conventions}
\label{app:conventions}

	In this appendix we present the conventions that are used throughout this work. 

\subsection{Metric and Clifford algebra}

	We use the following conventions for the Euclidean two-dimensional metric and space-time. We choose ${x^1 \in ( - \infty, + \infty )}$ as the spatial coordinate, while ${x^2 \in ( - \infty, + \infty )}$ is the Euclidean temporal coordinate. The corresponding Euclidean metric reads
		\begin{align}
			\eta = \mathrm{diag} ( + 1, + 1 ) \, .	\label{eq:euclidean_metric}
		\end{align}
	Furthermore, we define the Euclidean $\gamma$-matrices via the Clifford algebra and the anti-commutation relation for the chiral $\gamma$-matrix $\gamma_\mathrm{ch}$,
		\begin{align}
			& \big\{ \gamma^\mu ,\, \gamma^\nu \big\}_+ = 2 \, \eta^{\mu \nu} \, \openone_{d_\gamma} \, ,	&&	\big\{ \gamma^\mu ,\, \gamma_\mathrm{ch} \big\}_+ = 0 \, .\label{eq:app_clifford}
		\end{align}
	At various occasion we will use a short hand notation for the Einstein summation convention over repeated indices in flavor space, Dirac space, and space-time, \textit{e.g.},
		\begin{align}
			\bar{\psi} \, \psi \equiv \, & \bar{\psi}_{f a} \psi^{f a} =	\vphantom{\bigg(\bigg)}
			\\
			= \, & \bar{\psi}_{f_1 a_1} \, \tensor{( \openone )}{^{f_1}_{f_2}} \tensor{( \openone )}{^{a_1}_{a_2}} \, \psi^{f_2 a_2} \, ,	\vphantom{\bigg(\bigg)}	\nonumber
			\\
			\bar{\psi} \, \slashed{\partial} \, \psi \equiv \, & \bar{\psi}_{f a_1} \, \tensor{( \gamma^\mu )}{^{a_1}_{a_2}} \, \partial_\mu \psi^{f a_2} =	\vphantom{\bigg(\bigg)}
			\\
			= \, & \bar{\psi}_{f_1 a_1} \, \tensor{( \openone )}{^{f_1}_{f_2}} \tensor{( \gamma^\mu )}{^{a_1}_{a_2}} \, \partial_\mu \psi^{f_2 a_2} \, ,	\vphantom{\bigg(\bigg)}	\nonumber
		\end{align}
	where $\{ a_i \}$ are indices in Dirac space and $\{ f_i \}$ are indices in flavor space, while $\mu$ is the index of a vector in Euclidean space-time.\\
	
	When switching to compactified space-time, we still use the metric \eqref{eq:euclidean_metric}, because a cylinder is a flat space-time. However, the coordinates are replaced as follows
		\begin{align}
			x^1 \mapsto x \in ( -\infty, \infty ) \, ,	&&	x^2 \mapsto \tau \in  ( 0, \beta ) \, ,
		\end{align}
	where $\beta = \tfrac{1}{T}$ is the inverse temperature. For more academic details on the compactification at non-zero temperature, see App.~\ref{app:qft_on_a_cylinder}.

\subsection{Relation between the Minkowski functional integral and Euclidean partition function}

	The functional integral in two-dimensional Minkowski space-time is defined (up to normalization factors) as
		\begin{align}
			\mathcal{Z} \propto \int [ \mathrm{d} \bar{\psi} , \mathrm{d}\psi ] \, \mathrm{e}^{ \mathrm{i} \mathcal{S}_\mathrm{M} [ \bar{\psi} , \psi ] } \, .
		\end{align}
	Quantities in Minkowski space-time are indicated via the letter ``M'', whereas Euclidean quantities are labeled with ``E''. For Minkowski space-time, we use the metric
		\begin{align}
			\eta_\mathrm{M} = \mathrm{diag} ( - 1, + 1 ) \, ,
		\end{align}
	with Cartesian spatial coordinate ${x \equiv x^1_\mathrm{M} \in ( - \infty, \infty )}$ and temporal coordinate ${t \equiv x^2_\mathrm{M} \in ( - \infty, \infty )}$.
	
	We choose the following convention for the Wick rotation \cite{Wick:1954eu} to Euclidean space-time,
		\begin{align}
			&	x^2_\mathrm{M} = t_\mathrm{M}	&&	\mapsto	&&	x^2_\mathrm{E} = \mathrm{i} t_\mathrm{M} = \mathrm{i} x^2_\mathrm{M} \, ,	\vphantom{\bigg(\bigg)}
			\\
			&	\partial_{2 \mathrm{M}}			&&	\mapsto	&&	\partial_{2 \mathrm{E}} = - \mathrm{i} \partial_{2 \mathrm{M}} \, ,	\vphantom{\bigg(\bigg)}
			\\
			&	\gamma^1_\mathrm{M}				&&	\mapsto	&&	\gamma^1_\mathrm{E} = \mathrm{i} \gamma^1_\mathrm{M} \, .	\vphantom{\bigg(\bigg)}
		\end{align}
	This choice implies
		\begin{align}
			\gamma^\mu_\mathrm{M} \, \partial_{\mu \mathrm{M}} = \, & \gamma^1_\mathrm{M} \, \partial_{1 \mathrm{M}} + \gamma^2_\mathrm{M} \, \partial_{2 \mathrm{M}} =	\vphantom{\bigg(\bigg)}
			\\
			= \, & - \gamma^1_\mathrm{M}\, \partial^1_{\mathrm{M}} + \gamma^2_\mathrm{M}\, \partial^2_{\mathrm{M}} =	\vphantom{\bigg(\bigg)}	\nonumber
			\\
			= \, & \mathrm{i} ( \gamma^1_\mathrm{E} \, \partial^1_{\mathrm{E}} + \gamma^2_\mathrm{E} \, \partial^2_{\mathrm{E}} ) = \mathrm{i} \gamma^\mu_\mathrm{E} \, \partial_{\mu \mathrm{E}} \, .	\vphantom{\bigg(\bigg)}	\nonumber
		\end{align}
	The original Gross-Neveu action in Minkowski space-time, \textit{cf.} Eq.~(2.8) of Ref.~\cite{Gross:1974jv}, and the Euclidean action (studied in this work) are related as follows,
		\begin{align}
			&	\mathcal{S}_\mathrm{M} =	\vphantom{\bigg(\bigg)}
			\\
			= \, & \int \mathrm{d}^2 x_\mathrm{M} \, \big[ \bar{\psi} \, \mathrm{i} \gamma^\mu_\mathrm{M} \, \partial_{\mu \mathrm{M}} \, \psi + \tfrac{g^2}{2 N} ( \bar{\psi} \, \psi )^2 \big] =	\vphantom{\bigg(\bigg)}	\nonumber
			\\
			= \, & - \mathrm{i} \int \mathrm{d}^2 x_\mathrm{E} \, \big[ - \bar{\psi} \, \gamma^\mu_\mathrm{E} \, \partial_{\mu \mathrm{E}} \, \psi + \tfrac{g^2}{2 N} \,  ( \bar{\psi} \, \psi )^2 \big] =	\vphantom{\bigg(\bigg)}	\nonumber
			\\
			= \, & \mathrm{i} \mathcal{S}_\mathrm{E} \, ,	\vphantom{\bigg(\bigg)}	\nonumber
		\end{align}
	Without explicitly indicating Euclidean space-time anymore, we define the Euclidean action as follows,
		\begin{align}
			\mathcal{S}_\mathrm{E} \equiv \int \mathrm{d}^2 x \, \big[ \bar{\psi} \, \slashed{\partial} \, \psi - \tfrac{g^2}{2 N} \, ( \bar{\psi} \, \psi )^2 \big] \, .
		\end{align}
	For the functional integral of our model this results in the following translation,
		\begin{align}
			\mathcal{Z} = \, & \int [ \mathrm{d} \bar{\psi}, \mathrm{d} \psi ] \, \mathrm{e}^{ \mathrm{i} \mathcal{S}_\mathrm{M} } =	\vphantom{\bigg(\bigg)}	\nonumber
			\\
			= \, & \int [ \mathrm{d} \bar{\psi}, \mathrm{d} \psi ] \, \mathrm{e}^{ - \mathcal{S}_\mathrm{E} } =	\vphantom{\bigg(\bigg)}	\nonumber
			\\
			= \, &  \int [ \mathrm{d} \bar{\psi}, \mathrm{d} \psi ] \, \mathrm{e}^{ - \int \mathrm{d}^2 x \, [ \bar{\psi} \, \slashed{\partial} \, \psi - \frac{g^2}{2 N} \, ( \bar{\psi} \, \psi )^2 ] } \,.	\vphantom{\bigg(\bigg)}	\nonumber
		\end{align}

\subsection{Fourier transformation}
\label{app:fourier_transformation}

	Throughout this work we use the following conventions for Fourier transformations in the compactified Euclidean space-time
		\begin{align}
			\varphi ( \tau, x ) = \, & \frac{1}{\beta} \sum_n \int \frac{\mathrm{d} p}{( 2 \pi )} \, \tilde{\varphi} ( \omega_n, p ) \, \mathrm{e}^{+ \mathrm{i} ( \omega_n \tau + p \, x )} \, ,	\vphantom{\Bigg(\Bigg)}
			\\
			\tilde{\varphi} ( \omega_n, p ) = \, & \int_{0}^{\beta} \mathrm{d} \tau \int \mathrm{d} x \, \varphi ( \tau, x ) \, \mathrm{e}^{- \mathrm{i} ( \omega_n \tau + p \, x )} \, ,	\vphantom{\Bigg(\Bigg)}
			\\
			\psi ( \tau, x ) = \, & \frac{1}{\beta} \sum_n \int \frac{\mathrm{d} p}{( 2 \pi )} \, \tilde{\psi} ( \nu_n, p ) \, \mathrm{e}^{+ \mathrm{i} ( \nu_n \tau + p \, x )} \, ,	\vphantom{\Bigg(\Bigg)}
			\\
			\tilde{\psi} ( \nu_n, p ) = \, & \int_{0}^{\beta} \mathrm{d} \tau \int \mathrm{d} x \, \psi ( \tau, x ) \, \mathrm{e}^{- \mathrm{i} ( \nu_n \tau + p \, x )} \, ,	\vphantom{\Bigg(\Bigg)}
			\\
			\bar{\psi} ( \tau, x ) = \, & \frac{1}{\beta} \sum_n \int \frac{\mathrm{d} p}{( 2 \pi )} \, \tilde{\bar{\psi}} ( \nu_n, p ) \, \mathrm{e}^{- \mathrm{i} ( \nu_n \tau + p \, x )} \, ,	\vphantom{\Bigg(\Bigg)}
			\\
			\tilde{\bar{\psi}} ( \nu_n, p ) = \, & \int_{0}^{\beta} \mathrm{d} \tau \int \mathrm{d} x \, \bar{\psi} ( \tau, x ) \, \mathrm{e}^{+ \mathrm{i} ( \nu_n \tau + p \, x )} \, .	\vphantom{\Bigg(\Bigg)}
		\end{align}
	We use anti-periodic boundary conditions for fermions and periodic boundary conditions for bosons at $\tau = \beta$, which results in the discrete bosonic and fermionic Matsubara frequencies \cite{Matsubara:1955ws}
		\begin{align}
			&	\omega_n = \tfrac{2 \pi}{\beta} \, n \, ,	&&	\nu_n = \tfrac{2 \pi}{\beta} ( n + \tfrac{1}{2} ) \, .
		\end{align}
	Furthermore, we abbreviate Matsubara sums using,
		\begin{align}
			\sum_n \equiv \sum_{n = - \infty}^{\infty} \, .
		\end{align}

\section{Hubbard-Stratonovich Transformation and bosonization of the GN model}
\label{app:hubbard-stratonovich}

	To obtain a bosonized version of the GN model we use the Hubbard-Stratonovich transformation \cite{Stratonovich:1957,Hubbard:1959ub}. We introduce the Gaussian integral over a bosonic field $\xi$ using
		\begin{align}
			1 = \# \int [ \mathrm{d} \xi ] \, \mathrm{e}^{- \int \mathrm{d}^2 x \, \frac{\xi^2}{2g^2}} \, ,
		\end{align}
	where $\#$ is some normalization factor. Combining this with the purely fermionic grand-canonical partition function based on the action \eqref{eq:gn-model}, we find
		\begin{align}
			\mathcal{Z} \propto \, &  \int [ \mathrm{d} \bar{\psi}, \mathrm{d}\psi, \mathrm{d} \xi ] \, \mathrm{e}^{- \int \mathrm{d}^2 x \, [ \bar{\psi} \, \slashed{\partial} \, \psi - \frac{g^2}{2 N} \, ( \bar{\psi} \, \psi )^2 + \frac{\xi^2}{2g^2} ]} \, .
		\end{align}
	Next, we shift the bosonic field integration variable
		\begin{align}
			\xi = h \, \phi + \tfrac{g^2}{\sqrt{N}} \, \bar{\psi} \, \psi \, ,
		\end{align}
	where we introduced the Yukawa coupling constant $h$ in order to have bosonic fields $\phi$ with zero energy dimension which is natural in two dimensions. Using
		\begin{align}
			\tfrac{1}{2 g^2} \, \xi^2 = \tfrac{h^2}{2 g^2} \phi^2 + \tfrac{h}{\sqrt{N}} \, \bar{\psi} \, \phi \, \psi + \tfrac{g^2}{2 N} \, ( \bar{\psi} \, \psi )^2 \, ,
		\end{align}
	we can eliminate the four-Fermi interaction term in favor of a Yukawa interaction term,
		\begin{align}
			\mathcal{Z} \propto \, &  \int [ \mathrm{d} \bar{\psi}, \mathrm{d} \psi, \mathrm{d} \phi ] \, \mathrm{e}^{- \int \mathrm{d}^2 x \, [ \bar{\psi} \, ( \slashed{\partial} + \frac{h}{\sqrt{N}} \phi ) \, \psi + \frac{h^2}{2 g^2} \phi^2 ]} \, .
		\end{align}

\section{Quantum fields on a cylinder -- the thermal GNY model}
\label{app:qft_on_a_cylinder}

	This appendix is dedicated to a geometrical construction of the thermal GNY model and mainly addresses readers (especially students), who are not yet familiar with QFTs at non-zero temperature.
	
	Usually in lectures and books on thermal QFT the concept of temperature is introduced via the compactification of the temporal Euclidean space-time direction. Intuitively, this procedure makes sense: High temperatures correspond to a lot of thermal fluctuations, which means that the time, that is needed for a proper measurement of an expectation value for an observable, is very short. On the other hand, low temperatures correspond to little thermal fluctuations, such that a long measurement time is needed to trust a measurement of a statistical observable. Hence, temperature $T$ and inverse Euclidean time $\beta = \tfrac{1}{T}$ can be associated to each other and the conventional Euclidean path integral is the zero-temperature ($\beta \rightarrow \infty$) limit of the thermal partition function. Additionally, one has to specify (anti) periodicity conditions for the fields to implement their proper statistics.
	
	However, one may also reverse the construction process and define the QFT right from the beginning in a space-time with cylindrical topology and extract the infinite radius (zero-temperature) limit afterward. The argument is as follows: temperature sets an external (minimal) energy scale for the QFT. This is analogous to putting a quantum mechanical particle into a box of finite size, where the energy levels of the system form a discrete spectrum, which is measured in multiples of the inverse box size -- the external energy scale. A cylindrical topology of the space-time produces exactly the same result: a discrete spectrum and additional energy scale in the system. Hence, starting formally with classical fields on a two-dimensional cylinder has to result in the same action as the one obtained via the heuristic compactification procedure. (Anti-)periodic ``boundary'' conditions are rather natural, when starting on a compact manifold.
	
	Consider the metric and inverse metric of a cylinder with circumference $\frac{1}{T} = \beta = 2 \pi R$,
		\begin{align}
			&	( g_{\alpha \beta} ) =
			\begin{pmatrix}
				1	&	0
				\\
				0	&	R^2
			\end{pmatrix} \, ,
			&&	( g^{\alpha \beta} ) =
			\begin{pmatrix}
				1	&	0
				\\
				0	&	\tfrac{1}{R^2}
			\end{pmatrix} \, ,	\label{eq:metric_cylinder}
		\end{align}
	where the coordinates are  ${( y^\mu ) = ( z, \phi )}$. Hence, ${y^1 \equiv z \in ( - \infty, + \infty )}$ is identified with the axial coordinate of the cylinder and the azimuth ${y^2 \equiv \phi \in [ 0, 2 \pi )}$ parameterizes the circumference, namely the inverse temperature of the system. 
	
	Next, we study the generic manifest covariant action of a free bosonic particle -- here on our cylinder,
		\begin{align}
			& \int_\mathcal{V} \mathrm{d}^2 y \, \sqrt{ \det g } \, \tfrac{1}{2} \, g^{\alpha \beta} \, ( \nabla_\alpha \varphi ) \, ( \nabla_\beta \varphi ) =	\vphantom{\Bigg(\Bigg)}
			\\
			= \, &  \int \int_{0}^{2 \pi} \mathrm{d} x \,  \mathrm{d} \phi \, R \, \tfrac{1}{2} \, \big[ ( \partial_x \varphi )^2 + \tfrac{1}{R^2} \, ( \partial_\phi \varphi )^2 \big] =	\vphantom{\Bigg(\Bigg)}	\nonumber
			\\
			= \, & \int \int_{0}^{\beta} \mathrm{d} x \, \mathrm{d} \tau \, \tfrac{1}{2} \, \big[ ( \partial_x \varphi )^2 + ( \partial_\tau \varphi )^2 \big] =	\vphantom{\Bigg(\Bigg)}	\nonumber
			\\
			= \, & \int  \int_{0}^{\beta} \mathrm{d}^2 x \, \tfrac{1}{2} \, ( \partial_\mu \varphi )^2 \, .	\vphantom{\Bigg(\Bigg)}	\nonumber
		\end{align}
	In the first step, we used that the covariant derivative reduces to the partial derivative for scalar fields \cite{Carroll:1997ar}. From the second to the third line, we substituted $\tau = R \, \phi$. Lastly, we defined new coordinates ${( x^\mu ) = ( x, \tau )}$ with ${x^1 \equiv x \in ( - \infty, + \infty )}$ and ${x^2 \equiv \tau \in [ 0, \beta )}$.\\

	Analogously, we can consider fermions using the tetrad formalism \cite{Carroll:1997ar}
		\begin{align}
			g_{\alpha \beta} = \tensor{e}{_\alpha^a} \, \tensor{e}{_\beta^b} \, \delta_{a b} \, ,
		\end{align}
	where $g_{\alpha \beta}$ is given by \eqref{eq:metric_cylinder} and $\delta_{a b}$ is the metric of an Euclidean two-dimensional space. For the tetrads, we find
		\begin{align}
			&	( \tensor{e}{_\alpha^a} ) =
			\begin{pmatrix}
				1	&	0
				\\
				0	&	R
			\end{pmatrix} \, ,
			&&	( \tensor{e}{^\alpha_a} ) =
			\begin{pmatrix}
				1	&	0
				\\
				0	&	\tfrac{1}{R}
			\end{pmatrix} \, .
		\end{align}
	Note that this choice is not unique and only defined up to $O(2)$-rotations in the local Euclidean frame.
	
	The covariant Dirac operator is defined via
		\begin{align}
			&	\gamma^a \, \tensor{e}{^\mu_a} \, \nabla_\mu \, ,	&&	\nabla_\mu \equiv \partial_\mu - \tfrac{\mathrm{i}}{4} \, \tensor{\omega}{_\mu^a^b} \, \sigma_{a b} \, ,
		\end{align}
	where
		\begin{align}
			\sigma_{a b} \equiv \tfrac{\mathrm{i}}{2} \, \big[ \gamma_a, \, \gamma_b \big]_- \, ,
		\end{align}
	is the generator of Euclidean Poincar\'e transformations and
		\begin{align}
			\tensor{\omega}{_\mu^a^b} \equiv \tensor{e}{_\nu^a} \, \Gamma^\nu_{\sigma \mu} \, \tensor{e}{^\sigma^b} + \tensor{e}{_\nu^a} \, ( \partial_\mu \tensor{e}{^\nu^b} ) 
		\end{align}
	is the (torsion-free) spin connection and
		\begin{align}
			\Gamma^\nu_{\sigma \mu} = \tfrac{1}{2} \, g^{\nu \alpha} \, ( \partial_\sigma g_{\alpha \mu} + \partial_\mu g_{\sigma \alpha} - \partial_\alpha g_{\sigma \mu} )
		\end{align}
	are the Christoffel symbols of second kind \cite{Christoffel:1869}. It turns out, that all components of the Christoffel symbols as well as all components of the spin-connection vanish, because the metric does not explicitly depend on the coordinates -- the surface of a cylinder is flat.
	
	We conclude, that the kinetic part of the manifest covariant Dirac-action can be rewritten as follows
		\begin{align}
			&	\int_\mathcal{V} \mathrm{d}^2 y \, \sqrt{ \det g} \, \bar{\psi} \, \gamma^a \tensor{e}{^\mu_a} \, \nabla_\mu \, \psi =	\vphantom{\Bigg(\Bigg)}
			\\
			= \, &  \int \int_{0}^{2 \pi} \mathrm{d} x \,  \mathrm{d} \phi \, R \, \bar{\psi} \, \big( \gamma^1 \, \partial_x + \gamma^2 \, \tfrac{1}{R} \, \partial_\phi \big) \psi =	\vphantom{\Bigg(\Bigg)}	\nonumber
			\\
			= \, &  \int \int_{0}^{\beta} \mathrm{d} x \,  \mathrm{d} \tau \, \bar{\psi} \, \big( \gamma^1 \, \partial_x + \gamma^2 \, \partial_\tau \big) \psi =	\vphantom{\Bigg(\Bigg)}	\nonumber
			\\
			= \, & \int \int_{0}^{\beta} \mathrm{d}^2 x \, \bar{\psi} \, \gamma^\mu \partial_\mu \psi \, .	\vphantom{\Bigg(\Bigg)}	\nonumber
		\end{align}
	In total, we find -- as expected -- that QFT-theory on a cylinder is equivalent to QFT in flat Euclidean space-time with (anti-)periodic boundary conditions in the temporal direction.
	
	The limit $\frac{1}{T} = \beta \rightarrow \infty$, thus the zero-temperature-limit is trivial now, because by shifting the periodic $\tau$ coordinate by $- \frac{\beta}{2}$, the only $\beta$ dependent objects are the integral boundaries. We can send $\beta \rightarrow \infty$ to recover Euclidean space-time.

\section{The LPA flow equation at non-zero \texorpdfstring{$\mu$}{mu} and \texorpdfstring{$T$}{mu}}
\label{app:frg}

	In this appendix we present a derivation of the RG flow equation \eqref{eq:pdeq-U} of the effective potential $U ( t, \sigma )$ at non-zero $\mu$ and $T$ in LPA truncation. Similar derivations can be found elsewhere and ours is solely presented in this appendix for the sake of completeness. Throughout this appendix, we work in momentum space using the conventions from App.~\ref{app:fourier_transformation} and the corresponding Fourier-transformed scale-dependent effective average action \eqref{eq:ansatz}. (We do not indicate fields in momentum space with ``tilde'' symbol in this appendix.)\\
	
	The starting point of the derivation is the ansatz for $\bar{\Gamma}_t [ \bar{\psi}, \psi, \varphi ]$ in Eq.~\eqref{eq:ansatz}. To obtain a flow equation for the scale-dependent effective potential $U ( t, \varphi )$, we need to project $\bar{\Gamma}_t [ \bar{\psi}, \psi, \varphi ]$ onto $U ( t, \varphi )$. This is done by evaluating $\bar{\Gamma}_t [ \bar{\psi}, \psi, \varphi ]$ on a constant background field configuration $\varphi ( \tau, x ) = \sigma$, $\bar{\psi} ( \tau, x ) = 0$, and $\psi ( \tau, x ) = 0$. Applying this projection prescription to the ERG equation \eqref{eq:wetterich}, we find
		\begin{align}
			& \beta ( 2 \pi ) \, \delta ( 0 ) \, \partial_t U ( t, \sigma ) =	\vphantom{\bigg(\bigg)}	\label{eq:wetterich_background_field}
			\\
			= \, & \partial_t \bar\Gamma_t [ \bar{\psi}, \psi, \varphi ] \big|_{\varphi = \sigma, \bar{\psi} = 0, \psi = 0} =	\vphantom{\bigg(\bigg)}	\nonumber
			\\
			= \, & \mathrm{STr} \Big[ \big( \tfrac{1}{2} \, \partial_t R_t \big) \, \big( \bar{\Gamma}^{(2)}_t [ \Phi ] + R_t \big)^{-1} \Big]  \Big|_{\varphi = \sigma, \bar{\psi} = 0, \psi = 0} \, ,	\vphantom{\bigg(\bigg)}	\nonumber
		\end{align}
	where $\beta ( 2 \pi ) \, \delta ( 0 )$ is an infinite volume element, which will also appear in the last line of the equation and thus ultimately cancels.
	
	In order to evaluate the supertrace $\mathrm{STr}$, we have to specify the full inverse two-point function (the full propagator) in field space, evaluated on the background field configuration(!),
		\begin{align}
			& \big( \bar{\Gamma}_t^{(2)} [ \bar{\psi}, \psi, \varphi ] + R_t \big)^{-1} \big|_{\varphi = \sigma, \bar{\psi} = 0, \psi = 0} =	\vphantom{\Bigg(\Bigg)}	\label{eq:full_propagator}
			\\
			= \, &
			\begin{pmatrix}
				\big( \bar{\Gamma}_t^{\varphi \varphi} + R_t^{\varphi \varphi} \big)^{-1}	&	0										&	0
				\\
				0										&	0										&	\big( \bar{\Gamma}_t^{\bar{\psi} \psi} + R_t^{\bar{\psi} \psi} \big)^{-1}
				\\
				0										&	\big( \bar{\Gamma}_t^{\psi \bar{\psi}} + R_t^{\psi \bar{\psi}} \big)^{-1}	&	0
			\end{pmatrix} \, ,	\nonumber
		\end{align}
	as well as the regulator, which is diagonal in field space (off-diagonal for fermions),
		\begin{align}
			\partial_t R_t = \, &
			\begin{pmatrix}
				\partial_t R_t^{\varphi \varphi}	&	0										&	0
				\\
				0										&	0										&	\partial_t R_t^{\psi \bar{\psi}}
				\\
				0										&	\partial_t R_t^{\bar{\psi} \psi}	&	0
			\end{pmatrix} \, .
		\end{align}
	The entries of these matrices in field space are as follows: The bosonic regulator is diagonal in momentum space and proportional only to the spatial part of the kinetic term of the bosons,
		\begin{align}
			& R_t^{\varphi_2 \varphi_1} ( \omega_{n_2}, p_2; \omega_{n_1}, p_1 ) \equiv	\vphantom{\bigg(\bigg)}	\label{eq:reg_boson}
			\\
			= \, & \beta \, \delta_{- n_2, n_1} \, ( 2 \pi ) \, \delta ( p_2 + p_1 ) \, p_2^2 \, r_\mathrm{b} ( t, p_2 ) \, .	\vphantom{\bigg(\bigg)}	\nonumber
		\end{align}
	Here, $r_\mathrm{b} ( t, p )$ is a bosonic regulator shape function (see below), which suppresses quantum fluctuations below the RG scale $k ( t )$. For the fermions the regulator is also diagonal in momentum space and also proportional only to the spatial kinetic term, but acquires some additional diagonal structure in Dirac and flavor space,	
		\begin{align}
			& R_t^{\bar{\psi}_2 \psi_1} ( \nu_{n_2}, p_2; \nu_{n_1}, p_1 )\tensor{}{^{a_2}_{a_1}} \equiv	\vphantom{\bigg(\bigg)}	\label{eq:reg_fermion}
			\\
			= \, & - \beta \, \delta_{n_2, n_1} \, ( 2 \pi ) \, \delta ( p_2 - p_1 ) \, \mathrm{i} p_2 \, \tensor{(\openone)}{^{f_2}_{f_1}} \, \tensor{(\gamma^1)}{^{a_2}_{a_1}} r_\mathrm{f} ( t, p_2) \, .	\vphantom{\bigg(\bigg)}	\nonumber
		\end{align}
	A fermionic regulator shape function $r_\mathrm{f} ( t, p )$ is specified below and operates similarly to its bosonic counterpart.
	
	In order to obtain the field space entries of the full propagator \eqref{eq:full_propagator}, one has to calculate the full two-point functions in advance. For the bosons it reads,
		\begin{align}
			& \big( \bar{\Gamma}^{\varphi_2 \varphi_1}_t + R_t^{\varphi_2 \varphi_1} \big) ( \omega_{n_2}, p_2; \omega_{n_1}, p_1 ) =	\vphantom{\bigg(\bigg)}	\label{eq:two-point_function_boson}
			\\
			= \, & \beta \, \delta_{- n_2, n_1} \, ( 2 \pi ) \, \delta ( p_2 + p_1 ) \times	\vphantom{\bigg(\bigg)}	\nonumber
			\\
			& \times \big( \omega_{n_2}^2 + p_2^2 \, [ 1 + r_\mathrm{b} ( t, p_2) ] + \partial_\sigma^2 U ( t, \sigma ) \big) \, ,	\vphantom{\bigg(\bigg)}	\nonumber
		\end{align}
	while the fermionic full two-point function is  given by
		\begin{align}
			& \big( \bar{\Gamma}^{\bar{\psi}_2 \psi_1}_t + R_{\mathrm{f}, t}^{\bar{\psi}_2 \psi_1} \big) ( \nu_{n_2}, p_2; \nu_{n_1}, p_1 )\tensor{}{^{f_2}_{f_1}^{a_2}_{a_1}} =	\vphantom{\bigg(\bigg)}	\label{eq:two-point_function_fermion}
			\\
			= \, & - \beta \, \delta_{n_2, n_1} \, ( 2 \pi ) \, \delta ( p_2 - p_1 ) \, \tensor{(\openone)}{^{f_2}_{f_1}}\times	\vphantom{\bigg(\bigg)}	\nonumber
			\\
			& \times \big[ \mathrm{i} ( \nu_{n_2} + \mathrm{i} \mu ) \, \gamma^2 + \mathrm{i} p_2 \, \gamma^1 [ 1 + r_\mathrm{f} ( t, p_2) ] + \tfrac{h \sigma}{\sqrt{N}} \, \openone \big]\tensor{}{^{a_2}_{a_1}} \, .	\vphantom{\bigg(\bigg)}	\nonumber
		\end{align}
	Within the last lines we used the following short-hand notation for derivatives \textit{w.r.t.}\ fields, which are evaluated on the constant background field configuration,
		\begin{align}
			& \bar{\Gamma}_t^{\Phi_2 \Phi_1} ( \omega_{n_2}, p_2; \omega_{n_1}, p_1 )^{BA} \equiv	\vphantom{\Bigg(\Bigg)}
			\\
			\equiv \, & \frac{\beta^2 ( 2 \pi )^2 \delta^2 \bar{\Gamma}_t [ \Phi ]}{\delta \Phi_{2,B} ( \omega_{n_2 }, p_2) \, \delta \Phi_{1,A} ( \omega_{n_1 }, p_1)} \bigg|_{\varphi = \sigma, \bar{\psi} = 0, \psi = 0} \, .	\vphantom{\Bigg(\Bigg)}	\nonumber
		\end{align}
		
	Now it is straight forward to invert the full two-point functions \eqref{eq:two-point_function_boson} and \eqref{eq:two-point_function_fermion} separately in momentum, Dirac, and flavor space. The bosonic propagator reads
		\begin{align}
			& \big( \bar{\Gamma}^{\varphi_2 \varphi_1}_t + R_t^{\varphi_2 \varphi_1} \big)^{-1} ( \omega_{n_2}, p_2; \omega_{n_1}, p_1 ) =	\vphantom{\Bigg(\Bigg)}
			\\
			= \, & \beta \, \delta_{- n_2, n_1} \, ( 2 \pi ) \, \delta ( p_2 + p_1 ) \times	\vphantom{\Bigg(\Bigg)}	\nonumber
			\\
			& \times \frac{1}{\omega_{n_2}^2 + p_2^2 [ 1 + r_\mathrm{b} ( t, p_2) ] + \partial_\sigma^2 U ( t, \sigma )} \, .	\vphantom{\Bigg(\Bigg)}	\nonumber
		\end{align}
	For the fermionic propagator we find
		\begin{align}
			& \big( \bar{\Gamma}^{\bar{\psi}_2 \psi_1}_t + R_{\mathrm{f}, t}^{\bar{\psi}_2 \psi_1} \big)^{-1} ( \nu_{n_2}, p_2; \nu_{n_1}, p_1 )\tensor{}{^{f_2}_{f_1}^{a_2}_{a_1}} =	\vphantom{\Bigg(\Bigg)}
			\\
			= \, & \beta \, \delta_{n_2, n_1} \, ( 2 \pi ) \, \delta ( p_2 - p_1 )  \, \tensor{(\openone)}{^{f_2}_{f_1}} \times	\vphantom{\Bigg(\Bigg)}	\nonumber
			\\
			& \times \frac{\big[ \mathrm{i} ( \nu_{n_2} + \mathrm{i} \mu ) \, \gamma^2 + \mathrm{i} p_2 \, \gamma^1 [ 1 + r_\mathrm{f} ( t, p_2) ] - \tfrac{h \sigma}{\sqrt{N}} \, \openone \big]\tensor{}{^{a_2}_{a_1}}}{( \nu_{n_2} + \mathrm{i} \mu )^2 + p_2^2 [ 1 + r_\mathrm{f} ( t, p_2 ) ]^2 + \frac{( h \, \sigma )^2}{N}} \, .	\vphantom{\Bigg(\Bigg)}	\nonumber
		\end{align}
	For all fermionic objects with reverse order of field space indices $( \bar{\psi} \psi \leftrightarrow \psi \bar{\psi} )$, one has to use the corresponding transposed objects and take care of sign flips due to their Grassmann nature.
	
	Combining all previous results, we can evaluate the field space trace in Eq.~\eqref{eq:wetterich_background_field} as well as the traces in Dirac and flavor space. Traces in momentum space are given by one-dimensional momentum-space integrals and Matsubara sums. By comparing coefficients on the \textit{l.h.s.}\ and \textit{r.h.s.}\ we eliminate the infinite volume factor and the flow equation ultimately reads
		\begin{align}
			& \partial_t U ( t, \sigma ) =	\vphantom{\Bigg(\Bigg)}	\label{eq:flow_equation_after_traces}
			\\
			= \, & + \frac{1}{\beta} \sum_n \int \frac{\mathrm{d} p}{( 2 \pi )} \, \frac{p^2 \, [ \frac{1}{2} \, \partial_t r_\mathrm{b} ( t, p ) ]}{\omega_n^2 + p^2 [ 1 + r_\mathrm{b} ( t, p ) ] + \partial_\sigma^2 U ( t, \sigma )} -	\vphantom{\Bigg(\Bigg)}	\nonumber
			\\
			& - \frac{1}{\beta} \sum_n \int \frac{\mathrm{d} p}{( 2 \pi )} \, \frac{2 \, d_\gamma N \, p^2 \,  [ \frac{1}{2} \, \partial_t r_\mathrm{f} ( t, p ) ] [ 1 + r_\mathrm{f} ( t, p ) ]}{ ( \nu_n + \mathrm{i} \mu )^2 + p^2 [ 1 + r_\mathrm{f} ( t, p ) ]^2 + \frac{( h \, \sigma )^2}{N} } \, ,	\vphantom{\Bigg(\Bigg)}	\nonumber
		\end{align}
	with the number of flavors $N$ and the dimension ${ d_\gamma = \mathrm{tr} ( \openone_{d_\gamma} ) = 2 }$ of the matrix representation of the Clifford algebra \eqref{eq:app_clifford}. It remains to specify appropriate regulator shape functions $r_\mathrm{b}$ and $r_\mathrm{f}$.
	
	We use so-called one-dimensional LPA-optimized regulator shape functions \cite{Litim:2000ci,Litim:2001up} (sometimes also denoted as \textit{Litim} or \textit{flat} regulators). The corresponding bosonic regulator shape function is chosen to be
		\begin{align}
			r_\mathrm{b} ( t, p ) \equiv \Big[ \tfrac{k^2 ( t )}{p^2} - 1 \Big] \, \Theta \Big( \tfrac{k^2 ( t )}{p^2} - 1 \Big) \, ,	\label{eq:regulator_shape_function_boson}
		\end{align}
	while the fermionic regulator shape function is defined via its bosonic counterpart,
		\begin{align}
			1 + r_\mathrm{f} ( t, p ) \equiv \sqrt{ 1 + r_\mathrm{b} ( t, p ) } \, ,	\label{eq:regulator_shape_function_fermion}
		\end{align}
	where we used the notion of RG time and the relation \eqref{eq:def_rg_time} between it and the RG scale $k ( t )$. The advantage of this choice of the fermionic regulator shape function \eqref{eq:regulator_shape_function_fermion} is that the numerator of the fermionic loop contribution in Eq.~\eqref{eq:flow_equation_after_traces} can be simplified in terms of the bosonic regulator,
		\begin{align}
			2 \, [ \partial_t r_\mathrm{f} ( t, p ) ] \, [ 1 + r_\mathrm{f} ( t, p ) ] = \partial_t r_\mathrm{b} ( t, p ) \, ,	\label{eq:derivative_regulator_fermion}
		\end{align}
	Hence, the bosonic and fermionic loop-integrals in Eq.~\eqref{eq:flow_equation_after_traces} are regulated at the same RG-scales in a unified manner. Additionally, we need
		\begin{align}
			& \partial_t r_\mathrm{b} ( t, p ) =	\vphantom{\bigg(\bigg)}	\label{eq:derivative_regulator_boson}
			\\
			= \, & - 2 \, \tfrac{k^2 ( t )}{p^2} \, \Big[ \Theta \Big( \tfrac{k^2 ( t )}{p^2} - 1 \Big) + \Big( \tfrac{k^2 ( t )}{p^2} - 1 \Big) \, \delta \Big( \tfrac{k^2 ( t )}{p^2} - 1 \Big) \Big] \, .	\vphantom{\bigg(\bigg)}	\nonumber
		\end{align}
	Inserting Eqs.~\eqref{eq:regulator_shape_function_boson} - \eqref{eq:derivative_regulator_boson} in Eq.~\eqref{eq:flow_equation_after_traces}, we can analytically evaluate the momentum loop-integral and obtain
		\begin{align}
			\partial_t U ( t, \sigma ) = \, & - \frac{1}{\pi} \, \frac{1}{\beta} \sum_n \, \frac{k^3 ( t )}{\omega_n^2 + E_\mathrm{b}^2 ( t, \sigma )} +	\vphantom{\Bigg(\Bigg)}
			\\
			& + \frac{d_\gamma N}{\pi} \, \frac{1}{\beta} \sum_n \frac{k^3 ( t )}{ ( \nu_n + \mathrm{i} \mu )^2 + E_\mathrm{f}^2 ( t, \sigma )}	\vphantom{\Bigg(\Bigg)}.	\nonumber
		\end{align}
	The Matsubara sums can be evaluated analytically using standard techniques of contour integration. After the $\tfrac{1}{N}$-rescalings of Eq.~\eqref{eq:rescaling_with_n} we finally obtain RG flow equation \eqref{eq:pdeq-U} for the scale-dependent effective potential in LPA at non-zero $T$ and non-zero $\mu$.

\section{Zero temperature calculations}
\label{app:vac_flow_L2D}

	In this appendix we present results complementary to the ones discussed in Sub.Sec.~\ref{subsec:vacFiniteN_LD1}. Instead of using the spatial, one-dimensional LPA-optimized (Litim) regulator (L1D) we switch to a two-dimensional one for the vacuum flows here. We again focus our attention to $N=2$. In Figs.~\ref{fig:flow_L2D_N=2,T=0.0,mu=0.0} and \ref{fig:k_L2D_N=2,T=0.0,mu=0.0} we present vacuum RG flows obtained by using the two-dimensional LPA-optimized (Litim) regulator (L2D) and consequently the flow Eq.~\eqref{eq:2dim_litim_flow_equation} (the $\sigma$-derivative of it). 
	
	We use the same initial condition~\eqref{eq:initial-condition} which we have used in the rest of the paper also to initialize the flows in this appendix. While it is possible to repeat the construction leading to the initial condition~\eqref{eq:initial-condition} using a two-dimensional regulator in vacuum, we nevertheless stick to the version obtained using a one dimensional regulator. The reason for this approach is that we want to compute on constant lines of UV physics when comparing the flows obtained with the L1D and L2D regulators. 
	
	Comparing the L1D flows of Figs.~\ref{fig:flow_L1D_N=2,T=0.0,mu=0.0} and \ref{fig:k_L1D_N=2,T=0.0,mu=0.0} to the L2D flows of Figs.~\ref{fig:flow_L2D_N=2,T=0.0,mu=0.0} and \ref{fig:k_L2D_N=2,T=0.0,mu=0.0}. We note that the L2D flow has a very short apparent dynamical range of arguably only one order of magnitude starting at $k(t)\approx 10^0$ while the corresponding L1D flow shows dynamics over more than four orders of magnitude again staring at $k(t)\approx 10^0$, see the convergence rates $|\partial_t m_\sigma^2(k)|$ in Figs. \ref{fig:k_L1D_N=2,T=0.0,mu=0.0} and \ref{fig:k_L2D_N=2,T=0.0,mu=0.0}. The condensate is non-vanishing in the IR (at $k_\mathrm{IR}=10^{-4}$) in both cases and has the same order of magnitude as the mean-field value $\sigma_0=10^0$, where the value of ${\sigma_\mathrm{min}\approx0.757}$ from L2D the computation is slightly smaller than the one obtained with L1D namely $\sigma_\mathrm{min}\approx 0.907$. For the squared curvature mass $m_\sigma^2$ in the IR we observe a large quantitative difference: with L1D $m_\sigma^2= 1.04 \cdot 10^{-5}$ while computations with L2D yield $m_\sigma^2 = 2.69\cdot 10^{-1}$  which is four orders of magnitude larger.
	
	Quantitative differences between the L1D and L2D flows are to be expected when using truncated (here in LPA truncation) FRG flow equations, see also Ref.~\cite{Pawlowski:2017gxj}. The important statement for the main part of this work is however independent of the regularization: we find $\mathbb{Z}_2$ symmetry breaking in vacuum signaled by a non-vanishing $\sigma_\mathrm{min}$ with $m_\sigma^2 > 0$. The small dynamic range of the L2D flow indicate that $k_\mathrm{IR}=10^{-4}$ is sufficiently small to include all relevant long-range fluctuations. This is arguably not the case for the L1D flows.
	
	\begin{figure}
		\centering
		\includegraphics{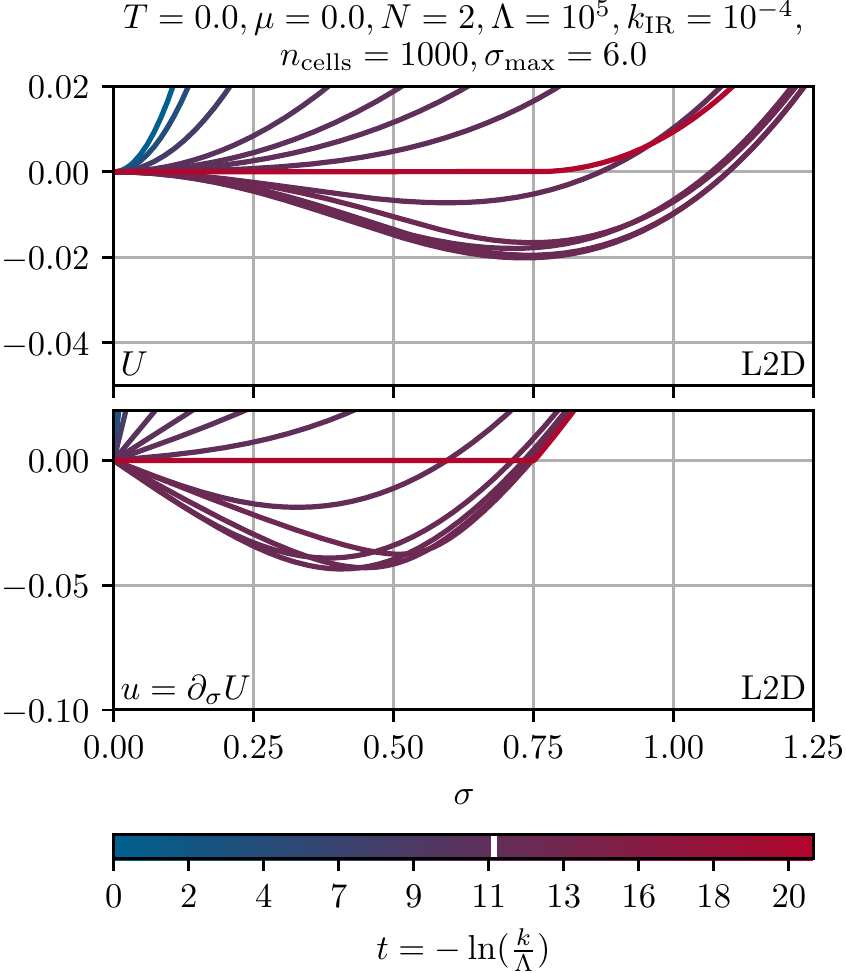}
		\caption{\label{fig:flow_L2D_N=2,T=0.0,mu=0.0}
			RG flow of the scale dependent effective potential $U ( t, \sigma )$ (upper panel) and its $\sigma$-derivative (the fluid) $u ( t, \sigma ) = \partial_\sigma U ( t, \sigma )$ (lower panel) from the UV ({blue}) to the IR ({red}) in vacuum ($T = \mu =0$). For the sake of simplicity (and using the (anti-)symmetry in $\sigma$) the functions $u ( t, \sigma )$ and $U ( t, \sigma )$ are plotted for positive $\sigma$ only. The different RG-times are encoded via the colored bar-legend below the plots. The white vertical line in the colored bar-legend denotes the RG time (scales) when the $\mathbb{Z}_2$ symmetry is broken (condensation). We do not find symmetry restoration in vacuum for $N = 2$ within the RG flow for $k\geq k_\mathrm{IR}=10^{-4}$. Results were obtained using the two-dimensional LPA-optimized (Litim) regulator. All other parameters are stated in the figure.
		}
	\end{figure}

	\begin{figure}
		\centering
		\includegraphics{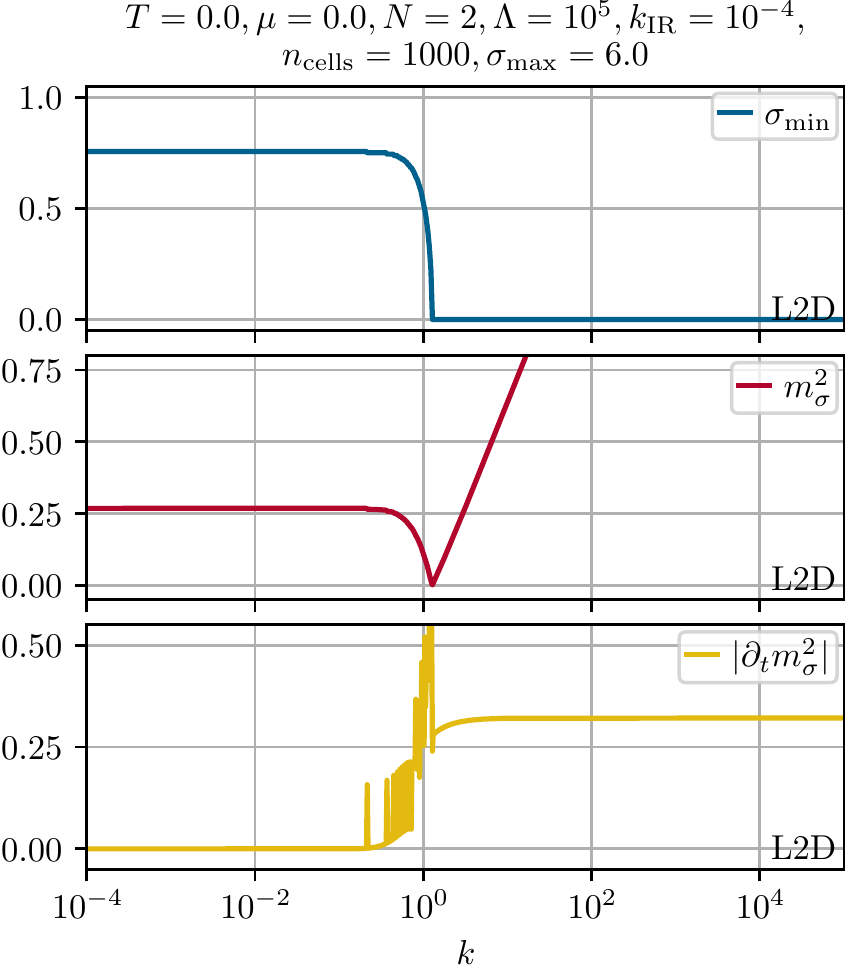}
		\caption{\label{fig:k_L2D_N=2,T=0.0,mu=0.0}
			RG flow of the minimum $\sigma_{\mathrm{min}} ( k )$ of the scale dependent effective potential $U ( k, \sigma )$ (upper panel), the squared curvature mass $m_\sigma^2 ( k ) = \partial_\sigma u ( t, \sigma )$ (middle panel), and the relative change of the squared curvature mass $| \partial_t m_\sigma^2 ( k ) |$ (lower panel) according to Eq.~\eqref{eq:changing_rate_mass} plotted as functions of the RG scale $k ( t )$ in vacuum ($T = \mu =0$). The plot corresponds to the RG flows of $u ( t, \sigma )$ and $U ( t, \sigma )$ of Fig.~\ref{fig:flow_L2D_N=2,T=0.0,mu=0.0}.
		}
	\end{figure}

\section{Numerical tests}
\label{app:numerical_tests}

	In this appendix, we present numerical tests for the choice of the (numeric) parameters, which are used for the calculations in the main part of this work. These analysis involve tests of the UV-cutoff independence (RG consistency) \cite{Braun:2018svj}, but also tests of numerical parameters like the size of the computational domain $[ 0, \sigma_\mathrm{max} ]$ as well as the spatial resolution $\Delta \sigma$ (the size of the volume cells in the finite volume scheme), compare Sub.Sec.~\ref{subsec:numerical_implementation} and Ref.~\cite{Koenigstein:2021syz}. These tests are crucial to ensure that (numeric) artifacts and (numerical) errors can be quantified and ultimately reduced as much as possible.
	
	Of course, we do not have any exact reference values at finite $N$, which can be used as benchmark results to precisely quantify relative (numerical) errors for the GNY model, similar to what is done in Refs.~\cite{Grossi:2019urj,Koenigstein:2021syz,Koenigstein:2021rxj,Steil:2021cbu,Steil:2021partIV,Keitel:2011pn}. Still, we can take reasonable IR observables, like the square of the $\sigma$-curvature mass $m_\sigma^2 = \partial_\sigma u ( t, \sigma ) \big|_{\sigma_\mathrm{min}}$ at the IR minimum $\sigma_\mathrm{min}$, and study their (in-)dependence of the (numerical) parameters. To this end, we fix all (numerical) parameters with the exception of one single parameter and steadily vary the latter. Thereby, we look out for parameter regions, where our observables do not change as functions of the parameter, which implies independence of the observable from the parameter. Hereby, we assume that correlation effects between the parameters are small in the respective regions. The same process is performed for the dependencies on $\Lambda$, $\Delta \sigma$, and $\sigma_\mathrm{max}$ and repeated at various points in the $\mu$-$T$-plane for $N = 2$.  Finally, we fix our numerical parameters for our main calculations, by choosing parameter sets, which are on plateaus, where changing the (numerical) parameters does not influence our IR observable anymore, for all tests.\\
	
	We are aware that this does not definitively ensures that all numerical errors are quantified for the problem and does not provide hard upper bounds on numerical errors. Still, it seems to be a reasonable procedure to increase the credibility of our findings and to give some insight into the achievable numerical precession of our setup. Furthermore, it may help our readers to judge the reliability and reproducibility of our results and is therefore presented in detail in this appendix.\\
	
	Within this work, we use the time-stepper \texttt{solve\_ivp} with the \texttt{LSODA} option using an Adams/BDF method with automatic stiffness detection and switching from the \textit{SciPy~1.0} library \cite{2020SciPy-NMeth}. We also crosschecked some of our results with the code-basis of Refs.~\cite{Koenigstein:2021syz,Koenigstein:2021rxj,Steil:2021cbu,Steil:2021partIV}, which is implemented with \textit{Wolfram Mathematica} \cite{Mathematica:12.1} using \texttt{NDSolve}. All results were obtained specifying relative and absolute tolerances of $10^{-10}$ ($10^{-12}$) for the solvers for in medium (vacuum) calculations.
	
\subsection{Size of the computational domain}
\label{subsec:test_computational_domain}

	The first test of our numerical parameters comprises the examination of the (in-)dependence of our IR results from the size of the computational domain $[ 0, \sigma_\mathrm{max} ]$. This test is important, because the RG flow equation is originally formulated as a PDE on an infinite spatial domain for $\sigma \in ( - \infty, \infty )$, which is a pure initial value problem. For practical purposes we argued in Sub.Sec.~\ref{subsec:boundary_conditions} that it is possible to (1.) formulate the problem only on the domain $[ 0, \infty )$ by employing the $\mathbb{Z}_2$ symmetry (including the introduction of an artificial new boundary condition at $\sigma = 0$) and (2.) to restrict the calculation to a compact domain $[0, \sigma_\mathrm{max}]$, because of the shape of $u ( t, \sigma )$ does not change for large $\sigma$ anyhow, and introducing another artificial boundary condition at $\sigma_\mathrm{max}$. Here, we demonstrate that it is possible to choose sufficiently large $\sigma_\mathrm{max}$ in order to get rid of errors due to the artificial large-$\sigma$ boundary conditions.
	
	For these tests, we choose the size of the volume cells (the spatial resolution) $\Delta \sigma = 0.006$, the UV cutoff $\Lambda = 10^5$, the IR-cutoff $k_\mathrm{IR} = 10^{-4}$. Within the next subsection we demonstrate that these are decent choices. We performed these tests at different points in the $\mu$-$T$-plane in regions, where we find symmetry breaking and restoration during the RG flow, in order to be sure that all kinds of effects are reasonably resolved by the choice of the numerical parameters.\\
	
\begin{figure}
			\centering
			\includegraphics{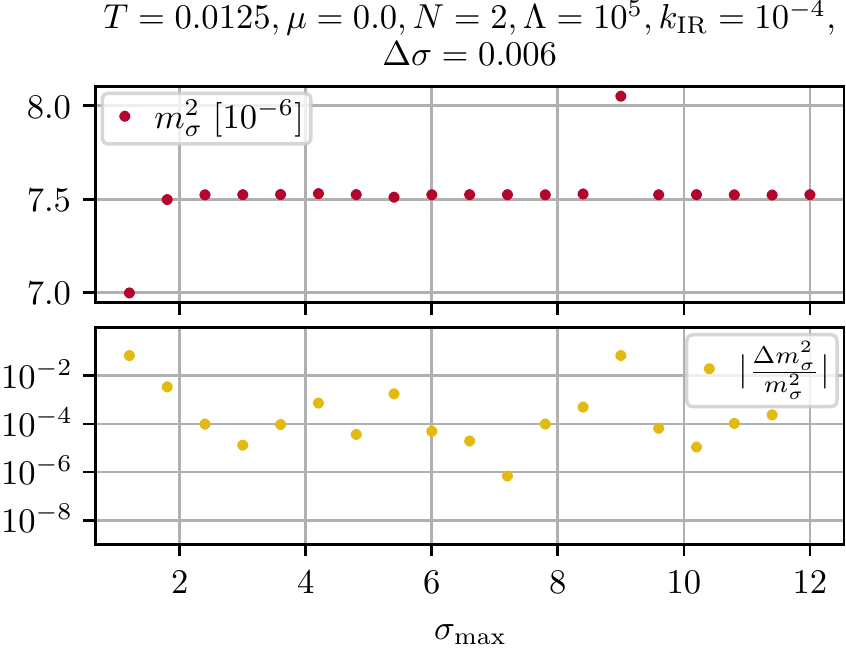}
			\caption{\label{fig:x_max_test_T=0.0125_mu=0.0}
				IR squared curvature mass $m_\sigma^2$ (upper panel) and the relative deviations of the IR curvature mass according to Eq.~\eqref{eq:mass_deviations_1} (lower panel) as functions of the size of the computational domain $\sigma_\mathrm{max}$. Tests were performed at $T = 0.0125$ and $\mu = 0.0$ with $N = 2$. All other numerical parameters are provided in the plots.
			}
		\end{figure}

\begin{figure}
			\centering
			\includegraphics{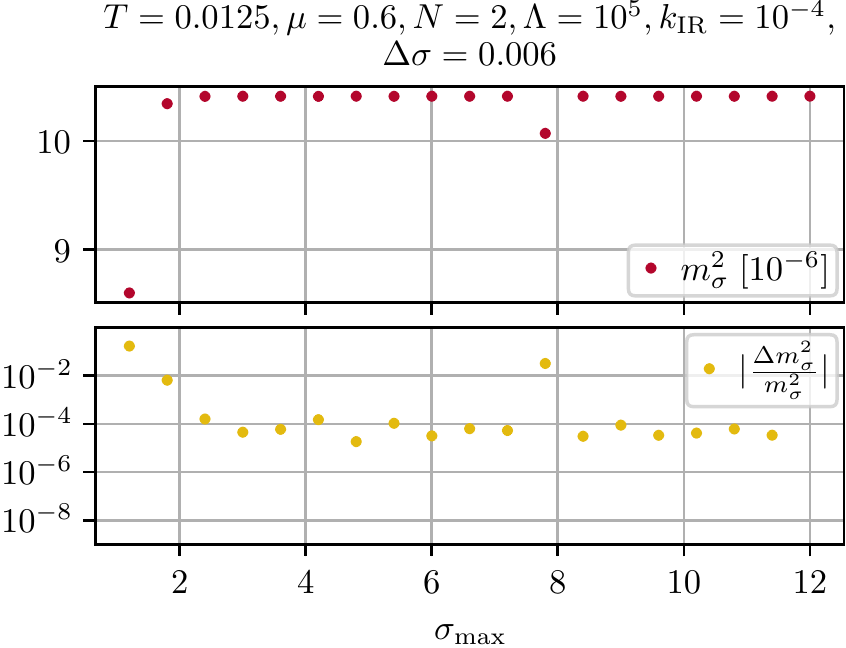}
			\caption{\label{fig:x_max_test_T=0.0125_mu=0.6}
				Same as Fig.~\ref{fig:x_max_test_T=0.0125_mu=0.0} but tests were performed at $T = 0.0125$ and $\mu = 0.6$.
			}
		\end{figure}

\begin{figure}
			\centering
			\includegraphics{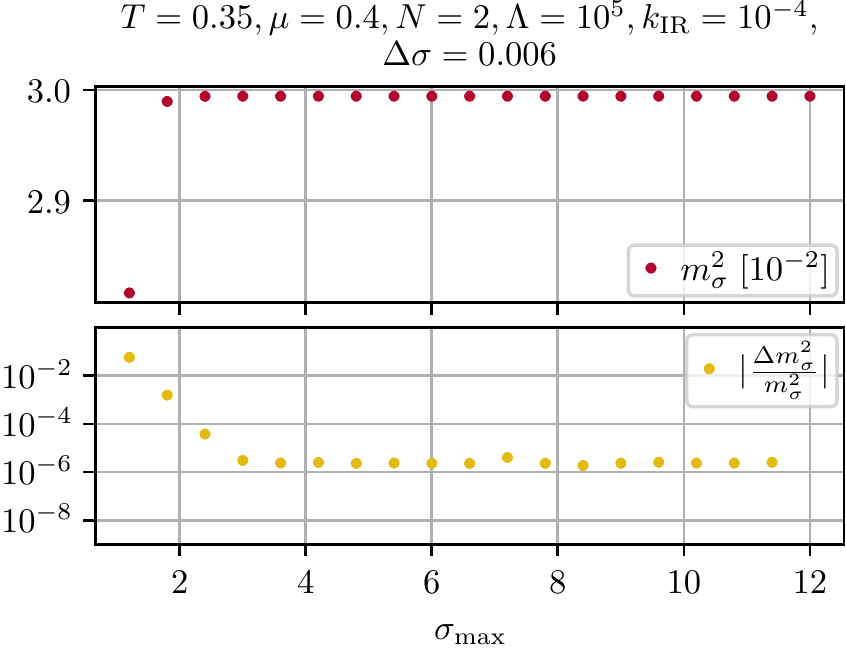}
			\caption{\label{fig:x_max_test_T=0.35_mu=0.4}
				Same as Fig.~\ref{fig:x_max_test_T=0.0125_mu=0.0} but tests were performed at $T = 0.35$ and $\mu = 0.4$.
			}
		\end{figure}

\begin{figure}
			\centering
			\includegraphics{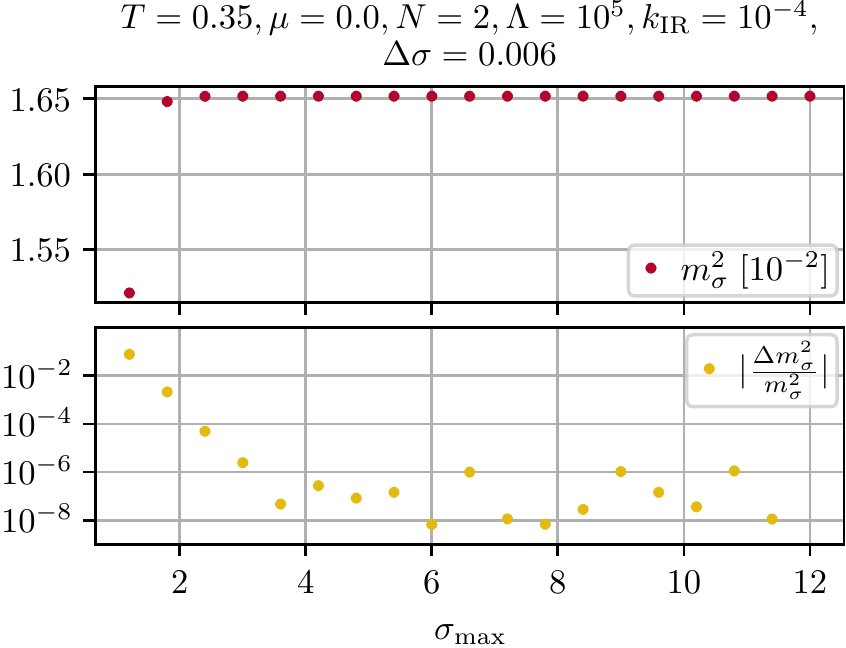}
			\caption{\label{fig:x_max_test_T=0.35_mu=0.0}
				Same as Fig.~\ref{fig:x_max_test_T=0.0125_mu=0.0} but tests were performed at $T = 0.35$ and $\mu = 0.0$.
			}
		\end{figure}

	In Figs.~\ref{fig:x_max_test_T=0.0125_mu=0.0}, \ref{fig:x_max_test_T=0.0125_mu=0.6}, \ref{fig:x_max_test_T=0.35_mu=0.4}, and \ref{fig:x_max_test_T=0.35_mu=0.0} we present the results of our tests. For all test points, we find that the IR minimum is always located exactly at $\sigma_\mathrm{min} = 0$ (in the volume cell $\sigma_0$). In the upper panels the square of the IR curvature mass $m_\mathrm{\sigma}^2$ at the IR minimum is plotted as a function of the computational interval size $\sigma_\mathrm{max}$. Each point represents a single RG flow for the corresponding parameter set. In the lower panels we plot the relative deviations of $m_\sigma^2$ for any $\sigma_\mathrm{max}$ from the $m_\sigma^2$ for the largest computational domain $\sigma_\mathrm{max} = 12$,
		\begin{align}
			\tfrac{\Delta m^2_\sigma}{m^2_\sigma} ( \sigma_\mathrm{max, i}) \equiv \bigg| \frac{m^2_\sigma ( \sigma_{\mathrm{max},i} ) - m^2_\sigma ( \sigma_\mathrm{max} = 12 )}{m^2_\sigma ( \sigma_\mathrm{max} = 12 )} \bigg| \, .	\label{eq:mass_deviations_1}
		\end{align}
	Hence, the lower panels can be seen as an rough estimate for the order of the relative errors -- at least in the regions where the points in the upper panels form a plateau. 
	
	Overall we find that it is possible to drastically reduce the influence of the size of the computational interval on the IR curvature mass by choosing sufficiently large $\sigma_\mathrm{max} \gtrsim 4$, where we reach the plateaus. We can infer from the plots that the relevant reference scale for a decent choice of $\sigma_\mathrm{max}$ seems to be the position of the minimum $\sigma_\mathrm{min} ( t )$ during the RG flow, which is maximally $\approx 1$ in the infinite-$N$ limit. From Figs.~\ref{fig:x_max_test_T=0.0125_mu=0.0}, \ref{fig:x_max_test_T=0.0125_mu=0.6}, \ref{fig:x_max_test_T=0.35_mu=0.4}, and \ref{fig:x_max_test_T=0.35_mu=0.0} we conclude that in order to reduce computational time\footnote{Keeping the resolution $\Delta \sigma$ fixed while increasing $\sigma_\mathrm{max}$ increases the number of volume cells $n$ and thus increases computational cost.} without loss of accuracy, we fix $\sigma_\mathrm{max} = 6$ as a rather conservative choice for all other calculations.
	
\subsection{Spatial resolution}

	Analogously to the previous paragraphs, we provide tests for the choice of a decent spatial resolution (the size of the finite volume cells) $\Delta \sigma$. Again, we keep all other parameters fixed, \textit{i.e.}, $\Lambda = 10^5$, $\sigma_\mathrm{max} = 6$, and $k_\mathrm{IR} = 10^{-4}$ and perform the tests at the same four points in the $\mu$-$T$-plane. The need for a proper spatial resolution is self-explanatory. Using too large $\Delta \sigma$ means that important effects during the RG flow cannot be resolved adequately and the precise extraction of IR observables is impossible. On the other hand, excessively small $\Delta \sigma$ might introduce artificial errors due to numerical operations to close to the machine or time-stepper precision. Furthermore decreasing $\Delta \sigma$ while keeping $\sigma_\mathrm{max} = 6$ fixed increases the number of volume cells and with it the computational time.

\begin{figure}[!htpb]
			\centering
			\includegraphics{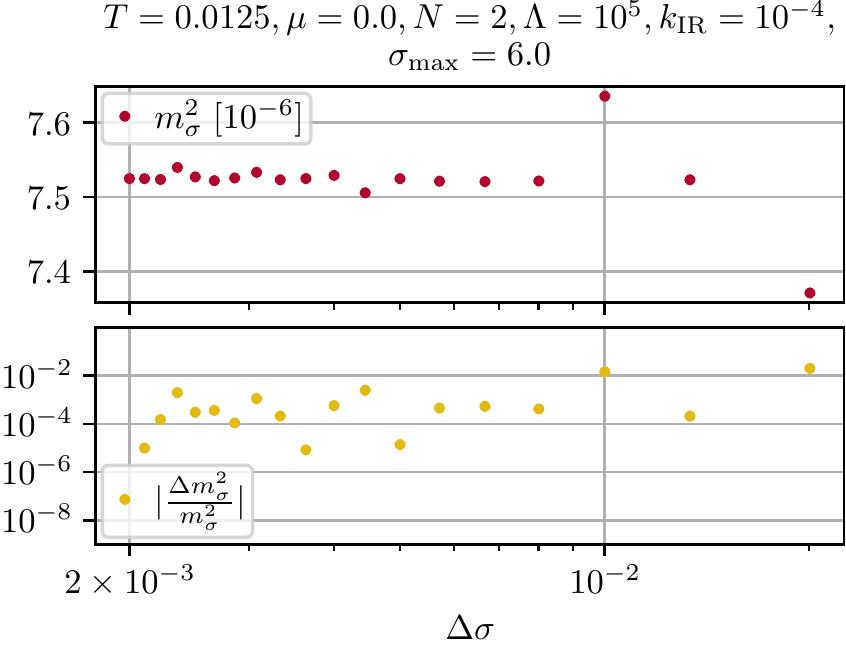}
			\caption{\label{fig:delta_x_test_T=0.0125_mu=0.0}
				IR squared curvature mass $m_\sigma^2$ (upper panel) and the relative deviations of the IR curvature mass according to Eq.~\eqref{eq:mass_deviations_2} (lower panel) as functions of the spatial resolution (the size of the volume cells) $\Delta \sigma$. Tests were performed at $T = 0.0125$ and $\mu = 0.0$ with $N = 2$. All other numerical parameters are provided in the plots.
			}
		\end{figure}

\begin{figure}[!htpb]
			\centering
			\includegraphics{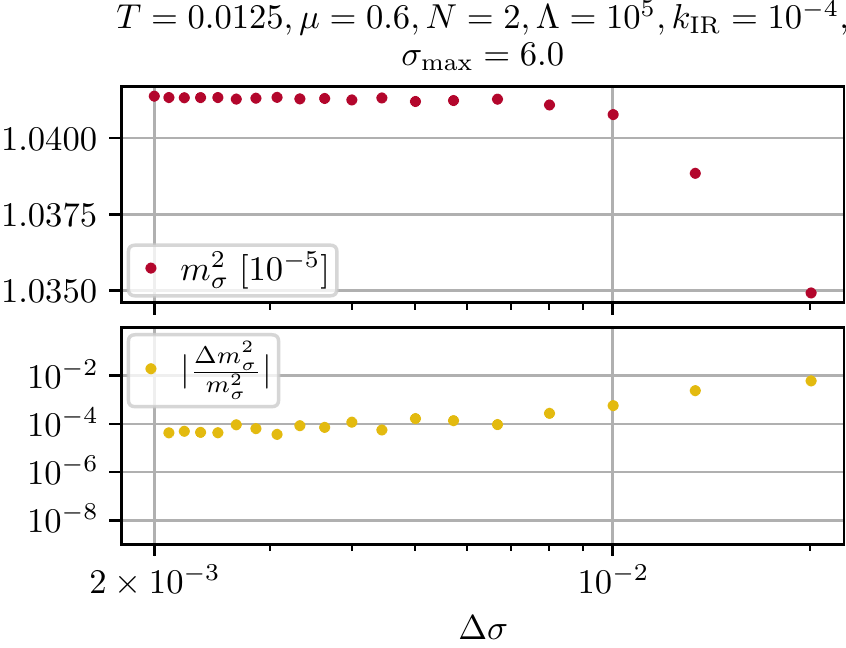}
			\caption{\label{fig:delta_x_test_T=0.0125_mu=0.6}
				Same as Fig.~\ref{fig:delta_x_test_T=0.0125_mu=0.0} but tests were performed at $T = 0.0125$ and $\mu = 0.0$.
			}
		\end{figure}

\begin{figure}[!htpb]
			\centering
			\includegraphics{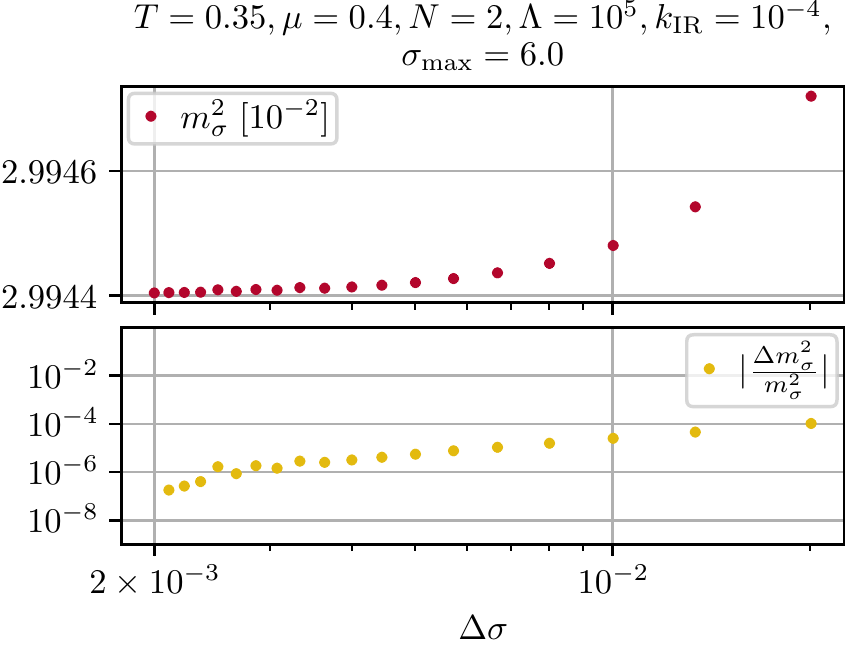}
			\caption{\label{fig:delta_x_test_T=0.35_mu=0.4}
				Same as Fig.~\ref{fig:delta_x_test_T=0.0125_mu=0.0} but tests were performed at $T = 0.35$ and $\mu = 0.4$.
			}
		\end{figure}

\begin{figure}[!htpb]
			\centering
			\includegraphics{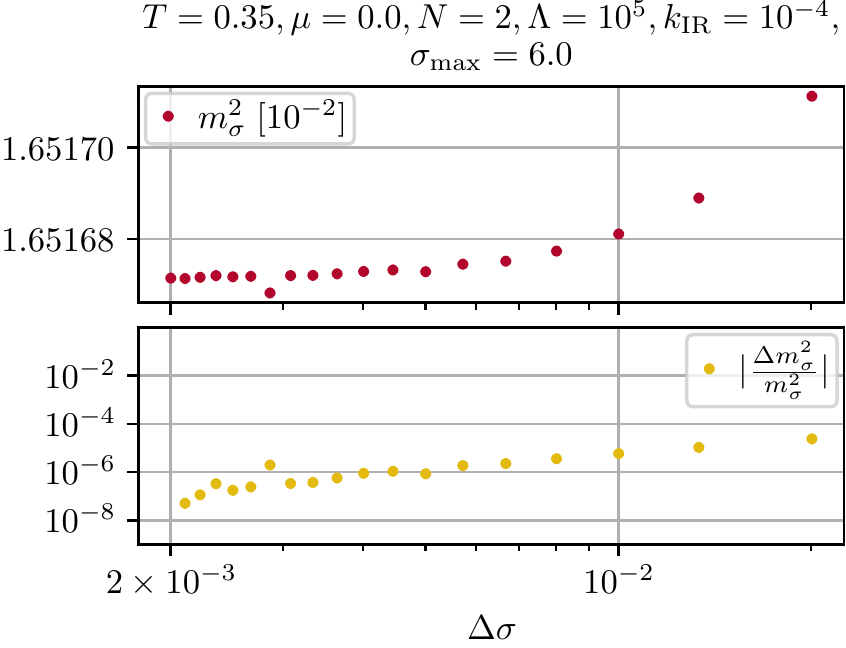}
			\caption{\label{fig:delta_x_test_T=0.35_mu=0.0}
				Same as Fig.~\ref{fig:delta_x_test_T=0.0125_mu=0.0} but tests were performed at $T = 0.35$ and $\mu = 0.0$.
			} 
		\end{figure}

	In Figs.~\ref{fig:delta_x_test_T=0.0125_mu=0.0}, \ref{fig:delta_x_test_T=0.0125_mu=0.6}, \ref{fig:delta_x_test_T=0.35_mu=0.4}, and \ref{fig:delta_x_test_T=0.35_mu=0.0} we present the results of our $\Delta \sigma$-dependence tests. The plots are analogous to the plots from the previous tests, except for the abscissa, which now labels $\Delta \sigma$. For the lowest panels we again used
		\begin{align}
			\tfrac{\Delta m^2_\sigma}{m^2_\sigma} ( \Delta \sigma_i ) \equiv \bigg| \frac{m^2_\sigma ( \Delta \sigma_i ) - m^2_\sigma ( \Delta \sigma_\mathrm{min} )}{m^2_\sigma ( \Delta \sigma_\mathrm{min} )} \bigg| \, ,	\label{eq:mass_deviations_2}
		\end{align}
	with
		\begin{align}
			\Delta \sigma_\mathrm{min} = \tfrac{\sigma_\mathrm{max}}{n_\mathrm{max} - 1} = \tfrac{6}{3000 - 1} \, .
		\end{align}
	
	From all tests we find, that it is generically possible to increase the precision of the results by lowering $\Delta \sigma$. Furthermore, we find that we do not run into problems of floating point arithmetic or insufficient time-stepper precision for all choices of $\Delta \sigma$. As compromise between run time and resolution, we choose $\Delta \sigma = \frac{6}{1000 - 1} \simeq 0.006$ for all other calculations. This choice presents as a very fine resolution which should be sufficient to captures all significant effects of the systems under consideration. 
		

\subsection{UV-cutoff independence}
\label{app:uv-cutoff_independence}

\begin{figure}[!htpb]
			\centering
			\includegraphics{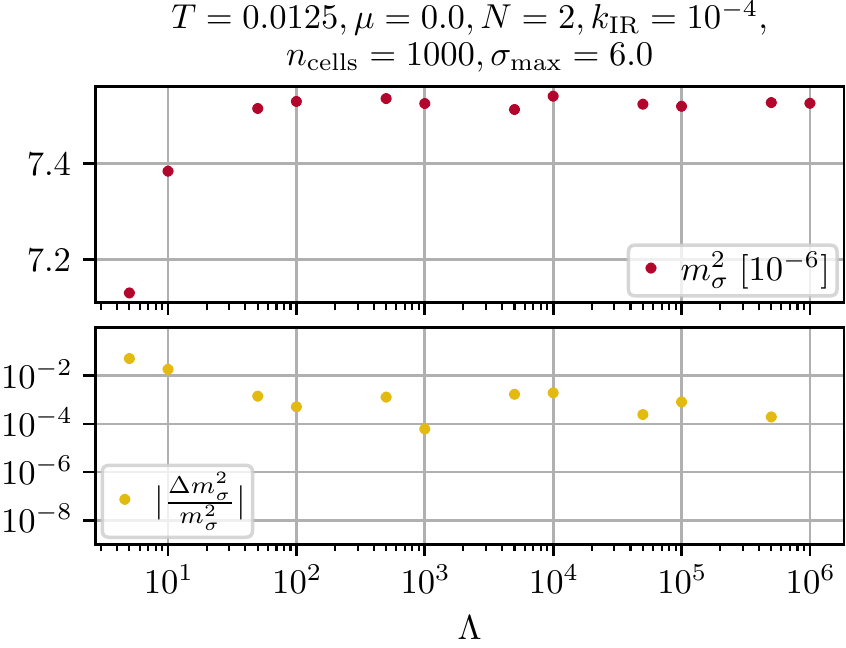}
			\caption{\label{fig:cutoff_test_T=0.0125_mu=0.0}
				IR squared curvature mass $m_\sigma^2$ (upper panel) and the relative deviations of the IR curvature mass according to Eq.~\eqref{eq:mass_deviations_3} (lower panel) as functions of the UV cutoff $\Lambda$. Tests were performed at $T = 0.0125$ and $\mu = 0.0$ with $N = 2$. All other numerical parameters are provided in the plots.
			}
		\end{figure}

\begin{figure}[!htpb]
			\centering
			\includegraphics{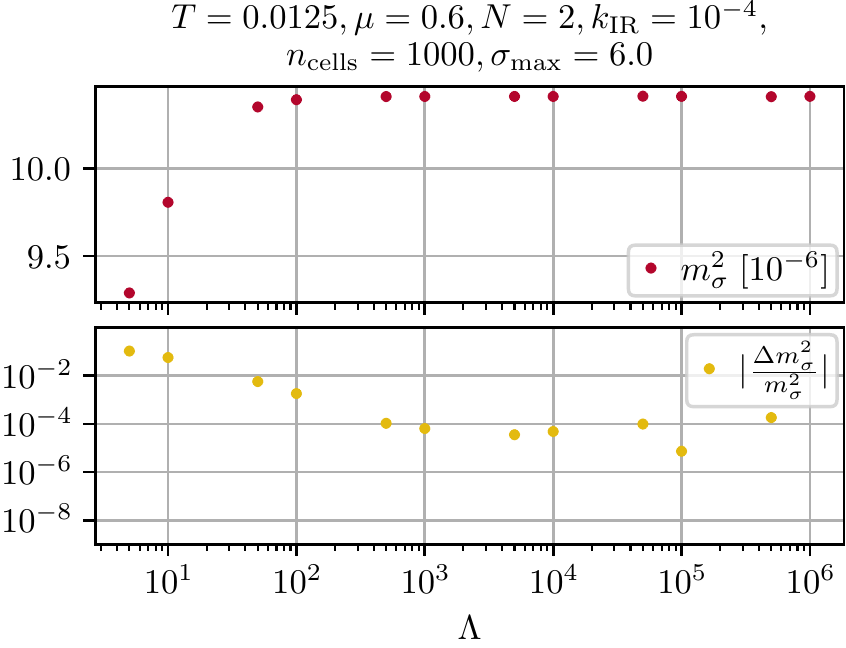}
			\caption{\label{fig:cutoff_test_T=0.0125_mu=0.6}
				Same as Fig.~\ref{fig:cutoff_test_T=0.0125_mu=0.0} but tests were performed at $T = 0.0125$ and $\mu = 0.0$.
			}
		\end{figure}

\begin{figure}[!htpb]
			\centering
			\includegraphics{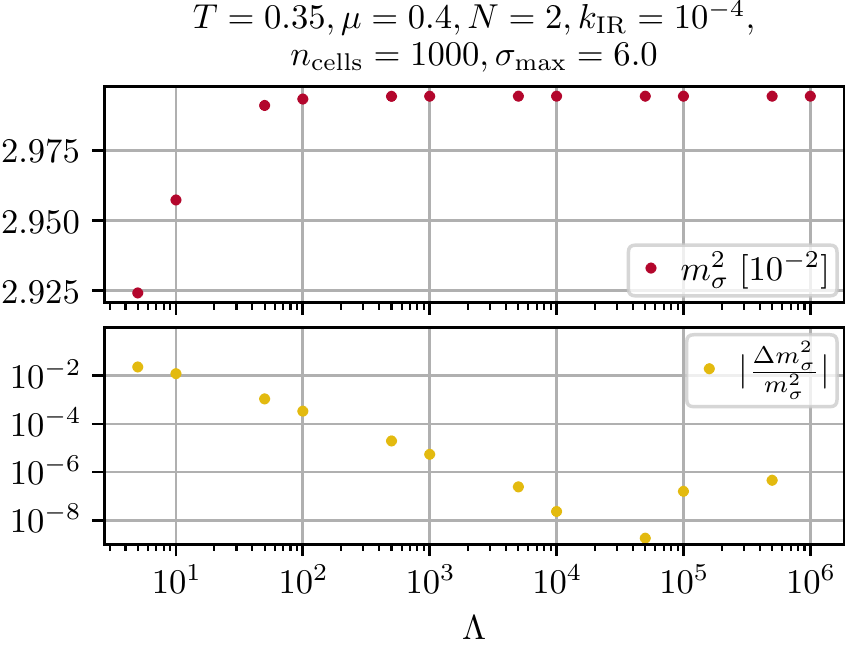}
			\caption{\label{fig:cutoff_test_T=0.35_mu=0.4}
				Same as Fig.~\ref{fig:cutoff_test_T=0.0125_mu=0.0} but tests were performed at $T = 0.35$ and $\mu = 0.4$.
			}
		\end{figure}

\begin{figure}[!htpb]
			\centering
			\includegraphics{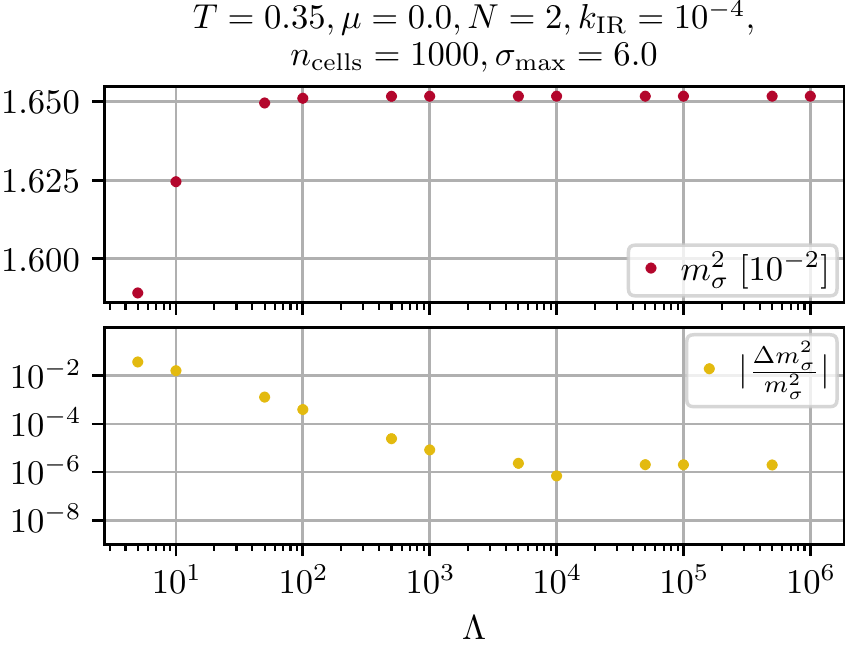}
			\caption{\label{fig:cutoff_test_T=0.35_mu=0.0}
				Same as Fig.~\ref{fig:cutoff_test_T=0.0125_mu=0.0} but tests were performed at $T = 0.35$ and $\mu = 0.0$.
			}
		\end{figure}

	The last tests within this appendix comprise the search for sufficient UV initial scales $\Lambda$. As discussed in detail in Refs.~\cite{Braun:2018svj,Koenigstein:2021syz} a sufficiently large choice of the UV initial scale is crucial to practically ensure RG consistency, hence $\Lambda$-independence of IR observables. To this end we used the same four points in the $\mu$-$T$-plane as before and gradually increased the UV cutoff $\Lambda$. Thereby we searched for values of $\Lambda$, where the IR observables, here the IR squared curvature mass $m_\sigma^2$, do not change anymore, hence
		\begin{align}
			\Lambda \, \tfrac{\mathrm{d}}{\mathrm{d}\Lambda} \, m_\sigma^2 ( \Lambda ) = 0 \, .	\label{eq:rg_consistency}
		\end{align}
	From Figs.~\ref{fig:cutoff_test_T=0.0125_mu=0.0}, \ref{fig:cutoff_test_T=0.0125_mu=0.6}, \ref{fig:cutoff_test_T=0.35_mu=0.4}, and \ref{fig:cutoff_test_T=0.35_mu=0.0} we can read of, that this seems to be the case for $\Lambda \gtrsim 10^4$, where either Eq.~\eqref{eq:rg_consistency} is fulfilled or at least other numerical errors completely dominate. For the calculations in the main part of this work we use $\Lambda = 10^5$.
	
	The lower panels of the plots are calculated via
		\begin{align}
			\tfrac{\Delta m^2_\sigma}{m^2_\sigma} ( \Lambda_i ) \equiv \bigg| \frac{m^2_\sigma ( \Lambda_i ) - m^2_\sigma ( \Lambda = 10^6 )}{m^2_\sigma ( \Lambda = 10^6 )} \bigg| \, .	\label{eq:mass_deviations_3}
		\end{align}
	These results are in accordance with the (RG) consistency tests of Sub.Sec.~\ref{subsec:numeric_consistency_check_mean-field}.


\section{Discretization schemes for the source/sink term}
\label{app:source_sink_implementation}

	In this appendix we discuss two possible discretization schemes for the source/sink term. To arrive at these schemes, we return to the initial idea of the finite volume discretization. Employing a finite volume discretization, \textit{cf.}\ Refs.~\cite{LeVeque:1992,LeVeque:2002,RezzollaZanotti:2013,Ames:1992} and references therein, involves the study of the integral form of the flow equation \eqref{eq:pdeq-u} for each volume cell separately. For the $i^\text{th}$ cell this reads
		\begin{align}
			& \int_{\sigma_i - \frac{\Delta \sigma}{2}}^{\sigma_i + \frac{\Delta \sigma}{2}} \mathrm{d} \sigma \, \partial_t u ( t, \sigma ) =	\label{finite_volume_scheme}
			\\
			= \, & \int_{\sigma_i - \frac{\Delta \sigma}{2}}^{\sigma_i + \frac{\Delta \sigma}{2}} \mathrm{d} \sigma \, \big[ \tfrac{\mathrm{d}}{\mathrm{d} \sigma} \, Q ( t, \partial_\sigma u ) + S ( t, \sigma ) \big] \, .	\nonumber
		\end{align}
	Keeping the control volumes fixed, one identifies the integral on the \textit{l.h.s.}\ with the $t$-derivative  of the cell averages of the fluid $\partial_t \bar{u}_i ( t )$ times $\Delta \sigma$. The integral over the $\sigma$-derivative of the diffusion flux on the \textit{r.h.s.} is approximated as stated in Sub.Sec.~\ref{subsec:numerical_implementation}. For the integral over the source/sink term $S ( t, \sigma )$ we basically consider two options.
	\begin{enumerate}
		\item	The first option (if $S ( t, \sigma )$ does not have any special structure) is an approximation. One approximates the source/sink term with $S ( t, \sigma_i )$ at the cell center times the cell-volume $\Delta \sigma$. Ignoring the diffusive contribution, the semi-discrete flow equation for the cell-averages $\bar{u}_i ( t )$ reads,
			\begin{align}
				\partial_t \bar{u}_i ( t ) = S ( t, \sigma_i ) \, ,	\label{eq:source_sink_1}
			\end{align}
		where we divided by $\Delta \sigma$.
		
		\item	The second option, which is due to a special feature of our RG flow equation, is, to make use of the fact that the source/sink term in the flow equation \eqref{eq:pdeq-u} presents as a spatial derivative of some function $s ( t, \sigma )$, which solely depends on $t$ and $\sigma$,
			\begin{align}
				S ( t, \sigma ) = \tfrac{\mathrm{d}}{\mathrm{d} \sigma} \, s ( t, \sigma ) \, .	\label{eq:source_2}
			\end{align}
		This means that the integral on the \textit{r.h.s.}\ of Eq.~\eqref{finite_volume_scheme} can be calculated exactly, by evaluating $s ( t, \sigma )$ on the cell surfaces $\sigma_{i + \frac{1}{2}}$. This results in,
			\begin{align}
				\partial_t \bar{u}_i ( t ) = \tfrac{1}{\Delta \sigma} \, \big[ s \big( t, \sigma_{i + \frac{1}{2}} \big) - s \big( t, \sigma_{i - \frac{1}{2}} \big) \big] \, ,	\label{eq:source_sink_2}
			\end{align}
		where we again ignored the diffusion flux and divided by $\Delta \sigma$.
	\end{enumerate}
	At first sight, it seems better to use the second and exact version. However, during our calculations, we did not experience any differences in precision between both versions for $T \neq 0$, as long as $\Delta \sigma$ is not too large. Nevertheless, concerning the runtime, the first version turned out preferable in practical computations. Eventually, this might be related to the fact, that, ignoring bosonic fluctuations (no diffusion), the first version reduces exactly to the mean-field calculation for $u ( t, \sigma )$ at positions $\sigma_i$, which can be directly seen from Eq.~\eqref{eq:source_sink_1}, where the PDE reduces into decoupled differential equations at the $\sigma_i$. For fluid dynamic problems of more than one spatial direction (more than one condensate in FRG), option one seems preferable anyhow, because it becomes more and more challenging to evaluate the source/sink on the higher-dimensional cell interfaces.
	
	Though, for $T = 0$ the analytic evaluation of the $\sigma$-derivative would produce Dirac-delta distributions via the Theta Heaviside function, see Eq.~\eqref{eq:zero_t_flow_equation}. Delta-peaks are extremely complicated to implement in a numerical setup, but they are important at $T=0$ and should not be disregarded. We therefore suggest to use the second version \eqref{eq:source_sink_2} (although it can only be used for FRG-mean-field calculations at $T=0$ for $u ( t, \sigma )$, which do not suffer from the problems described in Sub.Sec.~\ref{subsec:chemical_potential_shock_wave}).\\
	
	Nevertheless, we believe that there is some need for further investigations on the best discretization schemes for such fermionic contributions, see also Refs.~\cite{Grossi:2021ksl}.

\FloatBarrier

%\nocite{*}
%\bibliography{bib/general,bib/numerics,bib/rg,bib/gn} % Produces the bibliography via BibTeX -- separated bibfiles for working versions.
\bibliography{gross_neveu} % Produces the bibliography via BibTeX -- unified bibfile for online versions.

\end{document}